\documentclass{aastex61}

\shortauthors{Ebrahimi, Karami and Soler}

\usepackage{amsmath}
\usepackage{amssymb}
\usepackage{amsfonts}
\usepackage{latexsym}
\usepackage{mathtools}
\usepackage{empheq}
\usepackage{graphicx}

\begin{document}

\title{The effect of a twisted magnetic field on the phase mixing of the kink magnetohydrodynamic waves in coronal loops}

\correspondingauthor{Zanyar Ebrahimi}
\email{z.ebrahimi@uok.ac.ir}

\author{Zanyar Ebrahimi}
\affil{Department of Physics, University of Kurdistan, Pasdaran Street, P.O. Box 66177-15175, Sanandaj, Iran}

\author{Kayoomars Karami}
\affiliation{Department of Physics, University of Kurdistan, Pasdaran Street, P.O. Box 66177-15175, Sanandaj, Iran}

\author{Roberto Soler}
\affiliation{Departament de F\'{\i}sica, Universitat de les Illes Balears, E-07122, Palma de Mallorca, Spain}
\affiliation{Institut d'Aplicacions Computacionals de Codi Comunitari ($IAC^3$), Universitat de les Illes Balears, E-07122, Palma de Mallorca, Spain}

\begin{abstract}

There are observational evidences for the existence of twisted magnetic field in the solar corona. This inspires us to investigate the effect of a twisted magnetic field on the evolution of magnetohydrodynamic (MHD) kink waves in coronal loops. To this aim, we solve the incompressible linearized MHD equations in a magnetically twisted nonuniform coronal flux tube in the limit of long wavelengths. Our results show that a twisted magnetic field can enhance or diminish the rate of phase-mixing of the Alfv\'{e}n continuum modes and the decay rate of the global kink oscillation depending on the twist model and the sign of the longitudinal ($k_z$) and azimuthal ($m$) wavenumbers. Also our results confirm that in the presence of a twisted magnetic field, when the sign of one of the two wavenumbers $m$ and $k_z$ is changed, the symmetry with respect to the propagation direction is broken. Even a small amount of twist can have an important impact on the process of energy cascade to small scales.

\end{abstract}

\keywords{Sun: corona --- Sun:
magnetic fields --- Sun: oscillations}

\section{Introduction} \label{Introduction}
Transverse oscillations of the solar coronal loops are one of the greatest seismological tools to extract or approximate the unknown parameters of the solar corona such as the magnetic field, the plasma density and the transport coefficients. Aschwanden et al. (1999) and Nakariakov et al. (1999) were first to report the observation of the transverse oscillations in the coronal loops using the Transition Region and Coronal Explorer (TRACE) telescope on 1998 July 14 in the 171-{\AA} Fe IX emission lines. Nakariakov et al. (1999) indicated that the oscillations are strongly damped and the ratio of the damping time to the period of the oscillation is around 3-5. This observation was identified as a standing kink MHD wave in a magnetic flux tube. However, the mechanism proposed in Nakariakov et al. (1999) to explain the damping involved the assumption of unrealistically large diffusion coefficients. A more satisfactory physical interpretation of the damping was given by Ruderman \& Roberts (2002).

 Among the suggested mechanisms responsible for the strong damping of the coronal loop oscillations (e.g. Ruderman and Roberts 2002; Ofman 2005, 2009; Morton and Erd\'{e}lyi 2009), resonant absorption of the MHD waves, that was established first by Ionson (1978), is a strong candidate. Several works developed this theory (e.g. Davila 1987; Sakurai, Goossens \& Hollweg 1991a,b; Goossens et al. 1995; Goossens \& Ruderman 1995; Erd\'{e}lyi 1997; Cally \& Andries 2010). The necessary condition for the resonant absorption is a continuum of Alfv\'{e}n or slow frequency across the loop (Ionson 1978; Hollweg 1984, 1987; Davila 1987; Sakurai, Goossens \& Hollweg 1991a). Resonant absorption occurs when the frequency of the global MHD mode matches at least with one of the frequencies of the background Alfv\'{e}n or slow continuum at a location called resonance point. As a result, the energy of the global MHD mode transfers to the local Alfv\'{e}n modes in a layer around the resonance point, named resonance layer (Lee \& Roberts 1986; see also Goossens et al. 2013; Soler \& Terradas 2015). In the absence of dissipation mechanisms, the amplitude of the oscillations diverges at the resonance point. Dissipation is important in the resonance layer where the oscillations make large gradients. The background Alfv\'{e}n or slow continuum can be due to the variation of the plasma density (e.g. Davila 1987; Ofman, Davila \& Steinolfson 1994; Ruderman \& Roberts 2002; Terradas, Oliver \& Ballester 2006; Soler \& Terradas 2015), twisted magnetic field (Ebrahimi \& Karami 2016) or both of them together (Karami \& Bahari 2010; Giagkiozis et al. 2016). There are a variety of theoretical works related to the damping of the coronal loop oscillations based on the theory of resonant absorption of MHD waves (e.g. Ruderman \& Roberts 2002; Goossens, Andries \& Aschwanden 2002; Van Doorsselaere et al. 2004; Andries et al. 2005; Terradas, Oliver \& Ballester 2006; Goossens et al. 2009; Karami, Nasiri \& Amiri 2009; Karami \& Bahari 2010; Soler et al. 2013; Soler \& Terradas 2015; Ebrahimi \& Karami 2016; Jung Yu \& Van Doorsselaere 2016; Giagkiozis et al. 2016). For a good review about the theory of resonant absorption, see also Goossens et al. (2011).

Ruderman \& Roberts (2002) studied the resonant absorption of kink waves in coronal loops. They suggested that only the loops with transverse density inhomogeneities on a small scale compared to the loop thickness are able to support coherent oscillations and consequently become observable. Safari et al. (2006) investigated the resonant absorption of MHD waves in coronal loops and found that as the longitudinal mode number increases, the maximum value of the wave amplitude moves away from the inhomogeneous region towards the loop axis and as a result the efficiency of the process of resonant absorption decreases.

 Goossens et al. (2014) investigated the transverse and torsional motions of MHD kink waves in coronal loops. They showed that the kink waves are not just transverse motions of coronal loops, but the velocity field of the kink waves involves both the transverse and torsional motions. Soler \& Terradas (2015) (hereafter ST2015) investigated the evolution of the MHD kink wave in a coronal loop by solving an initial value problem. Inspired by Cally (1991), they showed that the MHD kink wave can be expressed as a superposition of Alfv\'{e}n continuum modes. They showed that in the presence of an Alfv\'{e}n frequency continuum made by the variation of the plasma density across the loop, the energy of the global kink wave transfers to the phase mixed azimuthal perturbations of the local Alfv\'{e}n waves in the inhomogeneous layer.

An interesting property of the coronal structures is that they can have a twisted magnetic field. Chae et al. (2000) stated that in order to have torsional motions in coronal loops, the magnetic field of the loop should be twisted around the loop axis. Chae \& Moon (2005) assumed that the
constriction of plasma (i.e. $\partial p/\partial r\neq 0$, where $p$ is the plasma pressure) is due to the magnetic tension of the
azimuthal component of the magnetic field. Using this, they found that for a specific observed coronal loop the magnetic twist on the loop axis is about $1.5 \pi$. The existence of magnetic field twist in coronal structures has been reported in several observations (e.g. Kwon \& Chae 2008; Aschwanden et al. 2012; Thalmann et al. 2014; Wang et al. 2015). For instance, Kwon \& Chae (2008) using the TRACE 171 {\AA} observations in several coronal loops,  reported that the number of twist turns, $N_{twist}$, have values in the range [$0.11,~0.87$]. These values, are small enough to let a typical coronal loop to be kink stable (for more details, see section \ref{model}).

There are ample theoretical works on the role of the magnetic twist in the MHD oscillations of the coronal loops (e.g. Bennett, Roberts \& Narain 1999; Erd\'{e}lyi \& Carter 2006; Erd\'{e}lyi \& Fedun 2006, 2007, 2010; Carter \& Erd\'{e}lyi 2008; Ruderman 2007, 2015; Karami \& Barin 2009; Karami \& Bahari 2010, 2012; Terradas \& Goossens 2012; Ruderman \& Terradas 2015; Ebrahimi \& Karami 2016).

Sakurai, Goossens \& Hollweg (1991a) investigated resonant absorption in twisted flux tubes and obtained jump conditions of the perturbations across the resonance layer. Using the jump conditions, there is no need to solve dissipative MHD equations in the resonance layer and one can connect ideal MHD solutions of the left and the right sides of the resonance layer (see also Sakurai, Goossens \& Hollweg 1991b; Goossens, Hollweg \& Sakurai 1992). Karami \& Bahari (2010) studied the effect of a twisted magnetic field on the resonant absorption of MHD waves in a coronal flux tube. They showed that when the amount of the magnetic twist is increased, the frequency, the damping rate and the ratio of the frequency to the damping rate increase and the period ratio of the fundamental mode to the first overtone mode decreases from its canonical value. Terradas \& Goossens (2012) investigated the MHD kink oscillations of the coronal loops in the presence of magnetic field twist. Solving the MHD equations numerically, they showed that in the presence of magnetic twist for a given value of longitudinal wavenumber, $k_z$, the quasi-mode frequency of MHD kink waves has different values for different signs of the azimuthal mode number ($m=\pm1$). They found that when $k_z>0$, the frequency for $m=1$ should be larger than that of the solution without twist. Conversely, the frequency for $m=-1$ should be smaller. They also showed that for a given value of the azimuthal mode number ($m=-1$ or $m=+1$), the frequency for different signs of $k_z$ has different values. Hence, the magnetic twist breaks the symmetry of the phase speed of the MHD kink wave with respect to the propagation direction. As a result, in the presence of magnetic twist, the standing MHD kink oscillation with line-tying boundary conditions at the footpoints of the loop cannot be Fourier-analyzed in azimuthal and longitudinal directions. Ruderman (2015) called the modes corresponding with $m=+1,~-1$ in a twisted flux tube, accelerated and decelerated kink wave, respectively. Ruderman \& Terradas (2015) investigated the standing MHD kink oscillations of thin twisted magnetic tubes. They found that depending on the value of the plasma density ratio of the interior and exterior of the loop, the period ratio of the first overtone to the fundamental kink mode can be increased or decreased by increasing the magnetic twist in the loop. They also showed that in the presence of magnetic twist, in general, the eigenmodes of the MHD kink oscillations have elliptical polarization. Recently, in the thin tube thin boundary (TTTB) approximation, Ebrahimi \& Karami (2016) analytically showed that the resonant absorption of kink MHD wave in a coronal flux tube with constant densities inside and outside the loop, can occur owing to the existence of a twisted magnetic field around the loop axis. They showed that when the ratio of the azimuthal to axial component of the background magnetic field increases, the frequency and the damping rate of the kink waves increase and the ratio of the frequency to the damping rate decreases. They found that with magnetic twist values in the range of observational values, the ratio of the damping time to the period of the oscillation is in good agreement with the observations.

Another consequence of existence of Alfv\'{e}n continuum across the loop is phase-mixing of Alfv\'{e}n waves, in which the oscillations of neighboring field lines become rapidly out of phase. This phenomenon leads to enhanced viscous and ohmic dissipations (Heyvaerts \& Priest 1983).
Phase-mixing may occur either spatially in a propagating wave or in time in a standing wave. In both cases, an Alfv\'{e}n wave is excited on each field line, which has an independent oscillation from its neighbors with a frequency in the Alfv\'{e}n continuum.
Phase-mixing is an essential ingredient of resonant absorption (Poedts 2002) that causes a cascade of energy to small length scales, where the dissipation mechanisms become more efficient. By studying an initial-value problem we can gain insight into some of the interesting features that a nonuniform medium brings. However, we do not consider dissipation in our work and only are interested in the initial stage of the phase-mixing before the dissipation becomes important.

The main goal of the current paper is to explore the temporal evolution of kink waves in twisted flux tubes. To do so, we add an azimuthal component to the background magnetic field of the model of ST2015 and compare the results obtained in the presence of the magnetic twist with the results of ST2015 (no magnetic twist). To achieve this aim, in section \ref{model} we introduce the equations of motion and the flux tube model. In section \ref{solution}, we solve an initial-value problem for the obtained equation of motion using the technique developed by ST2015. In section \ref{results}, we present numerical results. Finally, section \ref{Conclusions} is devoted to our conclusions.\\

\section{Equations of motion and model}\label{model}

The linearized ideal MHD equations for an incompressible plasma read
\begin{equation}\label{mhd1}
     \rho(r)\frac{\partial^2 \boldsymbol{\xi}}{\partial t^2}=-\nabla \delta p+\frac{1}{\mu_0}\{(\nabla\times\delta{\mathbf B})\times{\mathbf B} +(\nabla\times{\mathbf B})\times\delta{\mathbf B}\},
\end{equation}
\begin{equation}\label{mhd2}
    \mathbf{\delta B}=\nabla\times(\boldsymbol{\xi}\times\mathbf{B}),
\end{equation}
\begin{equation}\label{mhd3}
    \nabla\cdot\boldsymbol{\xi}=0,
\end{equation}
where $\boldsymbol{\xi}$ is the Lagrangian displacement of the
plasma, $\mathbf{\delta B}$ and $\delta p$ are the Eulerian perturbations of the
magnetic field and plasma pressure, respectively. Here $\mu_0$ is the magnetic permeability
of the free space. Note that Eq. (\ref{mhd3}) shows the incompressibility condition, which we adopt for the simplicity of calculations. Although the solar corona in general is a compressible medium, Goossens et al. (2009) elaborated that in the thin tube approximation that is applicable to the problem of long-wavelength transverse oscillations of coronal loops, kink waves are almost incompressible to a high degree of accuracy. They showed that the compressibility of the kink mode is proportional to $(k_z R)^2$. Hence, in the long-wavelength limit the frequency and damping rate of the kink mode are the same in both compressible and incompressible cases. The same result of Goossens et al. (2009) was previously obtained by Edwin \& Roberts (1983), who explained that the kink mode behaves as an incompressible wave in the slender tube limit. Therefore, in order to apply the results of incompressible kink waves to the corona in what follows, we restrict our calculations to the limit of long-wavelength kink modes.

Using Eq. (\ref{mhd3}), we can rewrite Eq. (\ref{mhd2}) as
\begin{equation}\label{deltaB}
    \mathbf{\delta B}=(\mathbf{B}\cdot\nabla)\boldsymbol{\xi}-(\boldsymbol{\xi}\cdot\nabla)\mathbf{B}.
\end{equation}
Putting Eq. (\ref{deltaB}) into (\ref{mhd1}) and doing some algebra yields
\begin{equation}\label{mhd4}
     \rho(r)\frac{\partial^2 \boldsymbol{\xi}}{\partial t^2}=-\nabla \delta P+\frac{1}{\mu_0}\left[(\mathbf{B}\cdot\nabla)(\mathbf{B}\cdot\nabla)\boldsymbol{\xi}-(\boldsymbol{\xi}\cdot\nabla)(\mathbf{B}\cdot\nabla)\mathbf{B}\right],
\end{equation}
where $\delta P=\delta p+(\delta \mathbf{B}\cdot \mathbf{B})/\mu_0$ is the Eulerian perturbation of the total (gas plus magnetic) pressure.

We model a typical coronal loop by a straight cylinder that has a
circular cross section of radius $R$. The background
plasma density in cylindrical coordinates ($r$, $\varphi$, $z$) is
assumed to be as follows
\begin{equation}\label{rho}
\rho(r)=\left\{\begin{array}{lll}
    \rho_{{\rm i}},&r\leqslant r_1,\\
    \frac{\rho_{{\rm i}}}{2}\left[\left(1+\frac{\rho_{{\rm e}}}{\rho_{{\rm i}}}\right)-\left(1-\frac{\rho_{{\rm e}}}{\rho_{{\rm i}}}\right)\sin\left(\frac{\pi}{l}(r-R)\right)\right],&r_1< r< r_2,\\
    \rho_{{\rm e}},&r\geqslant r_2,
      \end{array}\right.
\end{equation}
where $r_1=R-l/2$ and $r_2=R+l/2$. Here, $l=r_2-r_1$ is the characteristic length of
the radial variation of the background plasma density. The density ratio $\rho_{\rm i}/\rho_{\rm e}$ is very difficult to estimate from observations. Typical values of this parameter are believed to be in the range $\rho_i/\rho_e = 2-10$ (Aschwanden et al. 2003). The background magnetic field is assumed to be constant and aligned with the flux tube axis in the regions $r\leqslant r_1$ and $r\geqslant r_2$, but it is twisted around the $z$-axis in the region $r_1<r<r_2$,

\begin{equation}\label{B}
    \mathbf{B}(r)=\left\{\begin{array}{lll}
     B_{0z}\hat{z},&r\leqslant r_1,&\\
    B_{0\varphi}(r)\hat{\phi}+B_{0z}\hat{z},&r_1< r< r_2,&\\
    B_{0z}\hat{z},&r\geqslant r_2,&\\
      \end{array}\right.
\end{equation}
where $B_{0z}$ is constant. We should note that in order to satisfy the magnetohydrostatic equation of motion,
\begin{equation}
    \frac{1}{\mu_0}(\nabla\times{\mathbf B})\times{\mathbf B}-\nabla p=0,
\end{equation}
the background magnetic field must be non-force-free, i.e. $$(\nabla\times{\mathbf B})\times{\mathbf B}=\hat{r}\frac{B_{0\varphi}}{r}\frac{\partial}{\partial r}(r B_{0\varphi})\neq 0.$$ This yields
\begin{equation}
B_{0\varphi}\neq \frac{C}{r},
\end{equation}
where $C$ is a constant of integration. Therefore, in the model presented here, we can consider any profile for the azimuthal component of the background magnetic field other than $B_{0\varphi}\propto r^{-1}$.

Since the equilibrium quantities are only functions of $r$, the perturbations can be Fourier-analyzed with respect to the $\varphi$ and $z$ coordinates. Hence,
\begin{eqnarray}\label{pert}
\delta P=\delta P(r,t)~e^{i(m\varphi+k_z z)},\label{pert1}\\
\boldsymbol{\xi}=\boldsymbol{\xi}(r,t)~e^{i(m\varphi+k_z
z)}\label{pert2}\nonumber,
\end{eqnarray}
where $m$ and $k_z$ are the azimuthal and axial wavenumbers, respectively. Therefore, we are not considering the case of standing oscillations line-tied at the ends of the tube (see Terradas \& Goossens 2012). Instead, we are implicitly considering propagating waves with fixed values of $k_z$ and $m$. The case of line-lied oscillations is more difficult to tackle analytically, and a fully numerical approach is generally required. That is beyond the aim of the present work. Inserting perturbations (\ref{pert1}) into Eqs. (\ref{mhd3}) and (\ref{mhd4}) and eliminating $\xi_\varphi$ and $\xi_z$, gives $\delta P$ in terms of $\xi_r$ as
\begin{equation}\label{p}
    \delta P=\frac{1}{k_z^2+m^2/r^2}\mathcal{L}\left(-\frac{1}{r}\frac{\partial(r\xi_r)}{\partial r}\right)+\frac{m/r}{k_z^2+m^2/r^2}f(r)\xi_r,
\end{equation}
where
\begin{equation}\label{L}
  \begin{split}
     \mathcal{L}&\equiv\rho(r)\frac{\partial^2}{\partial t^2}+\frac{1}{\mu_0}\left(\frac{m}{r}B_{0\varphi}+k_z B_{0z}\right)^2\\
                &\equiv\rho(r)\left(\frac{\partial^2}{\partial t^2}+\omega_{A}^2(r)\right),
  \end{split}
\end{equation}
\begin{equation}\label{fr}
    f(r)\equiv\frac{2}{\mu_0}\left(m\frac{B_{0\varphi}^2}{r^2}+k_z\frac{B_{0\varphi} B_{0z}}{r}\right).
\end{equation}
Here, the operator $\mathcal{L}$ is the generalization of the Alfv\'{e}n operator
$\mathcal{L}_A=\rho(r)\frac{\partial^2}{\partial t^2}+\frac{k_z^2
B_{0z}^2}{\mu_0}$, Eq. (9) in ST2015, in the presence of magnetic
twist and
\begin{equation}\label{omegaA}
     \omega_{A}(r)\equiv\frac{1}{\sqrt{\mu_0 \rho(r)}}\left(\frac{m}{r}B_{0\varphi}(r)+k_zB_{0z}\right),
\end{equation}
is the background Alfv\'{e}n frequency.
Substituting Eq. (\ref{p}) in the $\varphi$ component of Eq. (\ref{mhd4}) and using Eqs. (\ref{mhd3}) and (\ref{B}) one can relate $\xi_\varphi$ to $\xi_r$ as follows
\begin{equation}\label{xiphi}
    \mathcal{L}\xi_\varphi=\frac{i m/r}{k_z^2+m^2/r^2}\mathcal{L}\left(\frac{1}{r}\frac{\partial(r\xi_r)}{\partial r}\right)\\
    +\frac{2i}{\mu_0}\frac{k_z^2}{k_z^2+m^2/r^2}\left(m\frac{B_{0\varphi}^2}{r^2}+k_z \frac{B_{0\varphi} B_{0z}}{r}\right)\xi_r.
\end{equation}

Defining $\alpha\equiv B_{0\varphi}(R)/B_{0z}$ as the twist parameter and $\epsilon\equiv k_z R$, one can obtain the orders of magnitude of the first and second terms on the right-hand side of Eq. (\ref{xiphi}) denoted by $T_0$ and $T_1$, respectively, as
\begin{eqnarray}\label{T0}
    &T_0&\equiv\frac{m/r}{k^2+m^2/r^2}\mathcal{L}\left(\frac{1}{r}\frac{\partial(r \xi_r)}{\partial r}\right)
    \simeq\left(\frac{B_{0z}^2}{\mu_0 R^2}\right)\frac{(\alpha+\epsilon)^2}{1+\epsilon^2}\xi_r,\\
    &T_1&\equiv\frac{k_z^2/\mu_0}{k^2+m^2/r^2}\left(2m\frac{B_{0\varphi}^2}{r^2}+2k_z\frac{B_{0\varphi} B_{0z}}{r}\right)\xi_r
    \simeq\left(\frac{B_{0z}^2}{\mu_0 R^2}\right)\epsilon^2\frac{\alpha^2+\alpha\epsilon}{1+\epsilon^2}\xi_r.
\end{eqnarray}
Therefore,
\begin{equation}\label{T1}
     \frac{T_1}{T_0}\sim\epsilon^2\frac{\alpha}{\alpha+\epsilon}\sim\left\{\begin{array}{lll}
      O(\epsilon^2), &\alpha\gtrsim\epsilon,&\\
      O(\alpha\epsilon), &\alpha\ll\epsilon.&
      \end{array}\right.
\end{equation}
It is clear that in the limit of long-wavelength ($\epsilon\ll1$), we can ignore $T_1$ against $T_0$ in Eq. (\ref{xiphi}). For instance, for $\alpha=0.01$ and $\epsilon=0.03$ we have $T_1/T_0\sim 10^{-4}$. Therefore, in the long wavelength limit ($\epsilon\ll1$), Eq. (\ref{xiphi}) takes the form
\begin{equation}\label{xiphi2}
    \xi_\varphi=\frac{i m/r}{k_z^2+\frac{m^2}{r^2}}\frac{1}{r}\frac{\partial (r\xi_r)}{\partial r}.
\end{equation}
Putting Eq. (\ref{xiphi2}) into (\ref{mhd3}) gives $\xi_{z}$ in
terms of $\xi_{r}$ as
\begin{equation}\label{xiz}
    \xi_z=\frac{i k_z}{k_z^2+\frac{m^2}{r^2}}\frac{1}{r}\frac{\partial (r\xi_r)}{\partial r}.
\end{equation}
Equations (\ref{xiphi2}) and (\ref{xiz}) show that for $\epsilon\ll1$, $\xi_\varphi$ and $\xi_z$ are not explicit functions of
the magnetic twist. However, the magnetic twist indirectly affects
$\xi_\varphi$ and $\xi_z$ by modifying the equation for $\xi_r$.

Eliminating $\delta P$, $\xi_\varphi$ and $\xi_z$ from Eqs.
(\ref{mhd3}) and (\ref{mhd4}), we obtain the following differential
equation for $\xi_r$ in the long wavelength limit ($\epsilon\ll 1$)
\begin{equation}\label{xir1}
    \mathcal{L}\mathcal{L}_s\xi_r+\left[\left(k_z^2+\frac{m^2}{r^2}\right)\frac{\partial\mathcal{L}}{\partial r}\left(\frac{1}{r}+\frac{\partial}{\partial r}\right)+\Phi(r)\right]\xi_r=0,
\end{equation}
which is the generalized form of Eq. (16) in ST2015, in the presence
of a twisted magnetic field in a thin flux tube ($\epsilon\ll 1$). Here, $\mathcal{L}_s$ is the surface wave operator defined as
\begin{equation}\label{Ls}
    \mathcal{L}_s\equiv\left(k_z^2+\frac{m^2}{r^2}\right)\frac{\partial^2}{\partial r^2}+\frac{1}{r}\left(k_z^2+\frac{3m^2}{r^2}\right)\frac{\partial}{\partial r}
    -\frac{1}{r^2}\left(k_z^2-\frac{m^2}{r^2}\right)-\left(k_z^2+\frac{m^2}{r^2}\right)^2,
\end{equation}
and
\begin{equation}\label{Phi}
    \Phi(r)\equiv g(r)\left(k_z^2+\frac{m^2}{r^2}\right)^2+\frac{2mk_z^2}{r^2}f(r)
    -\frac{m}{r}\left(k_z^2+\frac{m^2}{r^2}\right)\frac{ {\rm d}f(r)}{{\rm d} r},
\end{equation}
where
\begin{equation}\label{gr}
    g(r)\equiv\frac{2}{\mu_0}\left(-\frac{B_{0\varphi}^2}{r^2}+\frac{B_{0\varphi}}{r}\frac{{\rm d} B_{0\varphi}}{{\rm d} r}\right).
\end{equation}
Note that in the case of untwisted magnetic field (i.e. $B_{0\varphi}=0$), we have
$f(r)=g(r)=\Phi(r)=0$ and $\mathcal{L}=\mathcal{L}_A$. In this case,
Eq. (\ref{xir1}) reduces to Eq. (16) in ST2015.

Here, we should note that in the case of twisted magnetic tubes, to avoid the kink instability  the twist value defined as $\phi_{\rm twist}=(L/R)(B_{\varphi}/B_z)=2\pi N_{\rm twist}$ must not exceed a critical value $\phi_{\rm c}$ (see e.g. Shafranov 1957; Kruskal et al. 1958; Hood \& Priest 1979; Furno et al. 2006; Lapenta et al. 2006). Here, $N_{\rm twist}$ is the number of twist turns in the tube and $\phi_{\rm twsit}$ is the angle of rotation (in radians) of twisted magnetic field per length $L$ along the tube axis. According to the Kruskal-Shafranov analysis, for the particular case of a laboratory torus of major radius $R_0$, two points located an axial distance $L = 2\pi R_0$ apart refer to the same location on the torus, and $k$ equals $2\pi/L$. The kink instability is present in such a torus when $\phi_{\rm twist}>\phi_c=2\pi$. Hood \& Priest (1979) considered the effect of line-tying at the ends of a flux tube and showed that for force-free magnetic fields of uniform twist, a magnetic twist larger than $\phi_{c}= 3.3\pi$ leads to the kink instability. Furno et al. (2006) investigated the kink instability in a flux tube that is tied at one end and free at the other end and showed that for $\phi_{\rm twist}>\phi_c=\pi$ the flux tube is kink unstable (see also Lapenta 2006). Since we are interested in investigating the propagating waves, to avoid the kink instability, following Furno et al. (2006) we consider $\phi_{c}= \pi$ and take the length scale $L=2\pi/k_z$ in the longitudinal direction of the tube. Therefore, the constraint
\begin{equation}\label{alphamax}
    \phi_{\rm twist}=\frac{2\pi}{k_z R}\frac{B_{\varphi}}{B_z}=2\pi\frac{\alpha}{\epsilon}<\phi_{\rm c}=\pi,
\end{equation}
yields an upper limit for the twist parameter $\alpha_{\max}=\epsilon/2$ in our model. However, to derive the real threshold for instability we need to investigate a stability analysis of the partial differential equations of motion which is beyond the scope of the present work.

\section{Solution}\label{solution}

Solutions of $\xi_{r}$ representing the surface kink waves in the
constant density and untwisted regions $r<r_1$ and $r>r_2$ in the
TT approximation ($\epsilon\ll 1$) have been obtained by ST2015
as follows
\begin{eqnarray}
    &&\xi_{r \rm i}(r,t)\approx A_{\rm i}(t),~~~~~~~r\leqslant r_1,\label{xii}\\
    &&\xi_{r \rm e}(r,t)\approx A_{\rm e}(t)r^{-2},~~r\geqslant r_2\label{xie},
\end{eqnarray}
where $A_{\rm i}(t)$ and $A_{\rm e}(t)$ are the time-dependent amplitudes. In the nonuniform region $r_1<r<r_2$, following ST2015, we perform a modal expansion of the radial component of the Lagrangian displacement $\xi_r(r,t)$ as
\begin{equation}\label{me}
    \xi_{r}(r,t)=\sum_{n=1}^{\infty}a_n(t)\psi_n(r),
\end{equation}
where the eigenfunctions $\psi_n(r)$ satisfy the regular Sturm-Liouville system defined by the Bessel differential equation
\begin{equation}
\frac{{\rm d}^2\psi}{{\rm d}r^2}+\frac{1}{r}\frac{{\rm d}\psi}{{\rm d}r}+\left(\lambda^2-\frac{1}{r^2}\right)\psi=0.
\end{equation}
Also, the functions $\psi_n(r)$ have the following orthonormality relation
\begin{equation}\label{ort}
    \frac{1}{l}\int_{r_1}^{r_2}\psi_n(r)\psi_{n'}(r)r {\rm d}r=\delta_{nn'}.
\end{equation}
Following ST2015, using Eqs. (\ref{xii})-(\ref{me}) and applying the continuity of $\xi_r$ and its derivative with respect to $r$ at $r=r_1$ and $r=r_2$ one can obtain the boundary conditions governing $\psi_n(r)$ as follows
\begin{eqnarray}\label{bcpsi}
    \left.\frac{d \psi}{d r}\right|_{r=r_1}=0,\\
   \left.\left(\frac{2}{r}\psi+\frac{d \psi}{d r}\right)\right|_{r=r_2}=0.
\end{eqnarray}
The coefficient $a_n(t)$ is computed by solving the following generalized eigenvalue problem which is obtained by inserting Eq. (\ref{me}) into (\ref{xir1}) (see Cally 1991 and ST2015)
\begin{equation}\label{eig}
    \mathbb{H}~ \boldsymbol{a} = \omega^2 \mathbb{M}~ \boldsymbol{a}.
\end{equation}
Here $\omega^2$ and $\boldsymbol{a}$ are the eigenvalue and the eigenvector, respectively, and the square matrices $\mathbb{H}$ and $\mathbb{M}$ are as follows
\begin{eqnarray}
    &&\begin{split}
    &H_{nn'}=\frac{1}{l}\int_{r_1}^{r_2}\left[\frac{1}{\mu_0}\left(\frac{m}{r}B_{0\varphi}+k_z B_{0z}\right)^2\mathcal{L}_s\psi_{n'}(r)+\Phi(r)\psi_{n'}(r)\right.\\
    &\left.+\frac{2}{\mu_0}\left(k_z^2+\frac{m^2}{r^2}\right)\left(\frac{m}{r}B_{0\varphi}+k_zB_{0z}\right)\left(-\frac{m}{r^2}B_{0\varphi}+\frac{m}{r}\frac{{\rm d} B_{0\varphi}}{{\rm d} r}\right)\left(\frac{\psi_{n'}(r)}{r}+\frac{{\rm d}\psi_{n'}(r)}{{\rm d}r}\right)\right]\psi_n(r) r {\rm d}r,
    \end{split}\\\nonumber\\
    &&M_{nn'}=\frac{1}{l}\int_{r_1}^{r_2}\left[\rho(r)\mathcal{L}_s\psi_{n'}(r)+\frac{{\rm d}\rho}{{\rm d}r}\left(k_z^2+\frac{m^2}{r^2}\right)
    \left(\frac{\psi_{n'}(r)}{r}+\frac{{\rm d}\psi_{n'}(r)}{{\rm d}r}\right)\right]\psi_n(r) r {\rm d}r.
\end{eqnarray}
Following ST2015, the coefficients $a_n(t)$ are obtained as
\begin{equation}\label{ant}
    a_n(t)=\sum_{n'=1}^{\infty}\beta_{nn'}\left[c_{n'}\cos(\omega_{n'}t)+d_{n'}\sin(\omega_{n'}t)\right],
\end{equation}
where $\beta_{nn'}$ is the $n$th component of the $n'$th eigenvector and $\omega_{n'}$ is the $n'$th eigenvalue. Also, the coefficients $c_n$ and $d_n$ are obtained with the help of suitable initial conditions.
From Eqs. (\ref{me}) and (\ref{ant}), the expression for $\xi_r(r,t)$ in the region $r_1<r<r_2$ takes the form
\begin{equation}\label{xir2}
    \xi_r(r,t)=\sum_{n=1}^{\infty}\sum_{n'=1}^{\infty}\beta_{nn'}\left[c_{n'}\cos(\omega_{n'}t)+d_{n'}\sin(\omega_{n'}t)\right]\psi_n(r).
\end{equation}
Equation (\ref{xir2}) can be recast in the following form
\begin{equation}\label{xir3}
    \xi_r(r,t)=\sum_{n=1}^{\infty}\left[c_{n}\cos(\omega_{n}t)+d_{n}\sin(\omega_{n}t)\right]\phi_n(r),
\end{equation}
where
\begin{equation}\label{phin}
    \phi_n(r)\equiv\sum_{n'=1}^{\infty}\beta_{n'n}\psi_{n'}(r),
\end{equation}
is the $n$'s eigenfunction of Alfv\'{e}n discrete modes (see Cally 1991 and ST2015). These modes are a discretized version of the Alfv\'{e}n continuum. Hence, in the formalism of ST2015 the kink wave is not a global mode, but instead, it is built up as a superposition of Alfv\'{e}n continuum modes. Like ST2015, we take the following initial conditions
\begin{eqnarray}
    &&\xi_r(r,t=0)=\left\{\begin{array}{lll}
    \xi_0,&r\leqslant r_1,&\\
    \xi_0\frac{\psi_1(r)}{\psi_1(r_1)},&r_1<r< r_2,&\\
    \xi_0\frac{\psi_1(r_2)}{\psi_1(r_1)}\left(\frac{r_2}{r}\right)^2,&r\geqslant r_2,
      \end{array}\right.\label{init1}\\
    &&\frac{\partial\xi_r}{\partial t}\Big|_{(r,t=0)}=0,\label{init2}
\end{eqnarray}
where $\xi_0$ is a constant. Using Eqs. (\ref{me}), (\ref{ort}),
(\ref{ant}), (\ref{init1}) and (\ref{init2}) one can get
 \begin{eqnarray}
    &&c_n=\beta^{-1}_{n,1}\frac{\xi_0}{\psi_1(r_1)},\\
    &&d_n=0.
\end{eqnarray}
To solve Eq. (\ref{eig}) numerically, we must truncate the infinite series of
Eq. (\ref{me}) to a finite number $N$ of terms. This means that the Alfv\'{e}n continuum is discretized in $N$ different discrete modes. The bigger
the value of $N$ the larger the evolution time that we are allowed to
proceed before the energy in the $N$th Fourier mode becomes
significant and to the modal expansion starts to become inaccurate (for more details see Cally 1991). We shall consider a sufficiently large $N$ to make sure that the number of terms in the modal expansion is enough for the considered duration of the temporal evolution.

\section{Numerical results }\label{results}

In subsections \ref{discon} and \ref{con}, we consider two types of twisted magnetic field to see how different twist profiles affect the phase-mixing of kink MHD waves in comparison with the results of ST2015. To do so, following ST2015, we set $\rho_{\rm i}/\rho_{\rm e}=5$, $l/R=0.2,1$ and $k_zR=\pi/100$. Time is in units of the period of the kink oscillation in a thin and untwisted loop, $P_{\rm kink}=2\pi/\omega_{\rm kink}$, where
\begin{equation}\label{omegakink}
\omega_{\rm kink}=k_z\sqrt{\frac{\rho_{\rm i}v_{A_i}^2+\rho_{\rm e}v_{A_e}^2}{\rho_{\rm i}+\rho_{\rm e}}},
\end{equation}
is the so-called kink frequency. Here $v_{A_i}=B_{0z}/\sqrt{\mu_0\rho_{\rm i}}$ and $v_{A_e}=B_{0z}/\sqrt{\mu_0\rho_{\rm e}}$ are the interior and exterior Alfv\'{e}n speeds, respectively.

\subsection{Model I: discontinuous magnetic field}\label{discon}

Following Ebrahimi \& Karami (2016), we consider the azimuthal component of the magnetic field in the annulus region (i.e. $r_1<r<r_2$) as
\begin{equation}
    B_{0\varphi}(r)=Ar(r-r_1).\label{Bdis}
\end{equation}
Note that from Eq. (\ref{Bdis}), due to having a rotational discontinuity of the background
magnetic field at the location $r=r_2$, we have a delta-function
current sheet at $r=r_2$ in the axial direction. Note that in the presence of resistivity (which is absent in our model), the tearing mode instability can occur in this current sheet when
the driving force of the inflow exceeds the opposing Lorentz force. However, Ebrahimi \& Karami (2016) showed that even in the presence of resistivity by choosing an appropriate thickness for the current sheet, tearing
mode instability can be avoided in the model (\ref{Bdis}) during the kink oscillations.

The magnetohydrostatic equilibrium equation
takes the form
\begin{equation}\label{mse}
    \frac{{\rm d} }{{\rm d}r}\left(p+\frac{B_{0\varphi}^2+B_{0z}^2}{2\mu_0}\right)=-\frac{B_{0\varphi}^2}{\mu_0 r},
\end{equation}
where $p$ is the gas pressure. Using Eqs. (\ref{B}), (\ref{mse}) and
continuity of the total (magnetic plus gas) pressure across $r=r_1$ and $r=r_2$, we obtain the gas pressure as
\begin{equation}\label{gasp1}
    \small
    p(r)=\left\{\begin{array}{lll}
    p_0,&r\leqslant r_1,\\\\
    p_0-\frac{A^2}{\mu_0}\left(\frac{3}{4}r^4-\frac{5}{3}r^3r_1+r^2r_1^2-\frac{1}{12}r_1^4\right),&r_1<r< r_2,\\\\
    p_0+\frac{A^2}{\mu_0}\left(-\frac{1}{4}r_2^4-\frac{1}{2}r_1^2r_2^2+\frac{2}{3}r_1r_2^3+\frac{1}{12}r_1^4\right), &r\geqslant r_2,
    \end{array}\right.
    \normalsize
\end{equation}
where $p_0$ is a constant.

\begin{figure}
\centering
    \begin{tabular}{cc}
        \includegraphics[width=70mm]{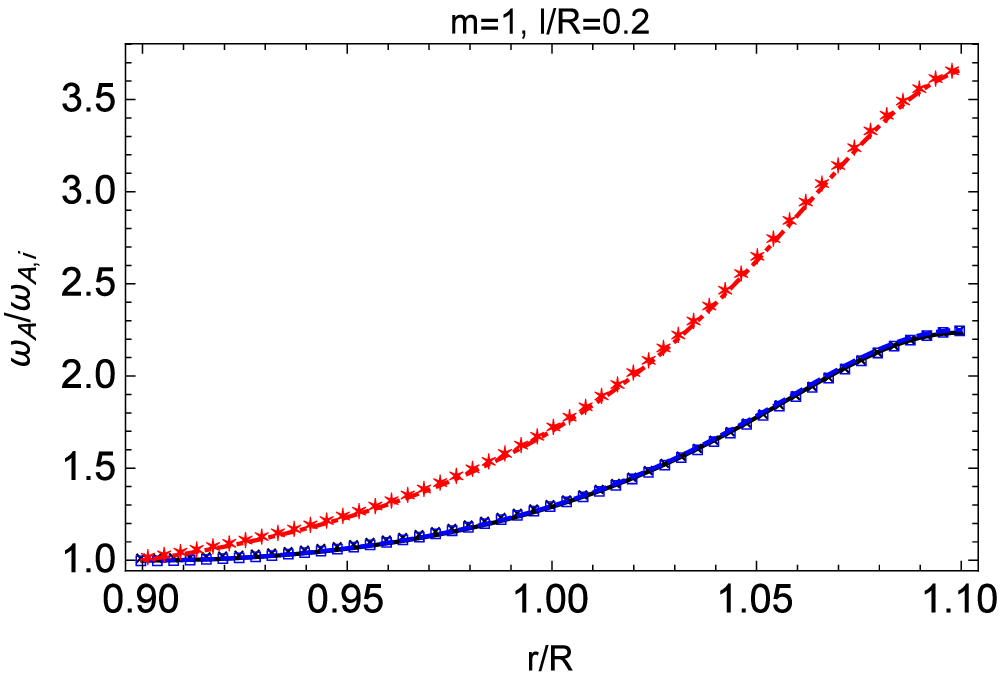}& \includegraphics[width=70mm]{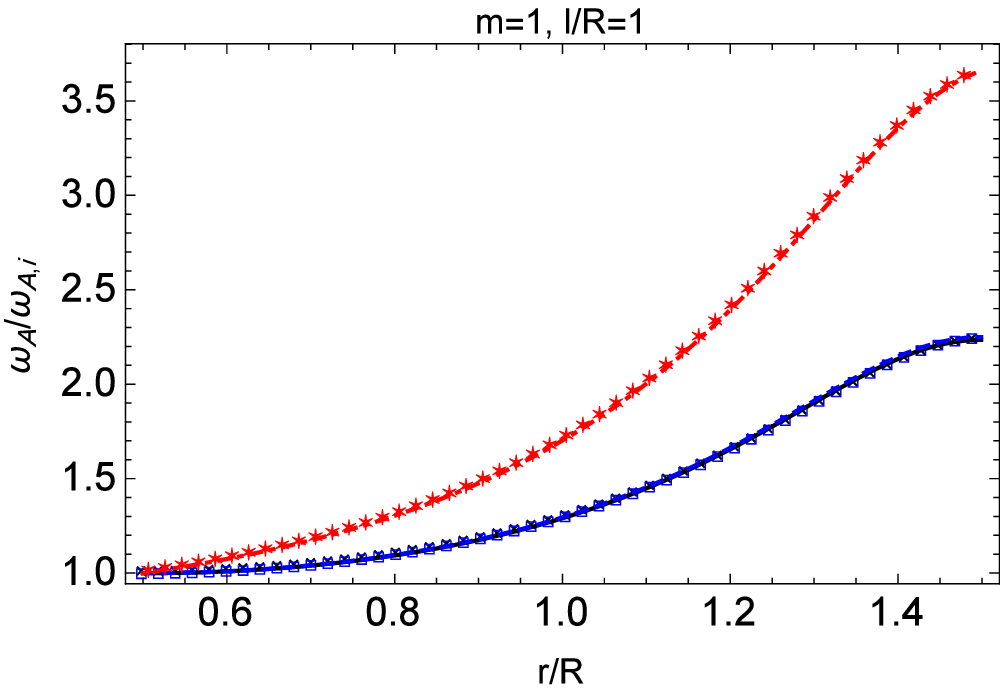}\\
        \includegraphics[width=70mm]{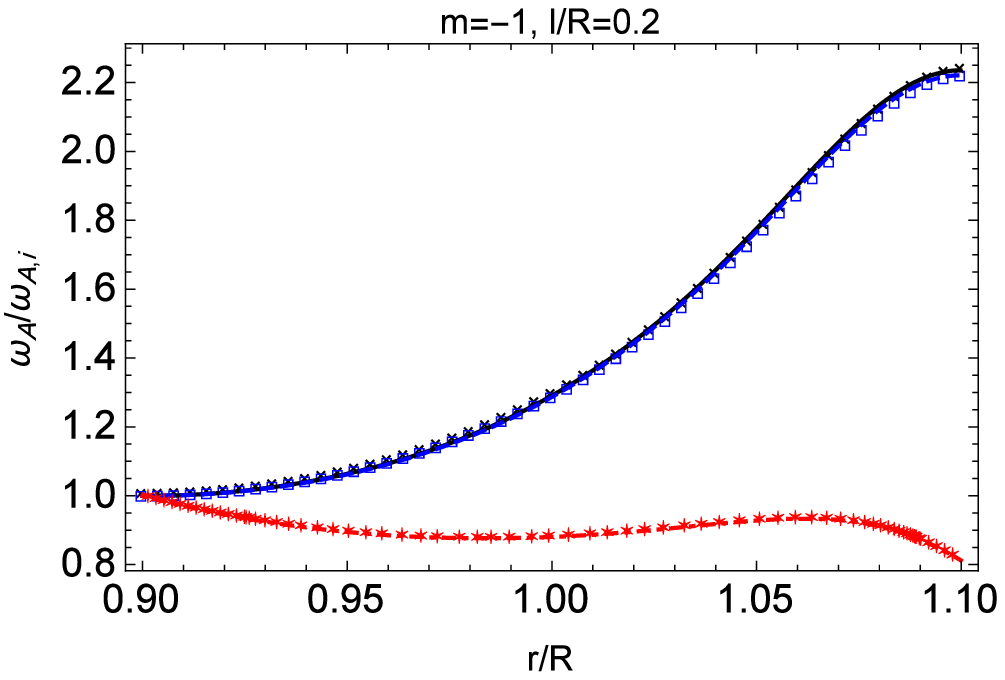}& \includegraphics[width=70mm]{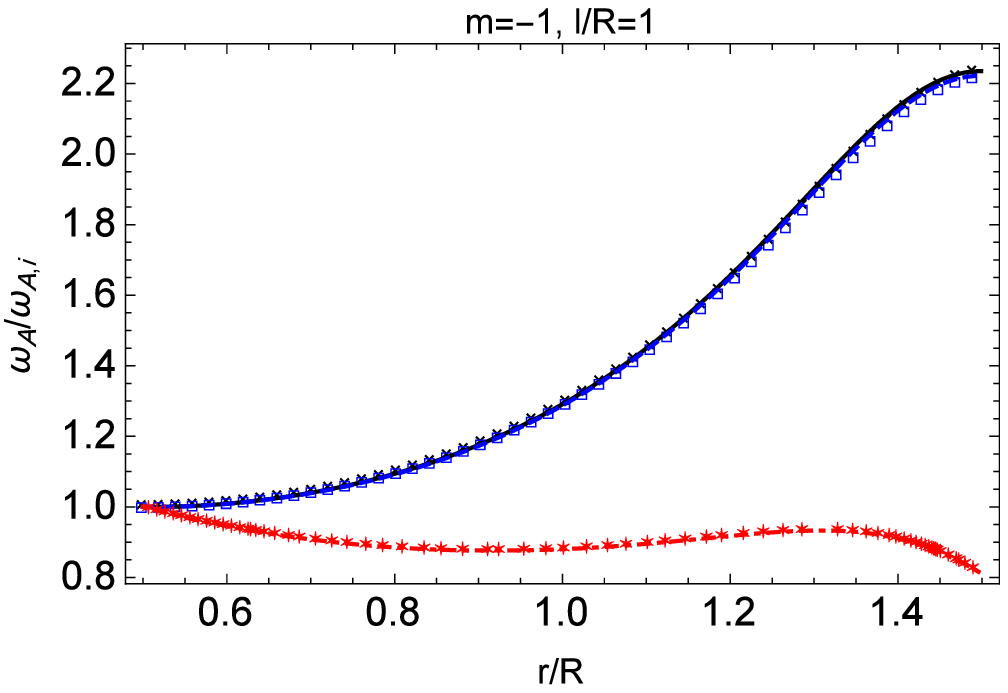}\\
    \end{tabular}
    \caption {Background Alfv\'{e}n frequency $\omega_A(r)$ in the annulus region ($r_1<r<r_2$) for the model I. Here, $l/R=0.2$ (left panels), $l/R=1$ (right panel), $m=+1$ (top panels), $m=-1$ (bottom panels), $k_z R=\pi/100$ and $\alpha=0$ (solid line); $\alpha=10^{-4}$ (dashed line); $\alpha=10^{-2}$ (dot-dashed line). The crosses, squares and asterisks, correspond to the discrete eigenfrequencies for $\alpha=0$, $10^{-4}$ and $\alpha=10^{-2}$, respectively. Here, $N=101$ for $m=+1$ and $N=300$ for $m=-1$ and $\alpha=10^{-2}$, but for convenience we show only multiples of 2 and 6 for $N=101,~300$, respectively. Note that the results for $\alpha=0$ and $10^{-4}$ overlap with each other.}
    \label{omega_MN}
 \end{figure}

\begin{figure}
\centering
    \begin{tabular}{cc}
       \includegraphics[width=70mm]{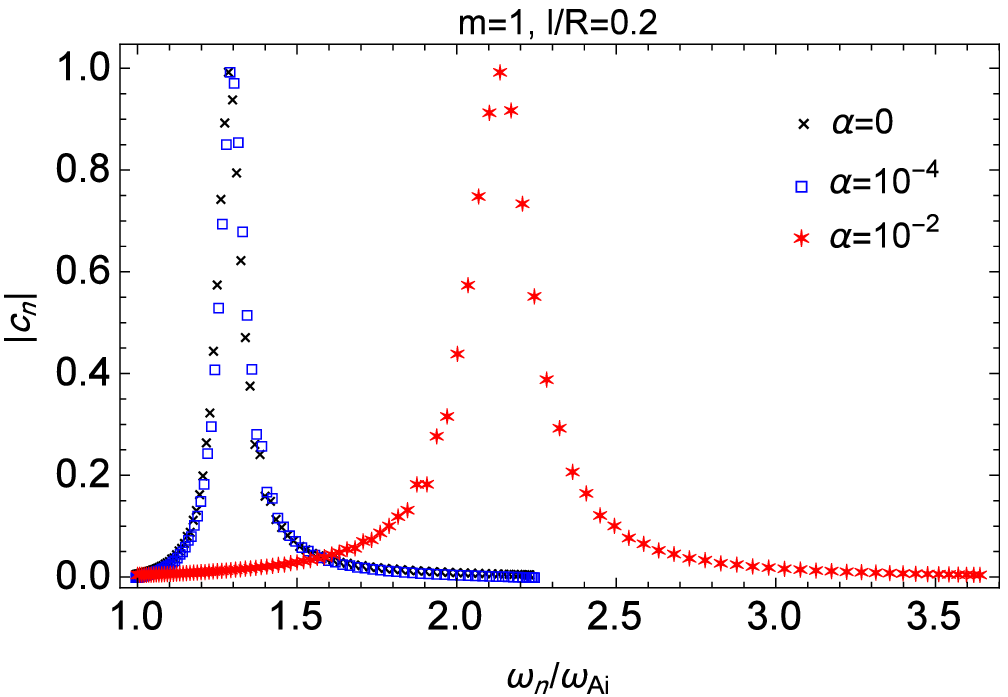}& \includegraphics[width=70mm]{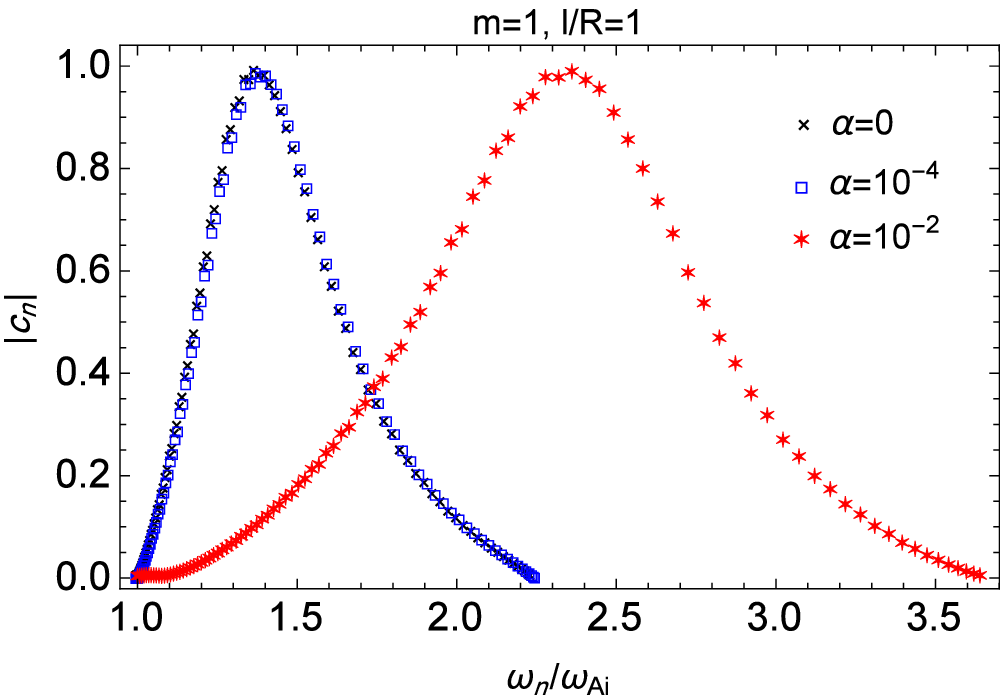}\\
       \includegraphics[width=70mm]{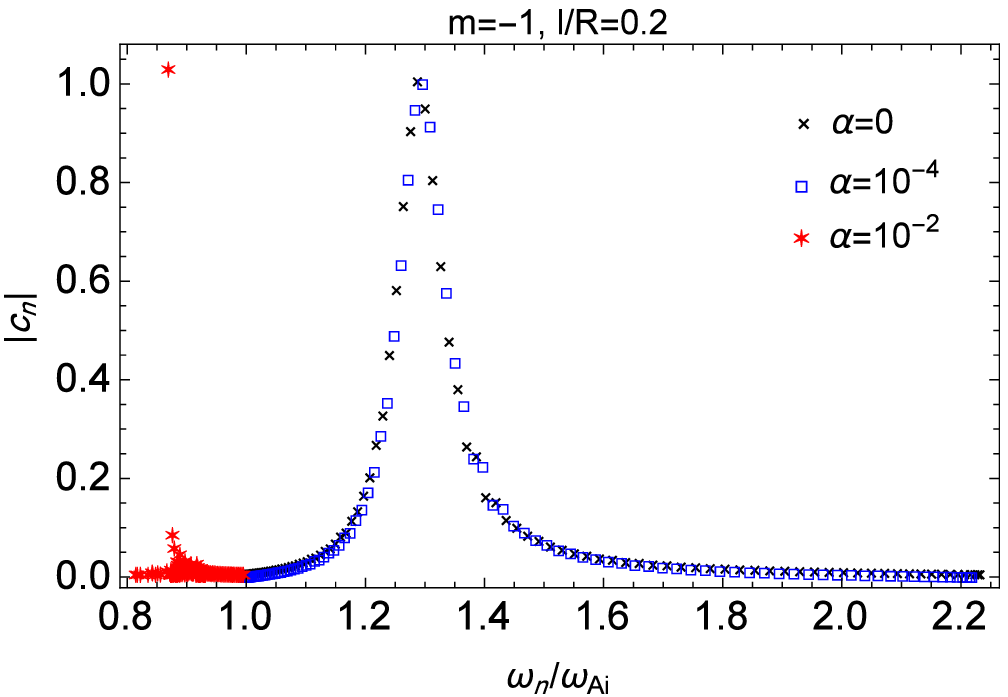}& \includegraphics[width=70mm]{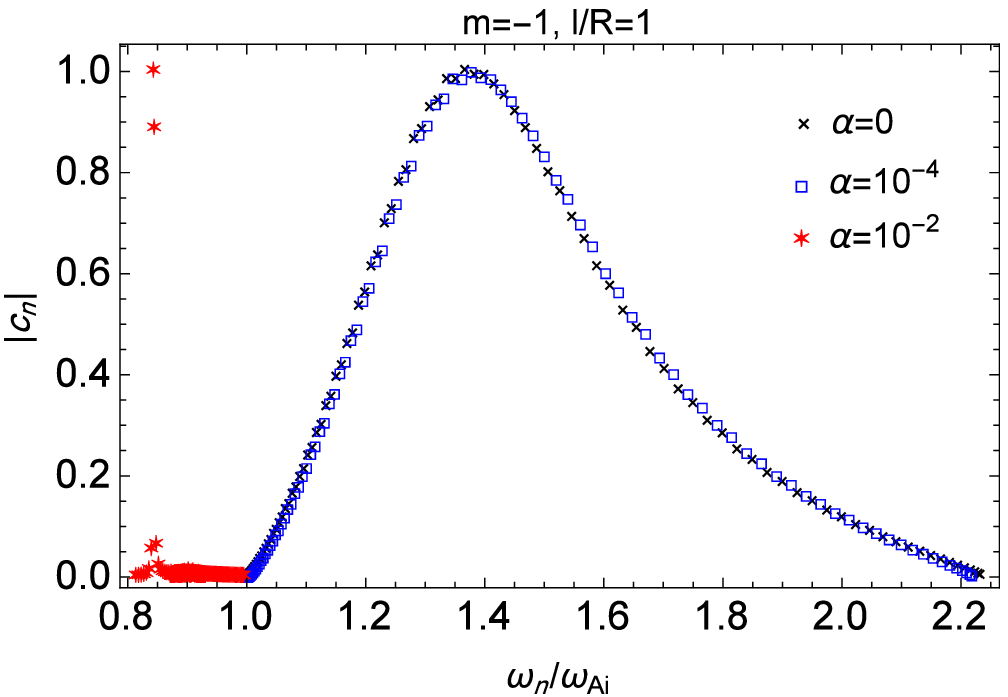}\\
    \end{tabular}
    \caption {Normalized values of $|c_n|$ versus their corresponding eigenfrequencies for the model I for $l/R=0.2$ (left panels), $l/R=1$ (right panels), $m=+1$ (top panel), $m=-1$ (bottom panels). Here, $\alpha=0$ (crosses), $\alpha=10^{-4}$ (squares), $\alpha=10^{-2}$ (asterisks). Other auxiliary parameters are as in Fig. \ref{omega_MN}. Note that the scale of the horizontal axis in the top and bottom panels is different. The results of $\alpha=0$ are exactly the same for $m=\pm 1$.}
    \label{cn_MN}
\end{figure}

\begin{figure}
  \centering
  \begin{tabular}{ccc}
    \includegraphics[width=50mm]{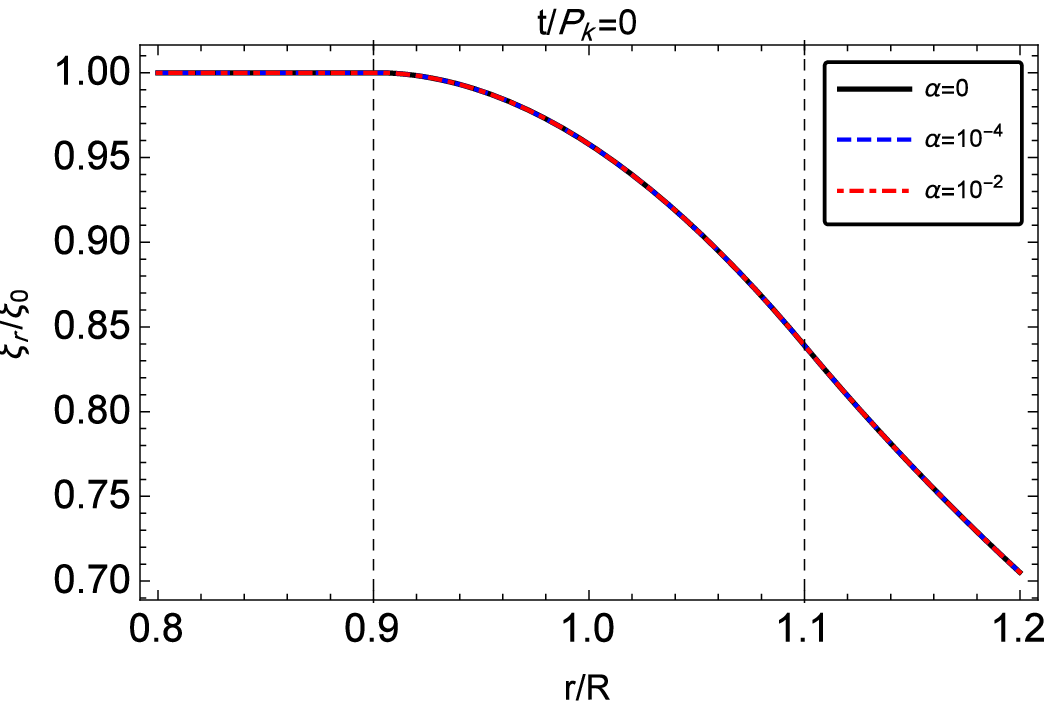}& \includegraphics[width=50mm]{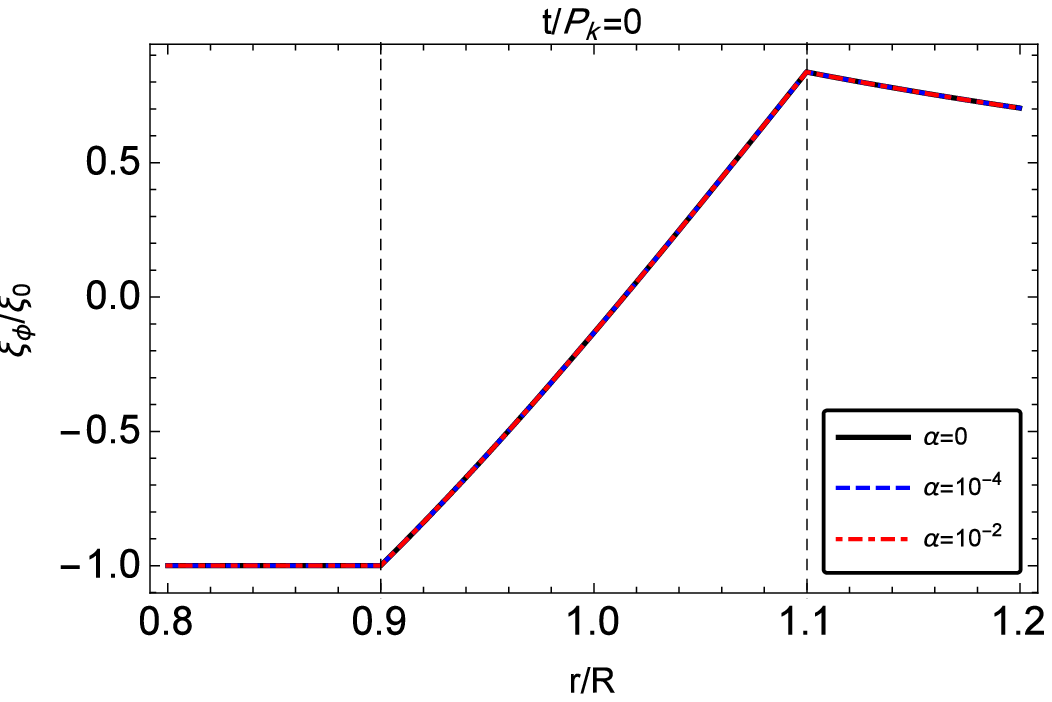}& \includegraphics[width=50mm]{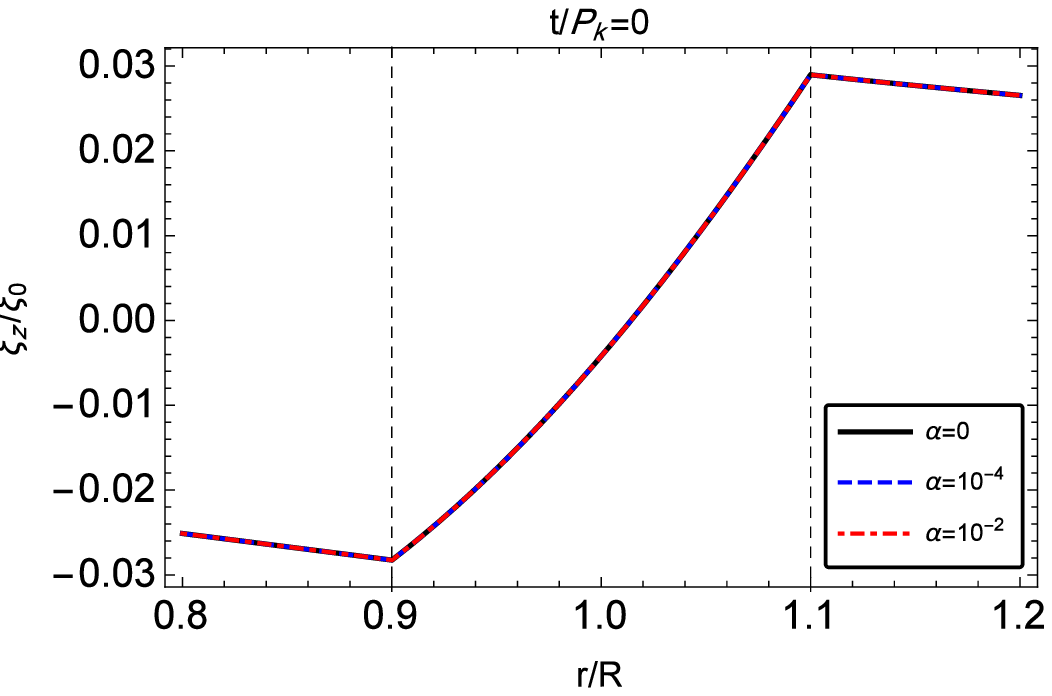}\\
    \includegraphics[width=50mm]{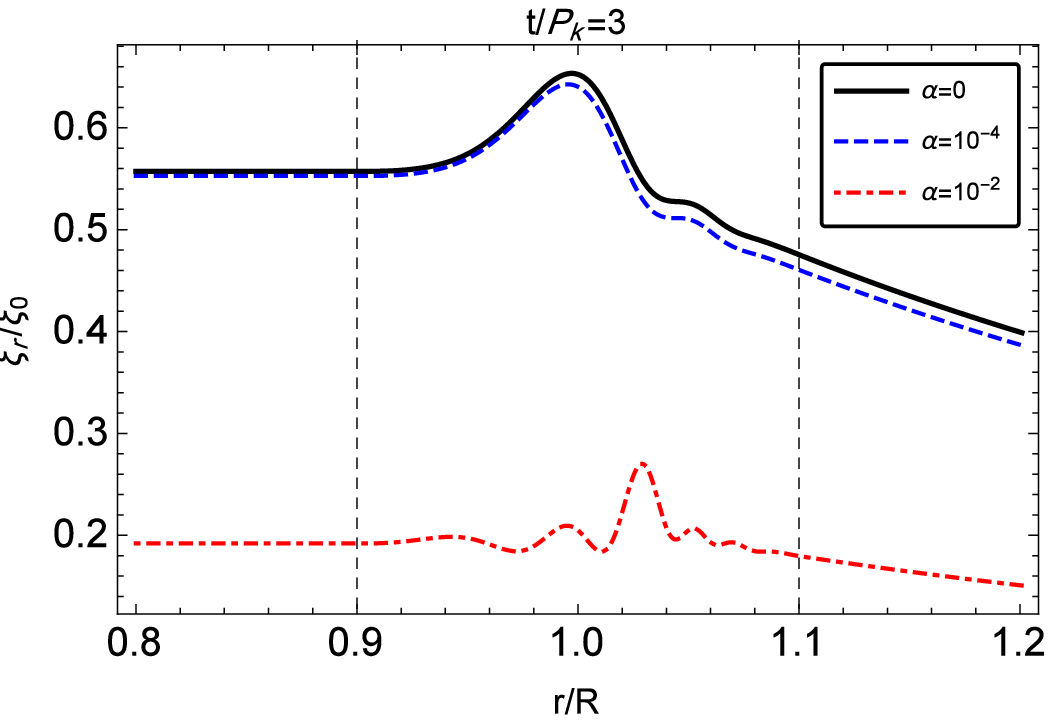}& \includegraphics[width=50mm]{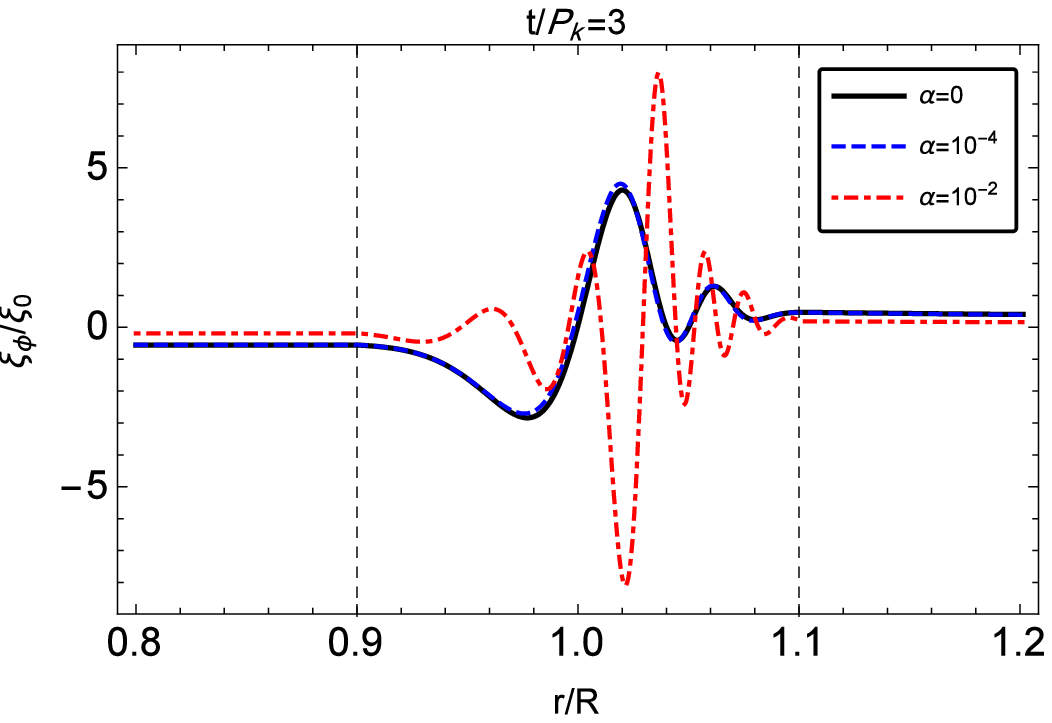}& \includegraphics[width=50mm]{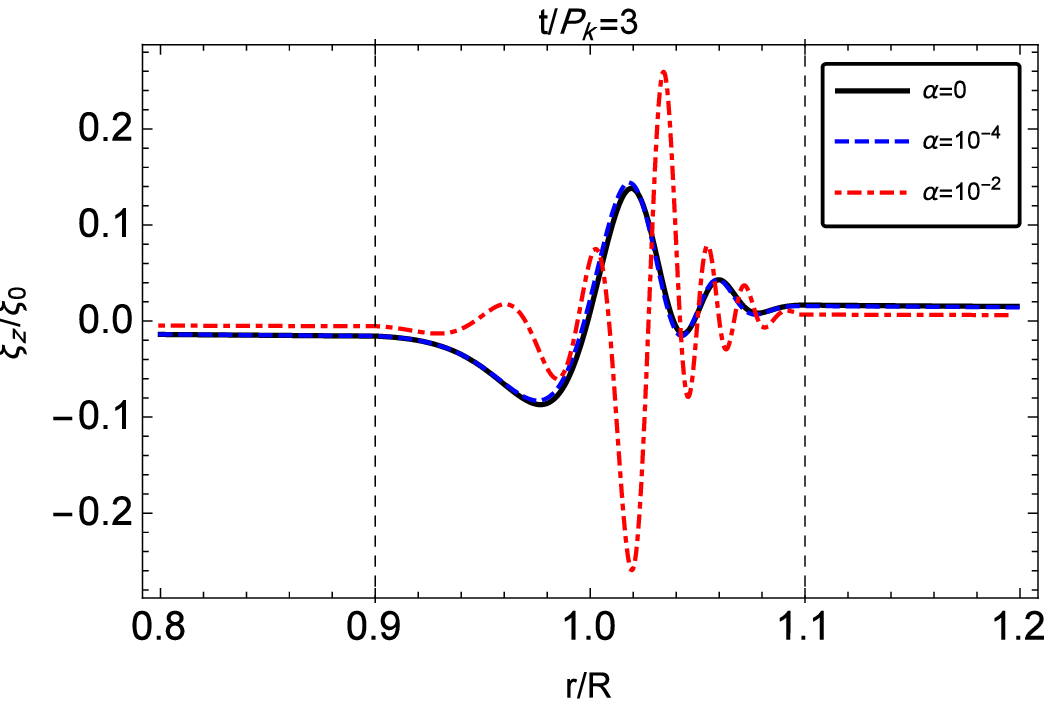}\\
    \includegraphics[width=50mm]{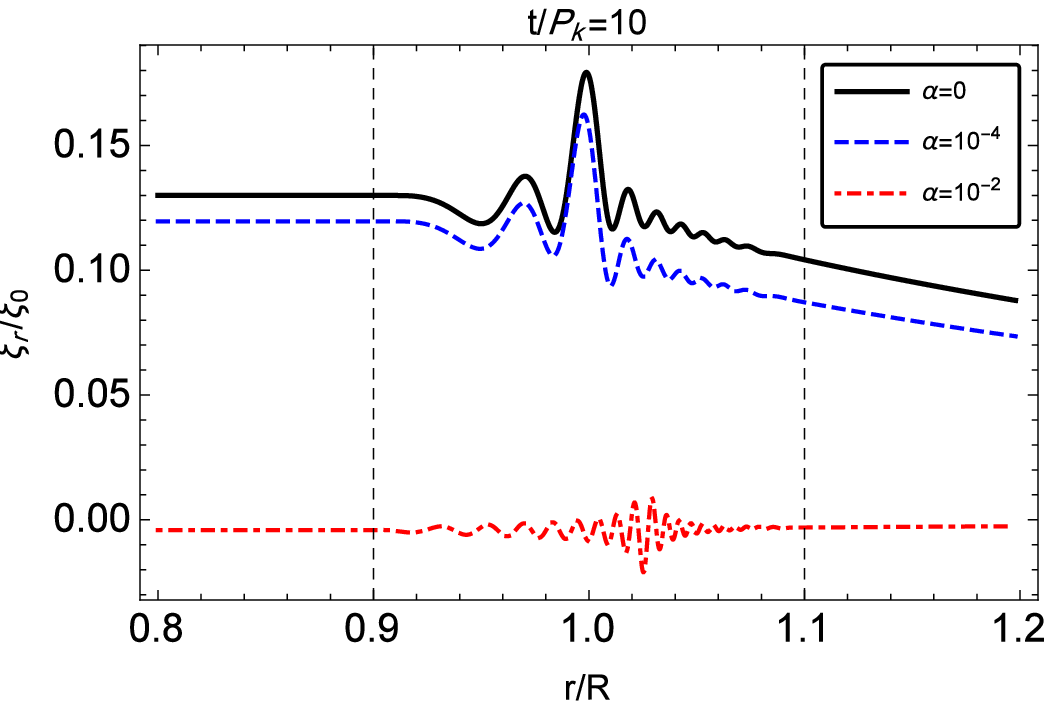}& \includegraphics[width=50mm]{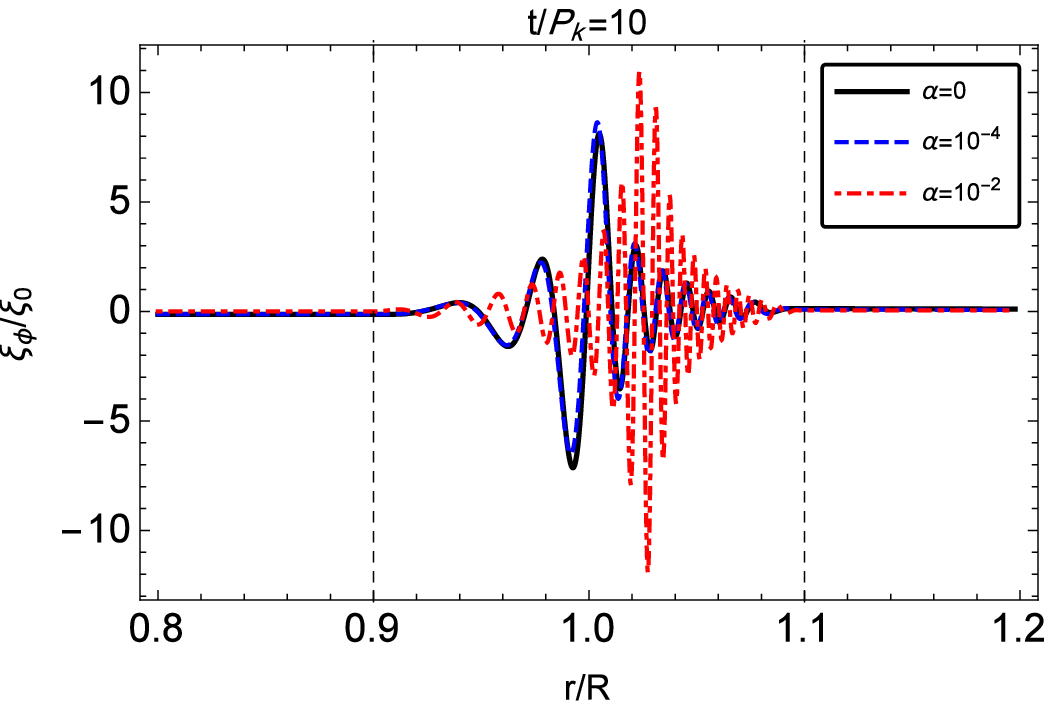}& \includegraphics[width=50mm]{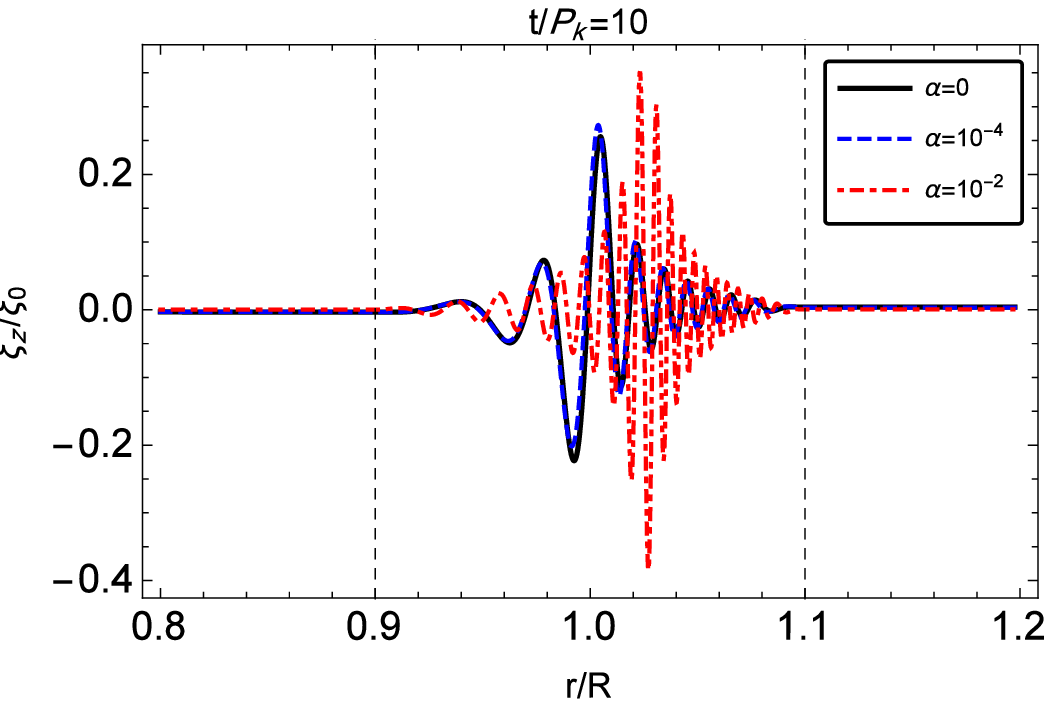}\\
  \end{tabular}
\caption {Temporal evolution of different components of the Lagrangian displacement, $\xi_r$ (left), $\xi_\varphi$ (middle)
and $\xi_z$ (right) for $\alpha=0$ (solid line), $\alpha=10^{-4}$ (blue dashed line) and $\alpha=10^{-2}$ (red dot-dashed line) for the model I with $l/R=0.2$ and $m=+1$. Here $t/P_k=0$ (top), 3 (middle) and 10 (bottom). The left and right vertical dashed lines denote $r_1$ and $r_2$, respectively. Other auxiliary parameters are as in Fig. \ref{omega_MN}.}
    \label{xi_MN_m(1)_lbyR(0.2)}
\end{figure}

\begin{figure}
  \centering
  \begin{tabular}{ccc}
    \includegraphics[width=50mm]{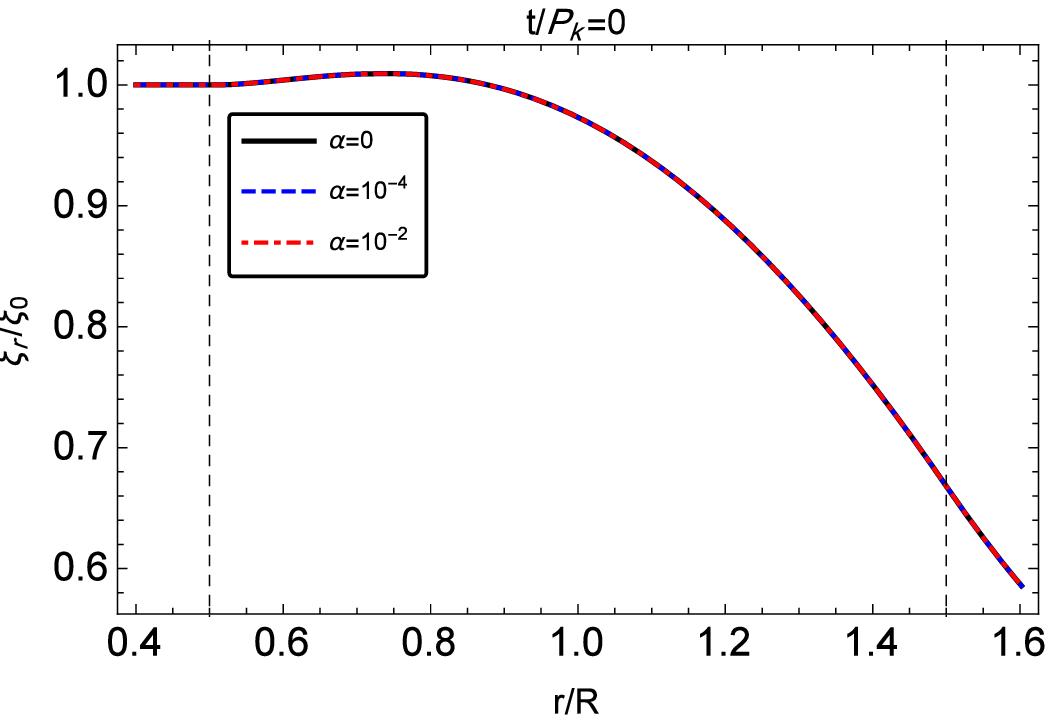}& \includegraphics[width=50mm]{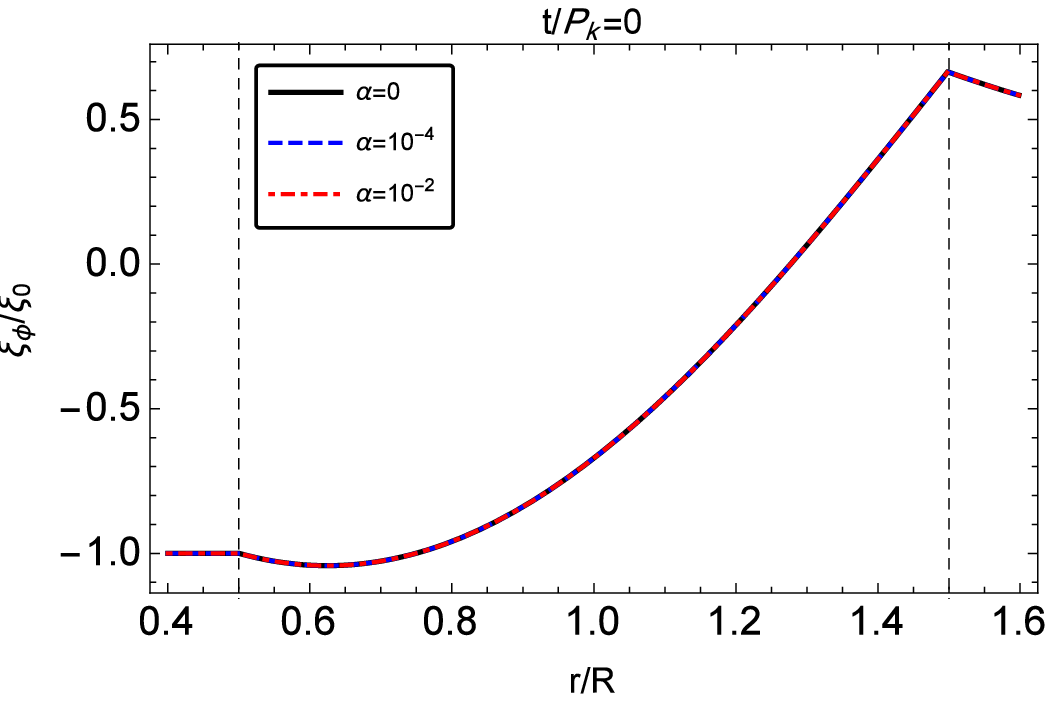}& \includegraphics[width=50mm]{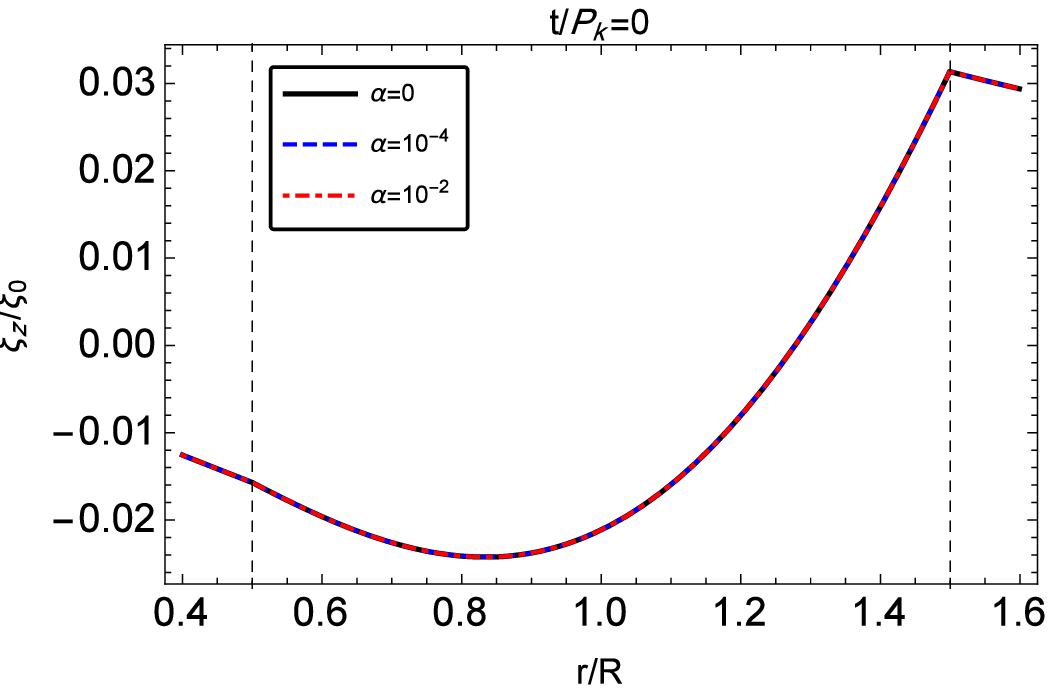}\\
    \includegraphics[width=50mm]{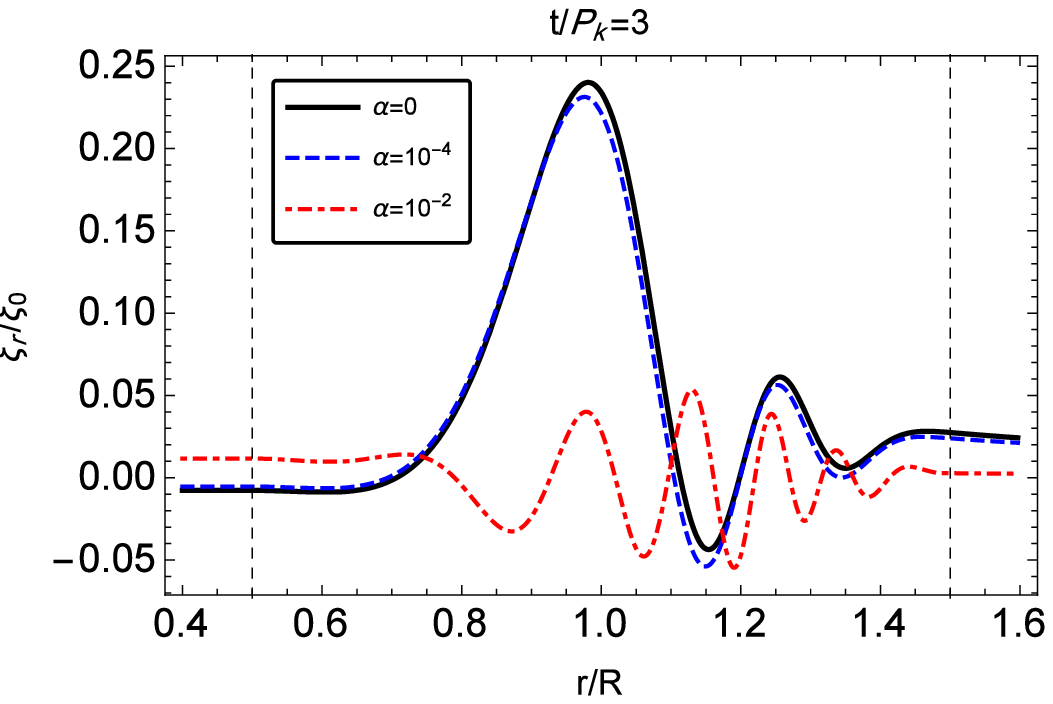}& \includegraphics[width=50mm]{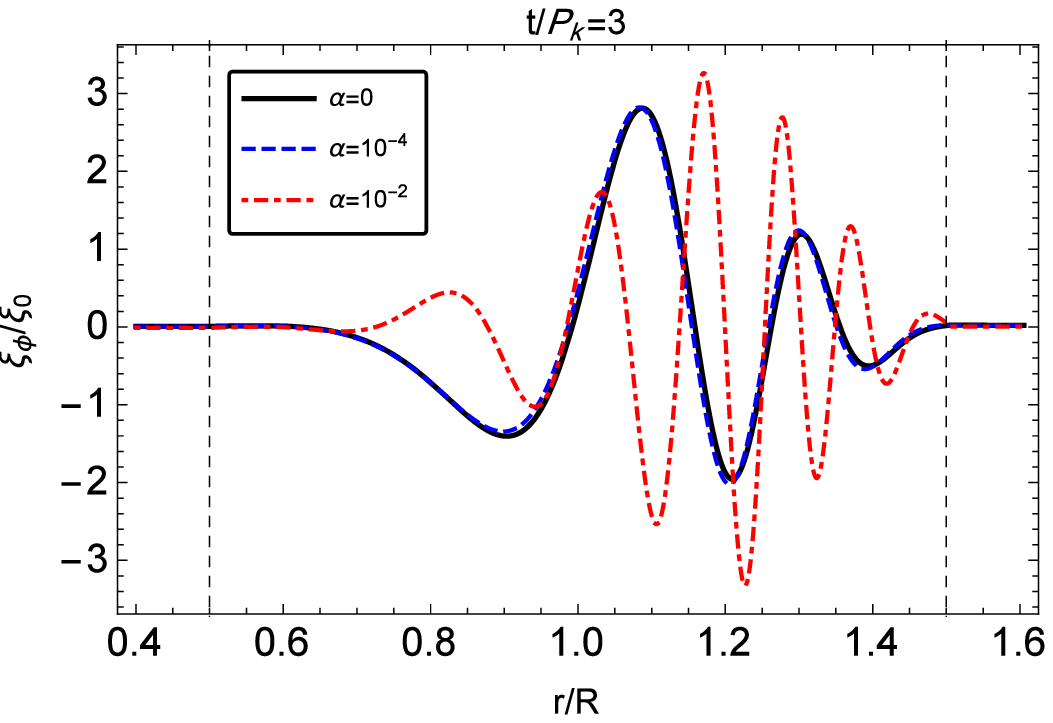}& \includegraphics[width=50mm]{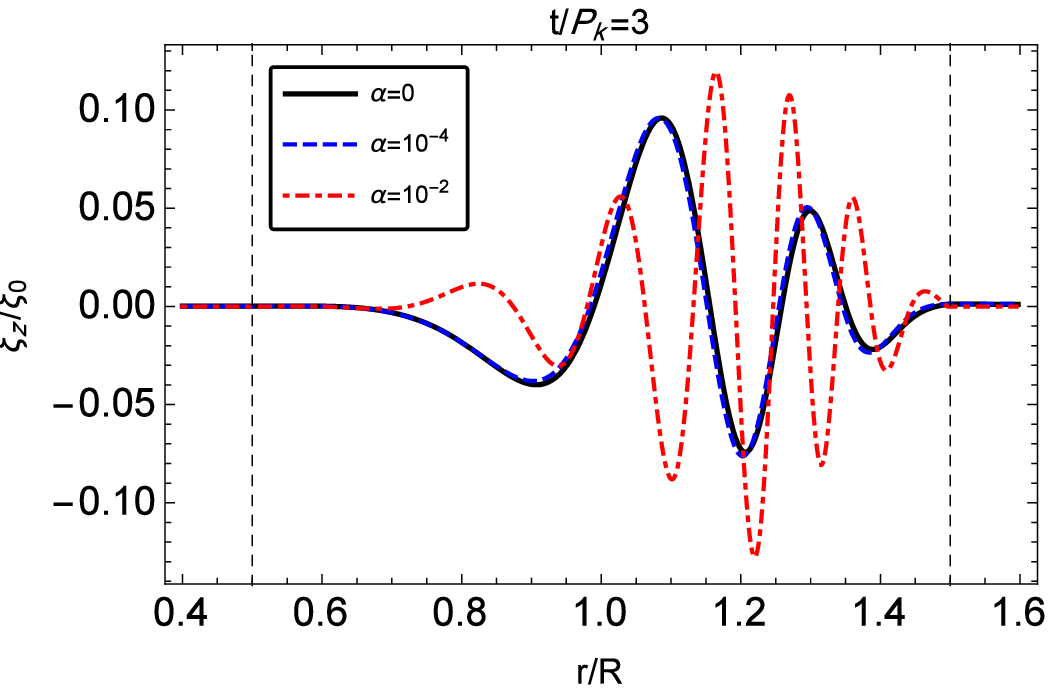}\\
    \includegraphics[width=50mm]{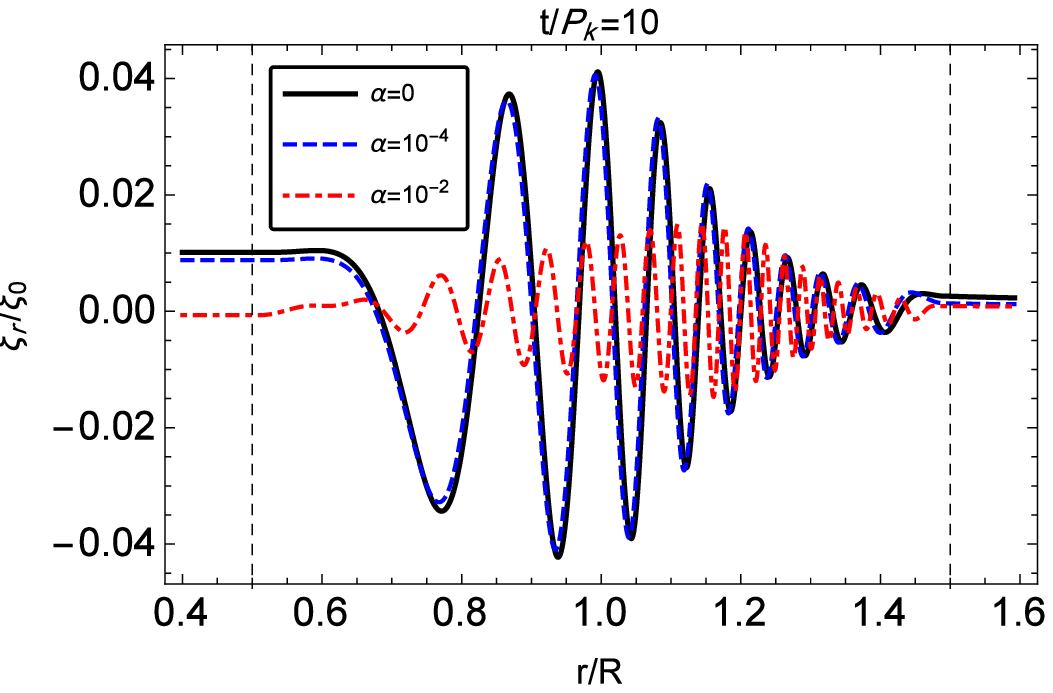}& \includegraphics[width=50mm]{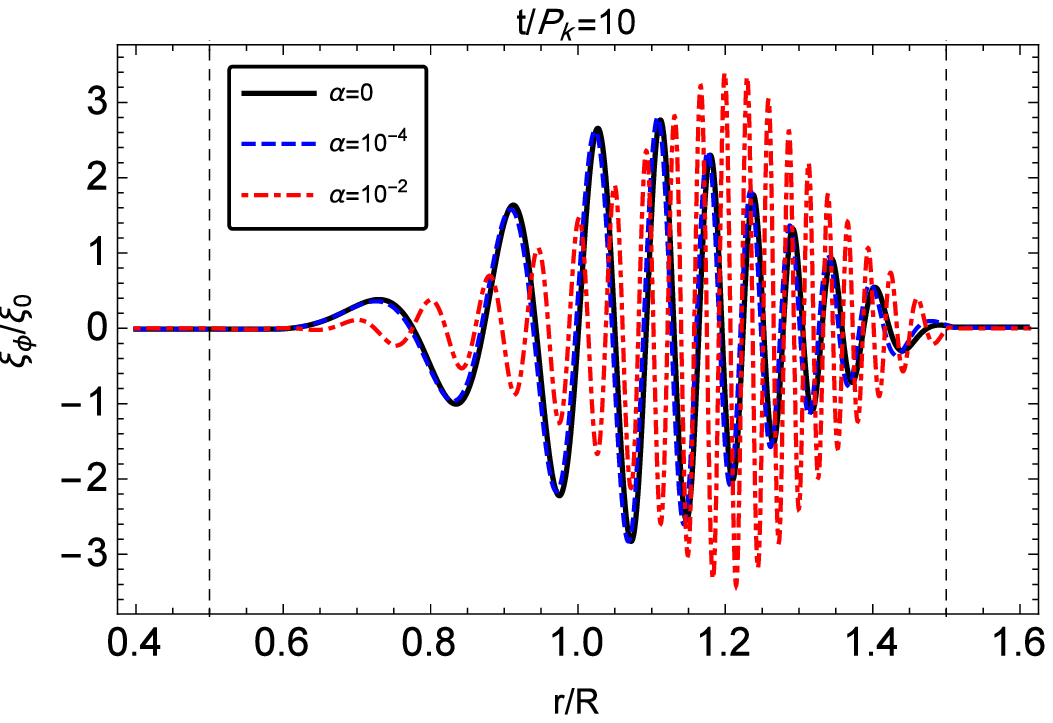}& \includegraphics[width=50mm]{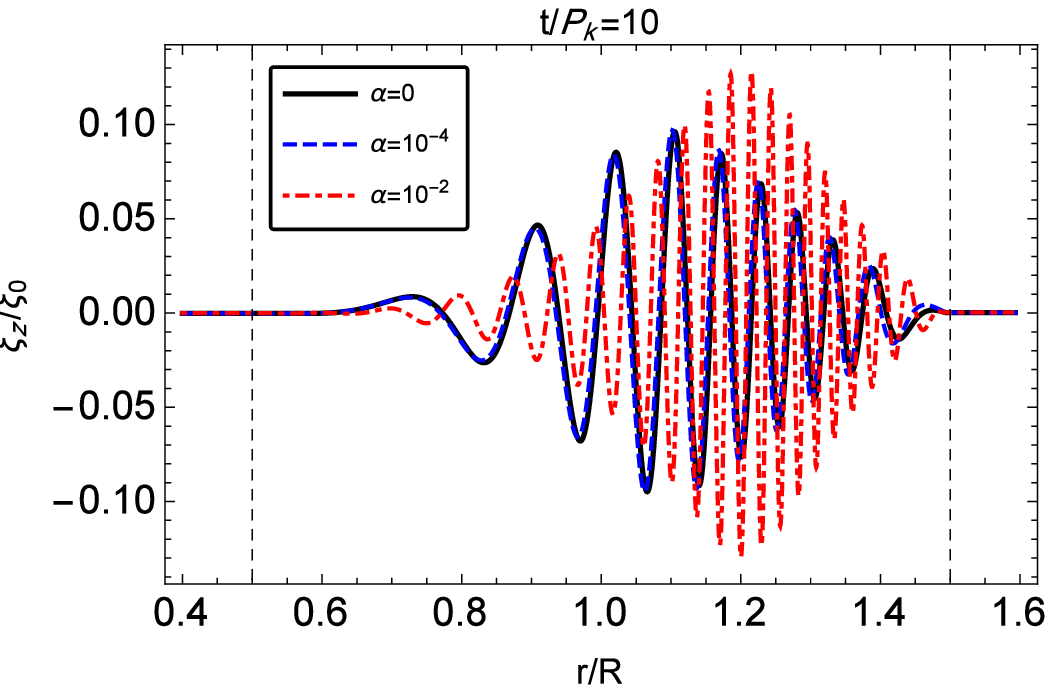}\\
  \end{tabular}
\caption {Same as Fig. \ref{xi_MN_m(1)_lbyR(0.2)}, but for $l/R=1$.}
    \label{xi_MN_m(1)_lbyR(1)}
\end{figure}

\begin{figure}
  \centering
  \begin{tabular}{ccc}
    \includegraphics[width=50mm]{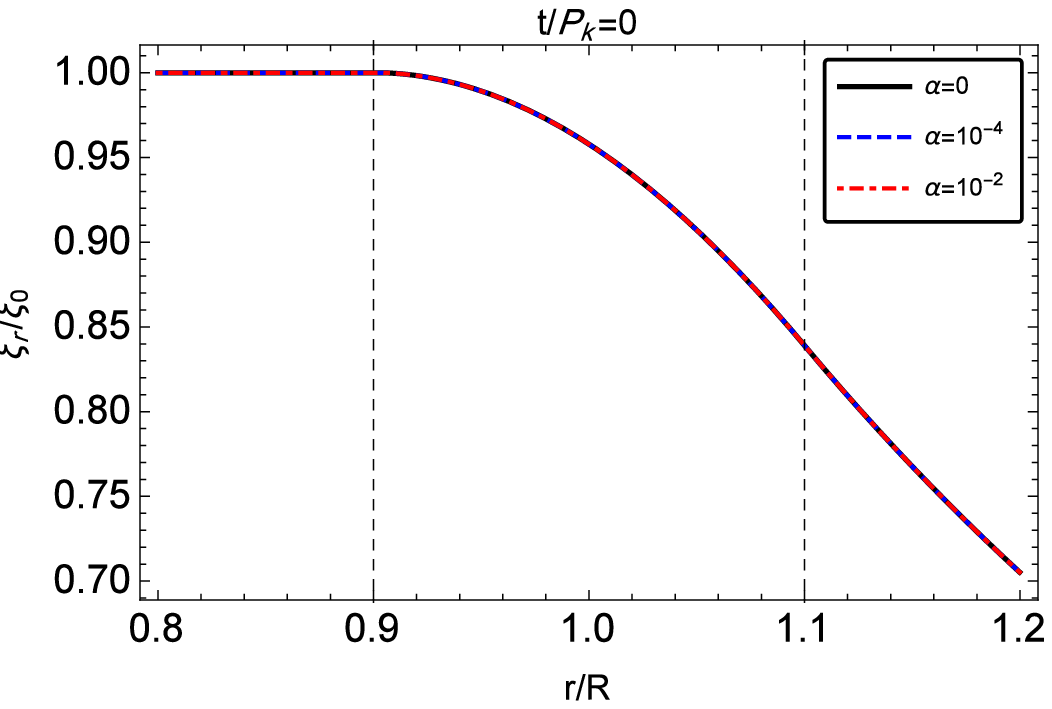}& \includegraphics[width=50mm]{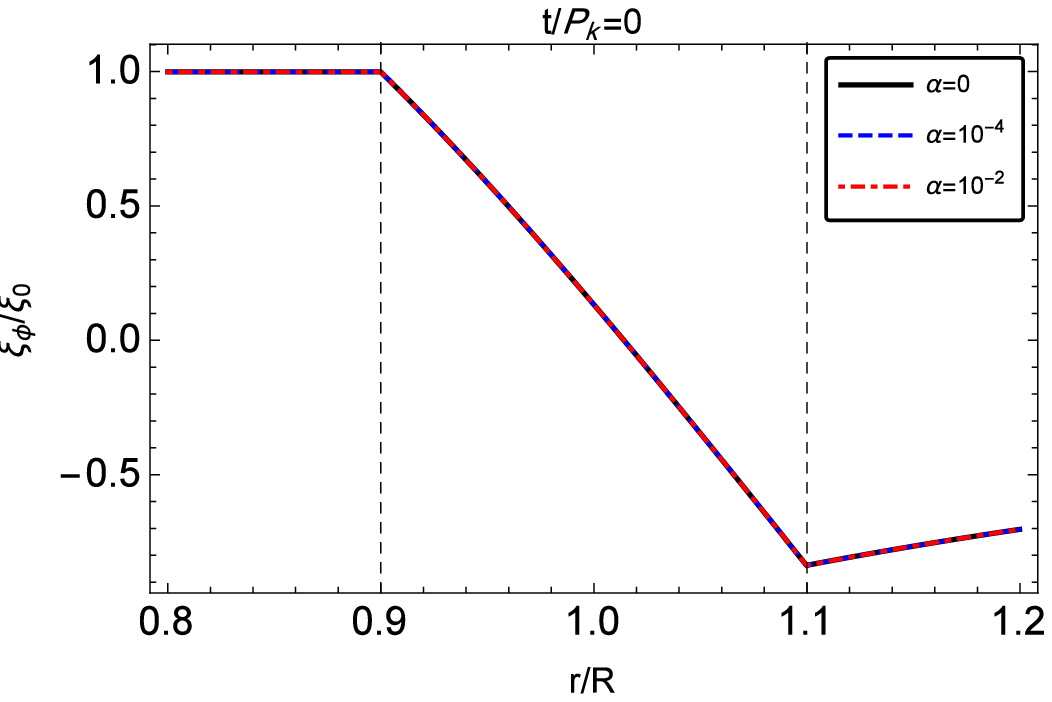}& \includegraphics[width=50mm]{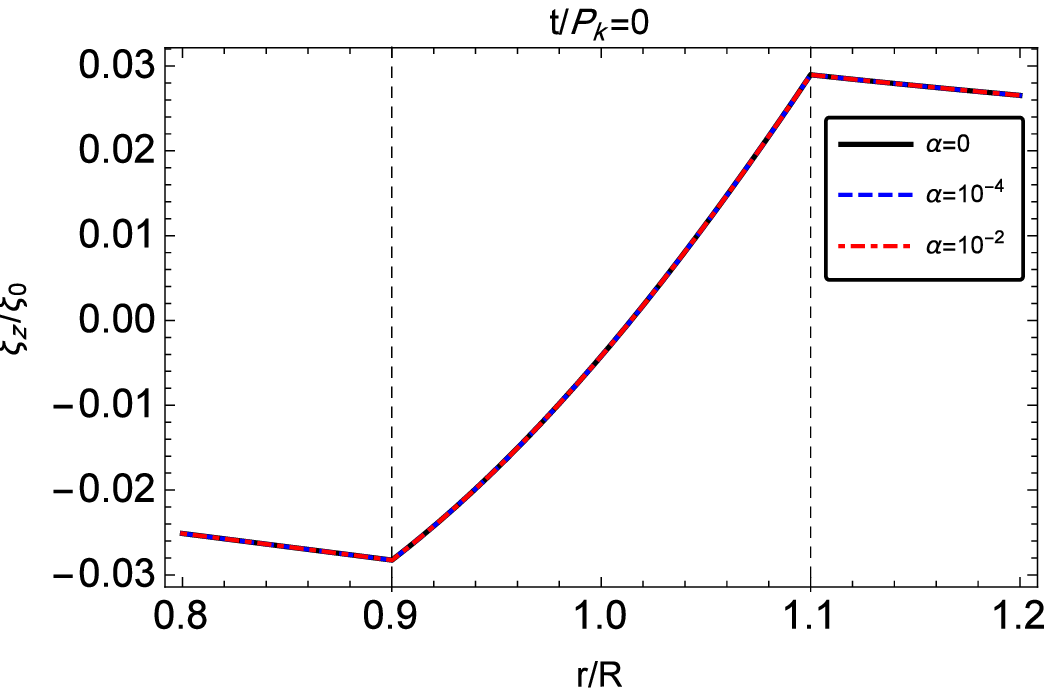}\\
    \includegraphics[width=50mm]{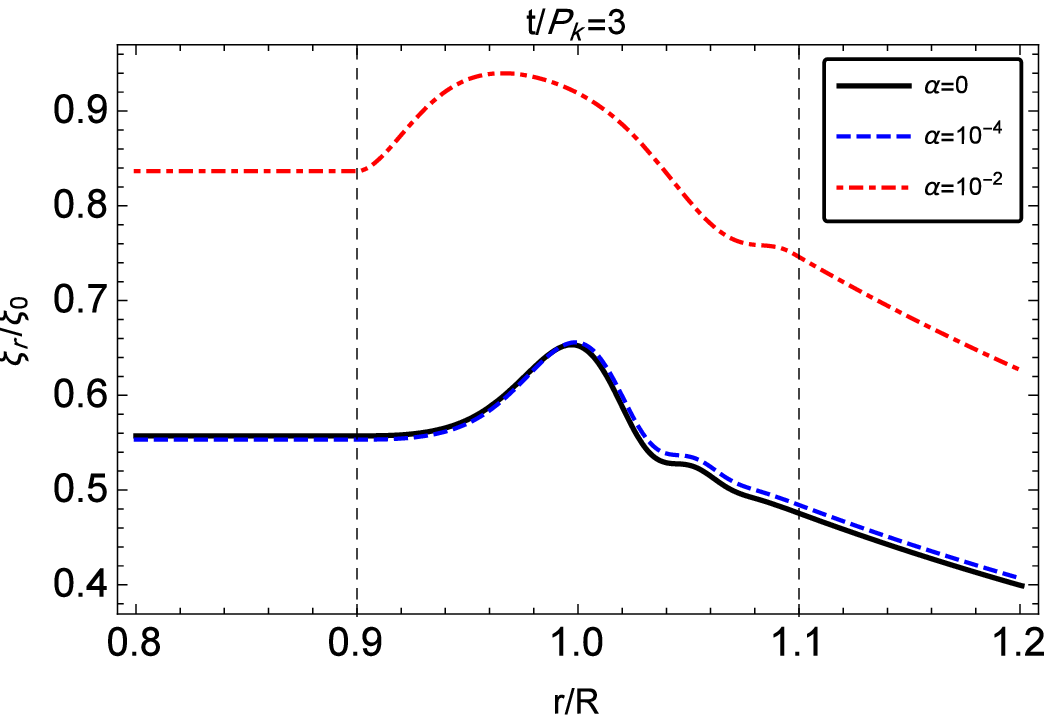}& \includegraphics[width=50mm]{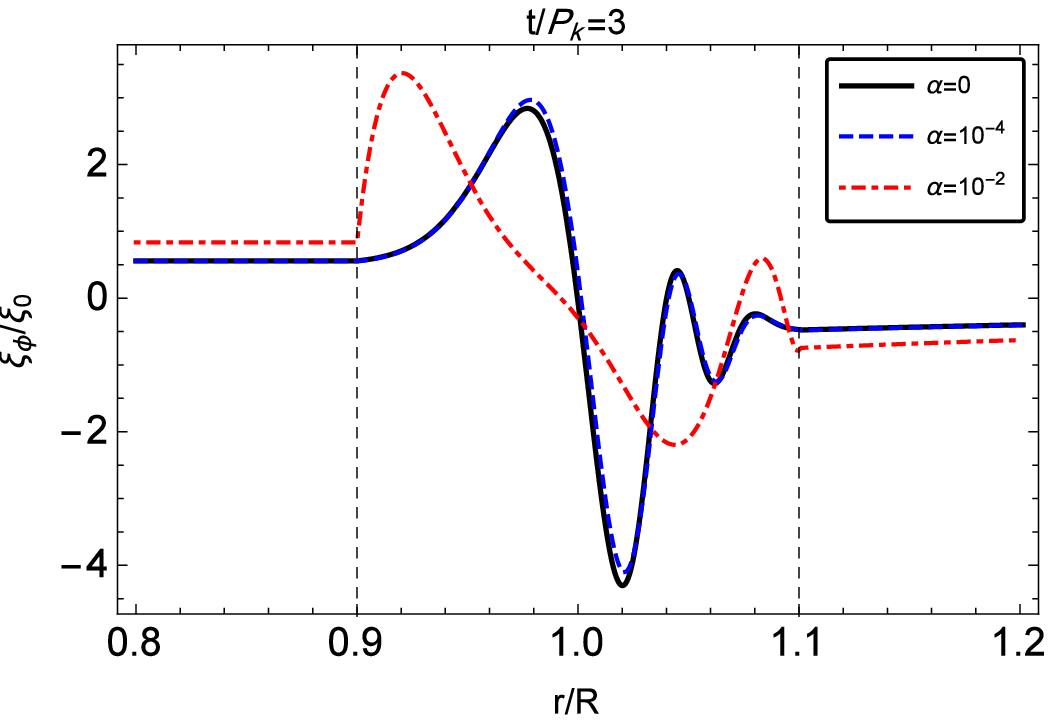}& \includegraphics[width=50mm]{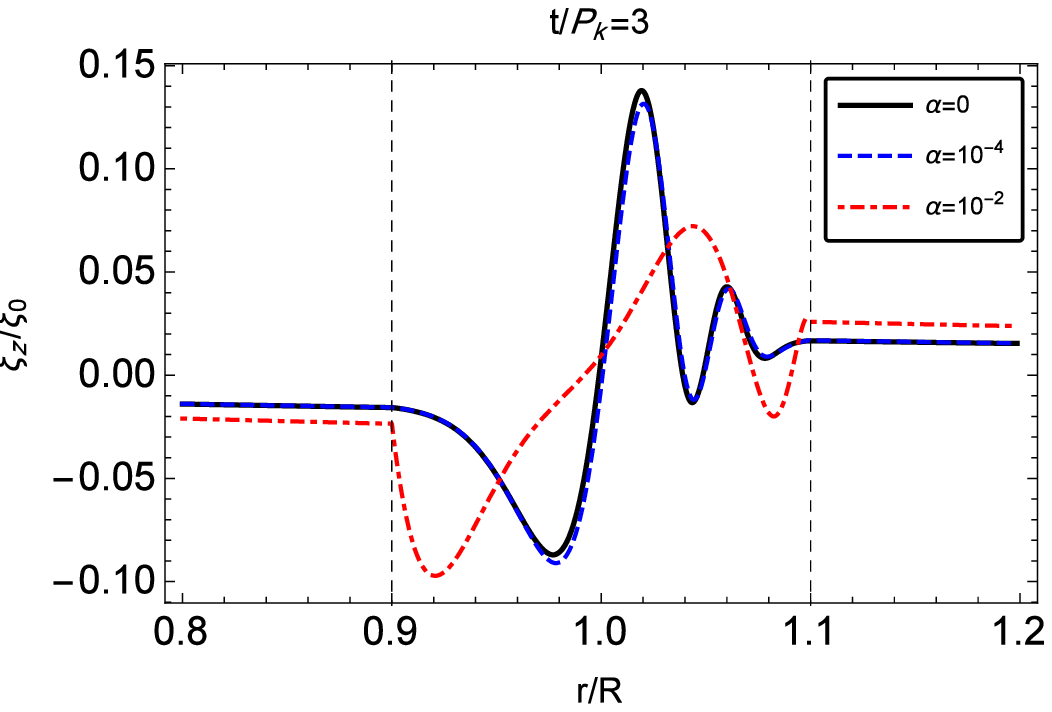}\\
    \includegraphics[width=50mm]{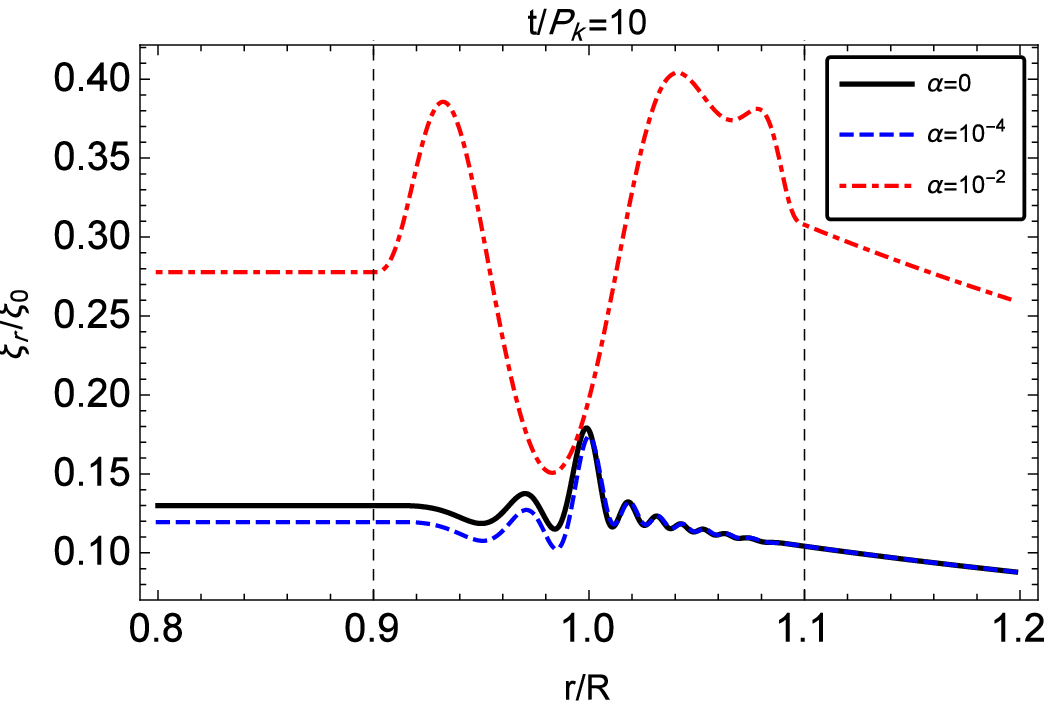}& \includegraphics[width=50mm]{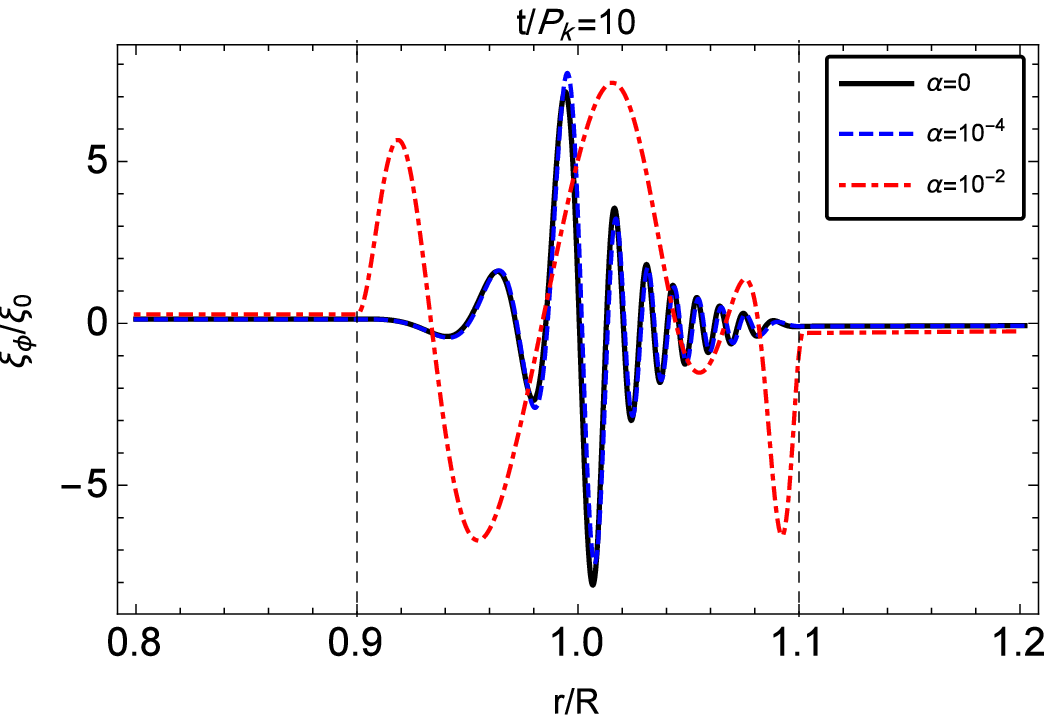}& \includegraphics[width=50mm]{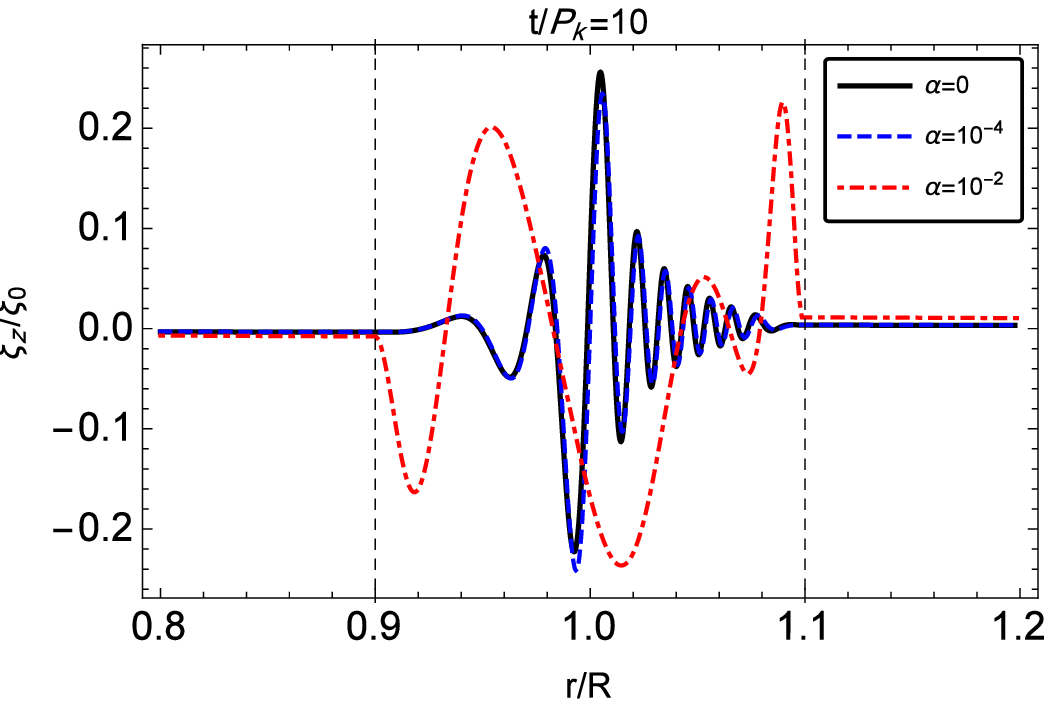}\\
  \end{tabular}
    \caption {Same as Fig. \ref{xi_MN_m(1)_lbyR(0.2)}, but for $m=-1$.}
    \label{xi_MN_m(-1)_lbyR(0.2)}
\end{figure}

\begin{figure}
  \centering
  \begin{tabular}{ccc}
    \includegraphics[width=50mm]{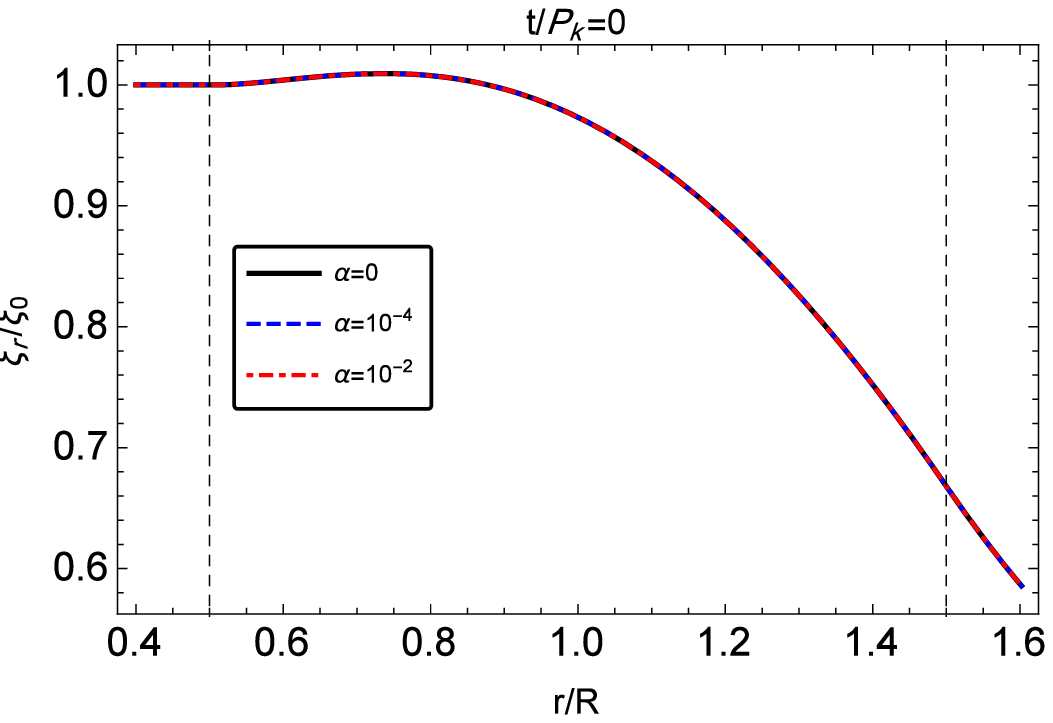}& \includegraphics[width=50mm]{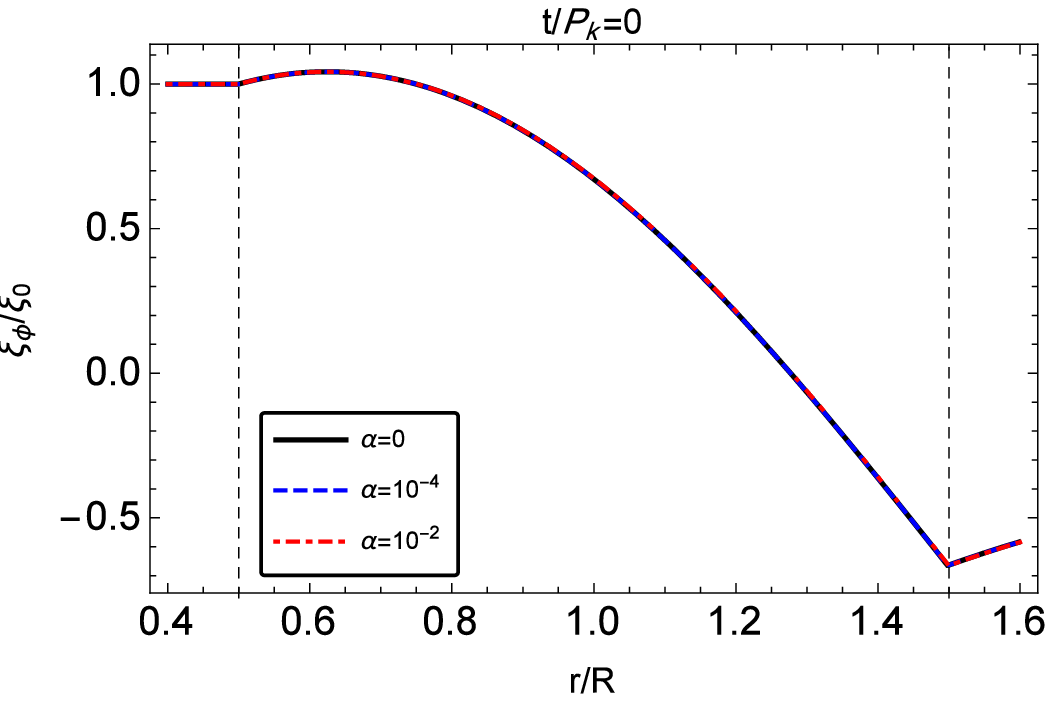}& \includegraphics[width=50mm]{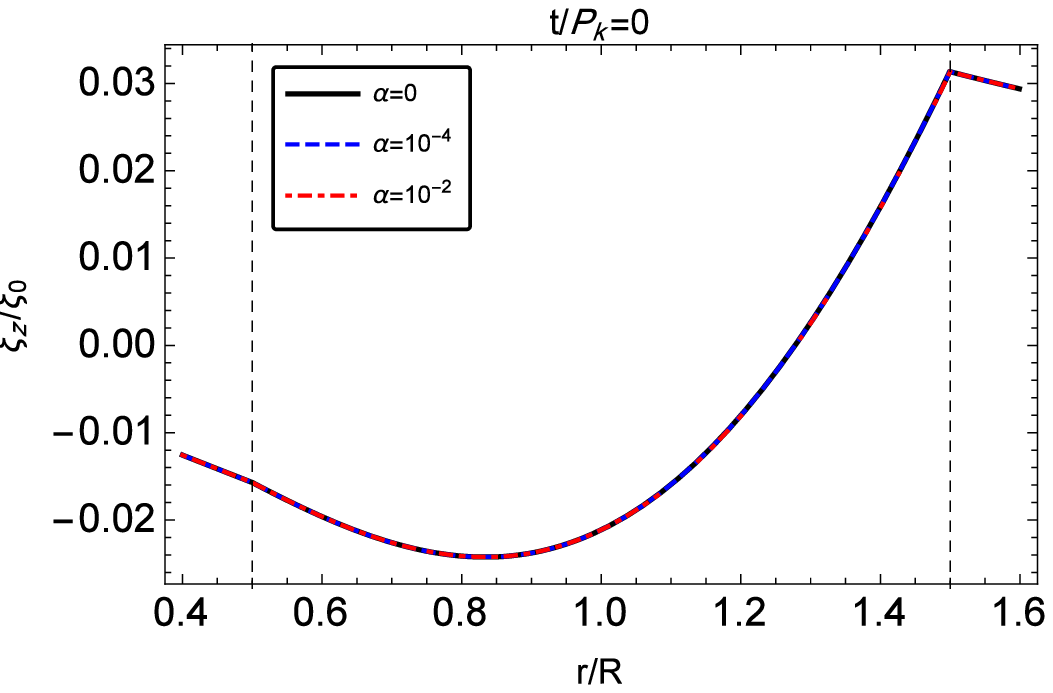}\\
    \includegraphics[width=50mm]{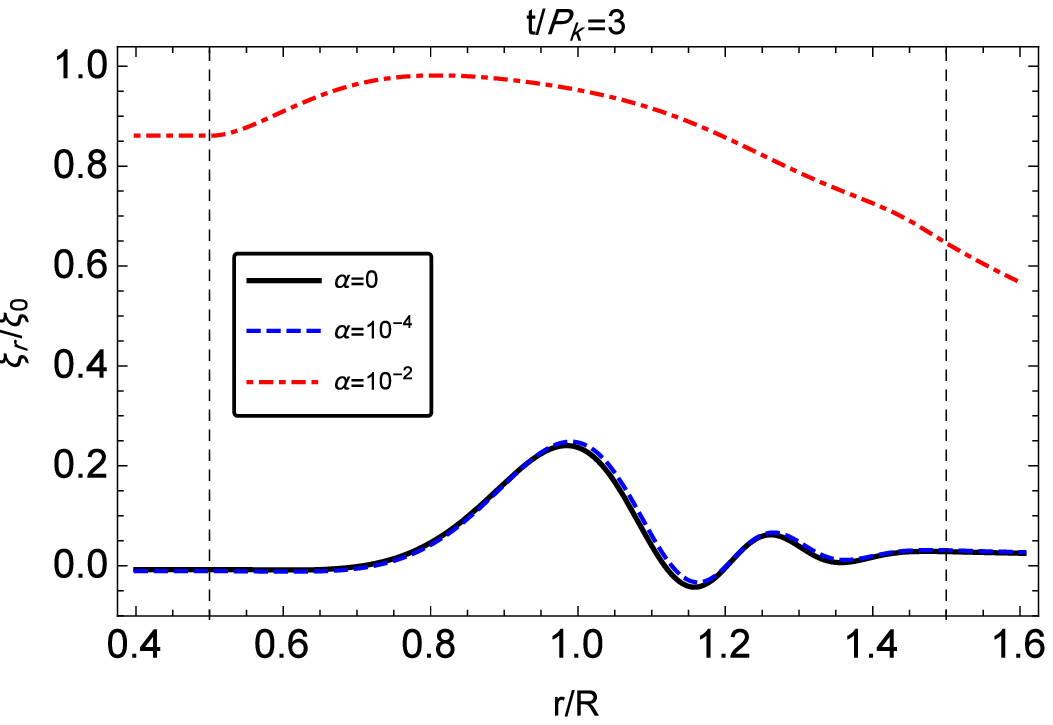}& \includegraphics[width=50mm]{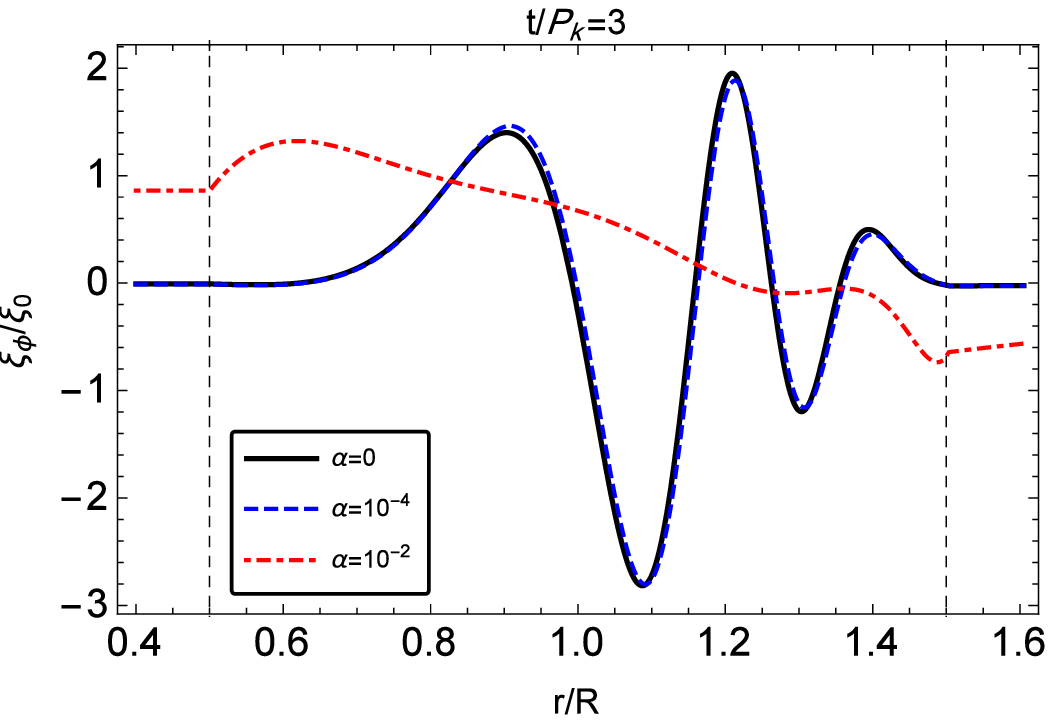}& \includegraphics[width=50mm]{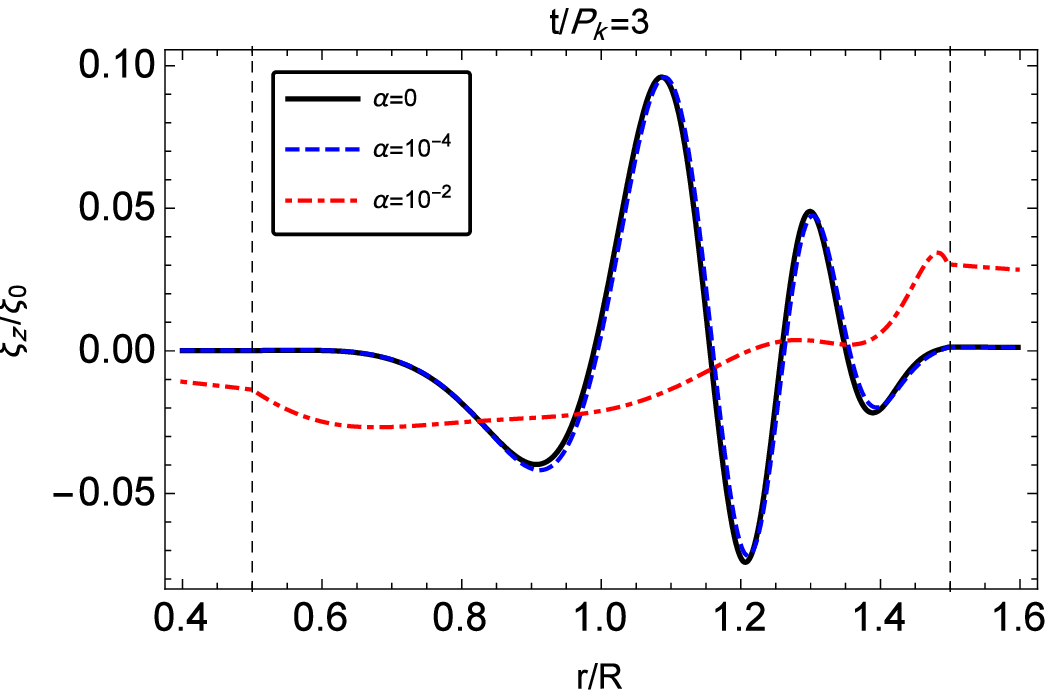}\\
    \includegraphics[width=50mm]{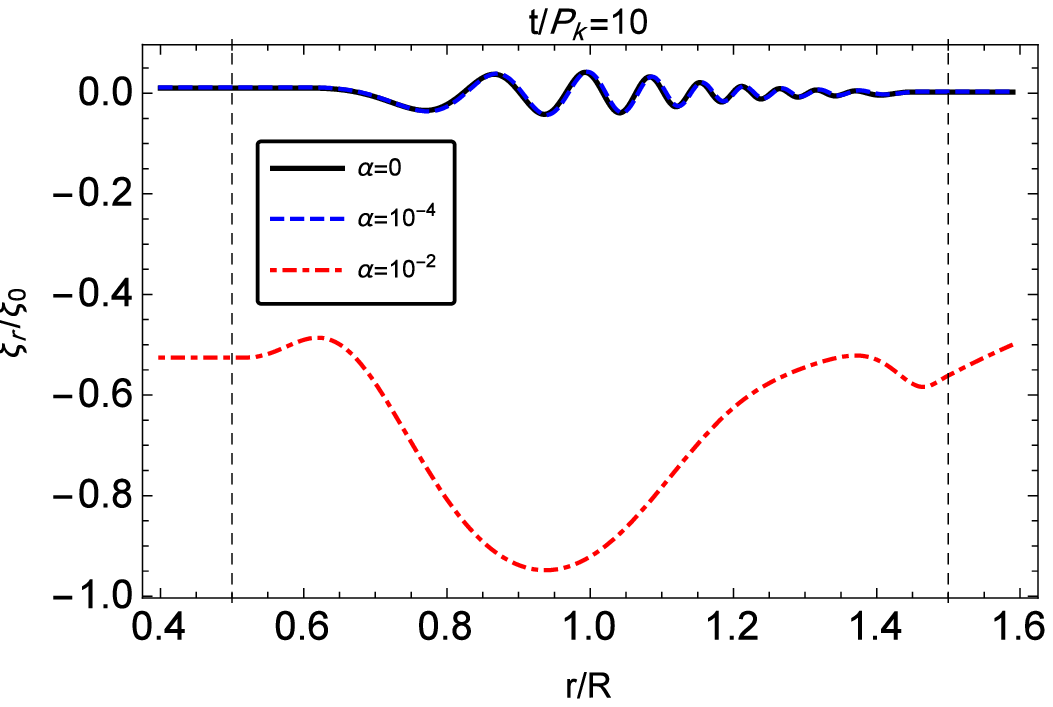}& \includegraphics[width=50mm]{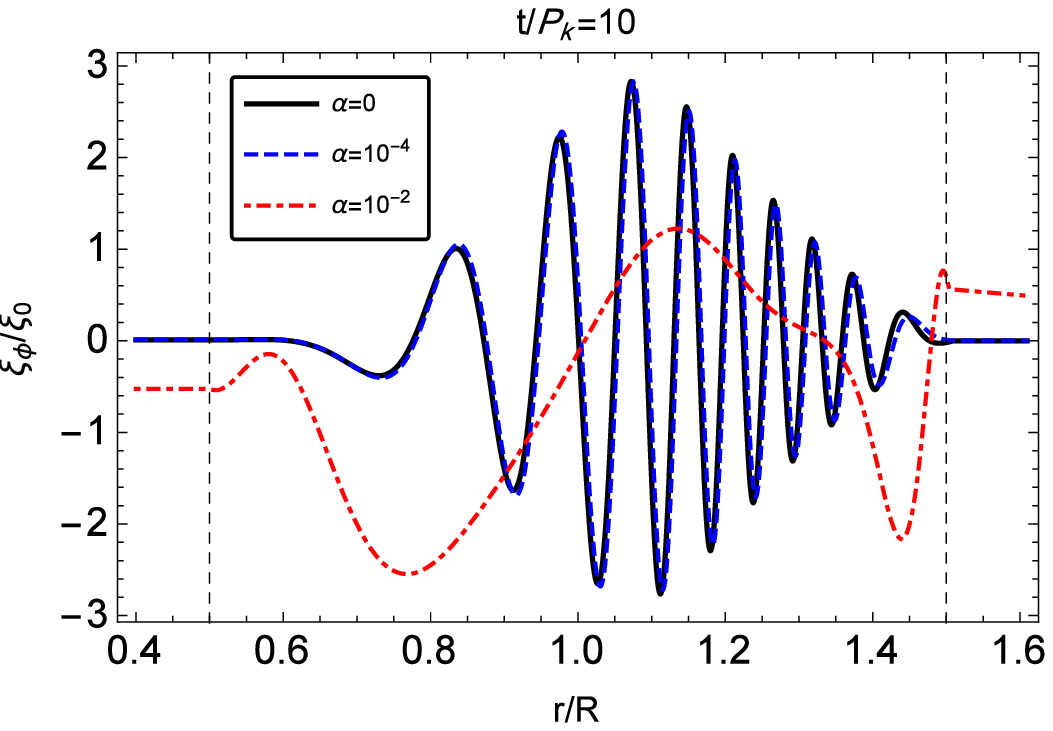}& \includegraphics[width=50mm]{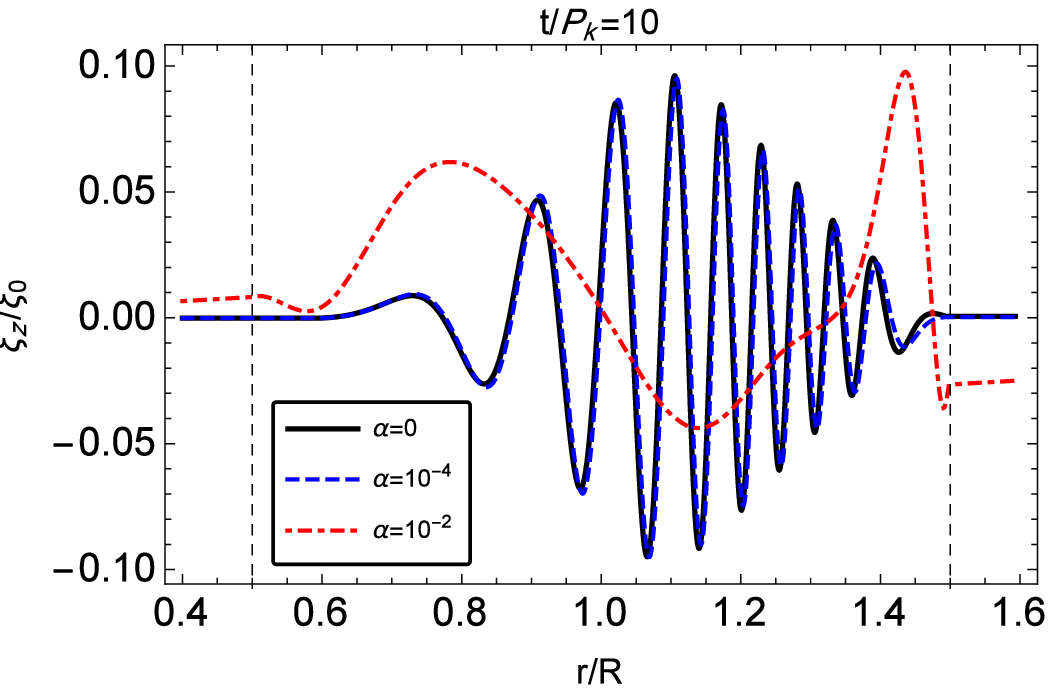}\\
  \end{tabular}
\caption {Same as Fig. \ref{xi_MN_m(1)_lbyR(0.2)}, but for $m=-1$ and $l/R=1$.}
    \label{xi_MN_m(-1)_lbyR(1)}
\end{figure}

\begin{figure}
\centering
    \begin{tabular}{cc}
        \includegraphics[width=75mm]{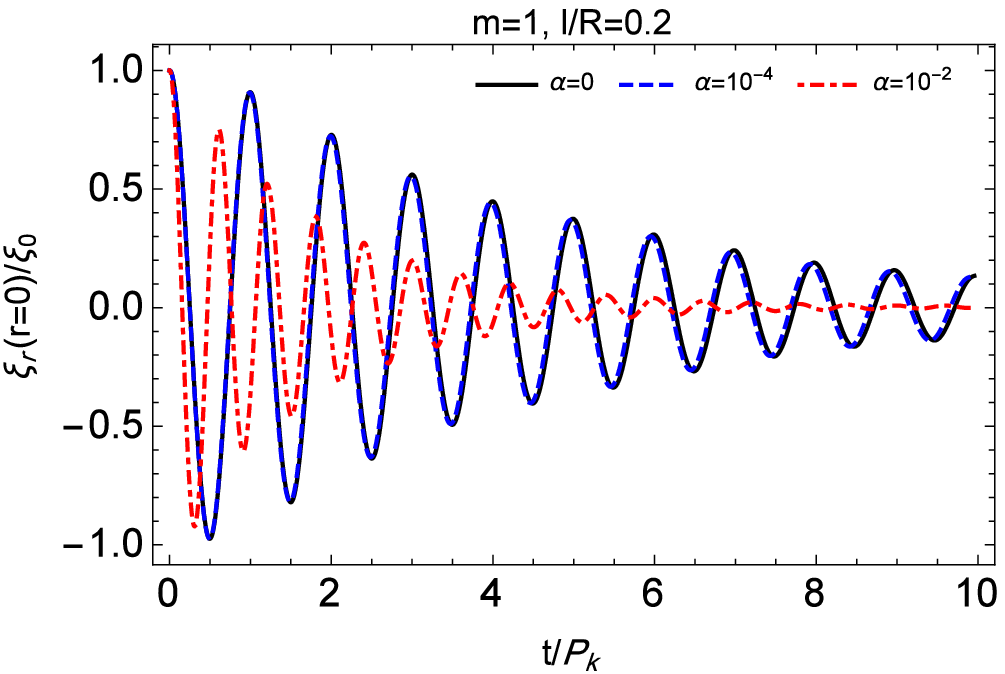}& \includegraphics[width=75mm]{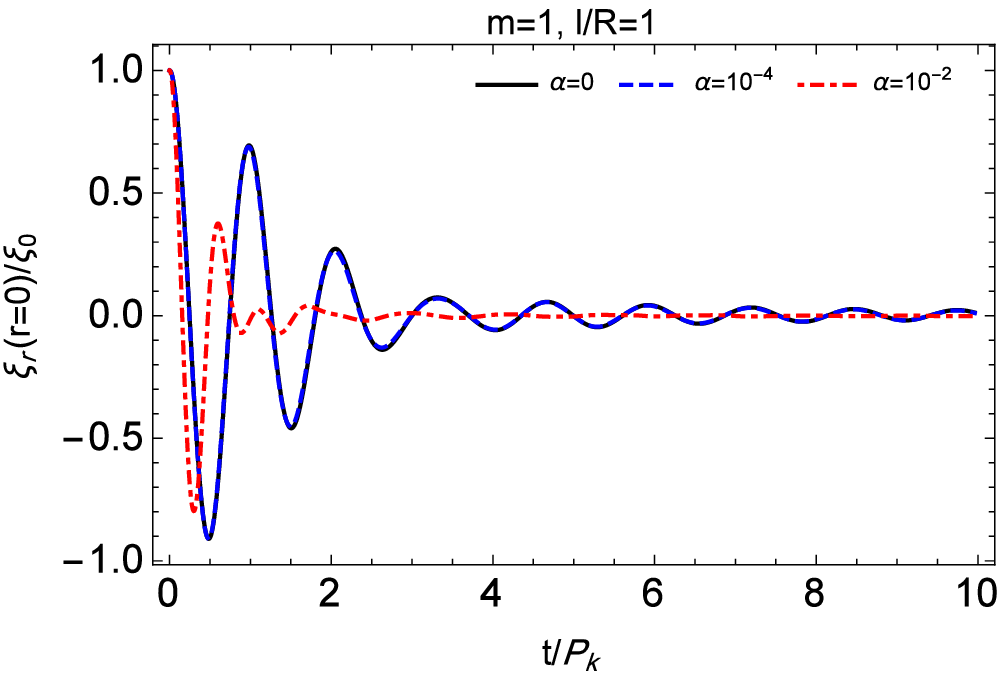}\\
        \includegraphics[width=75mm]{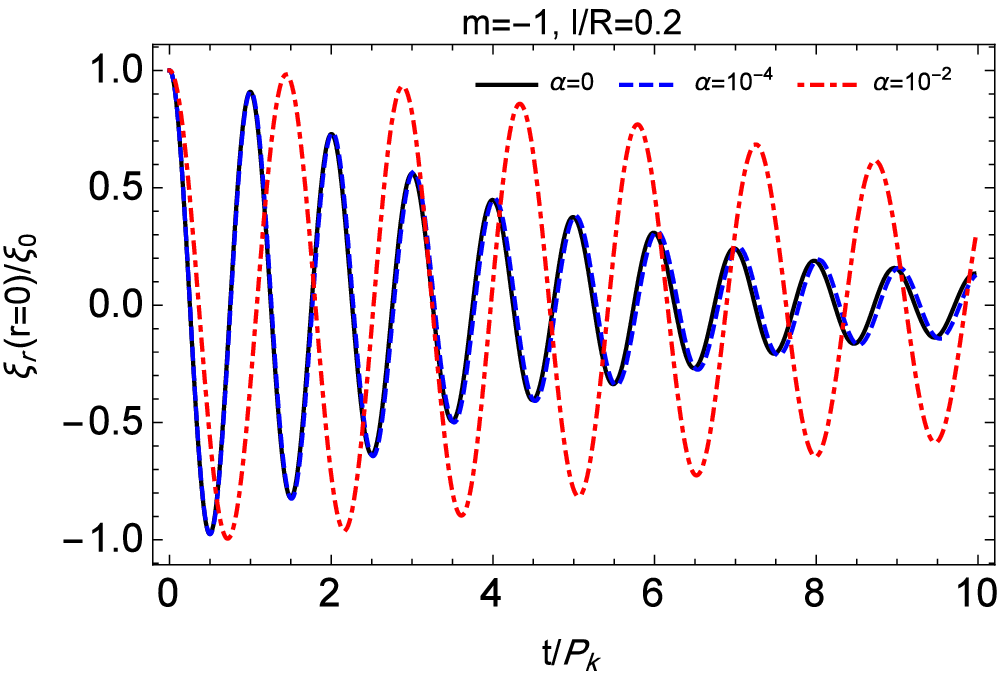}& \includegraphics[width=75mm]{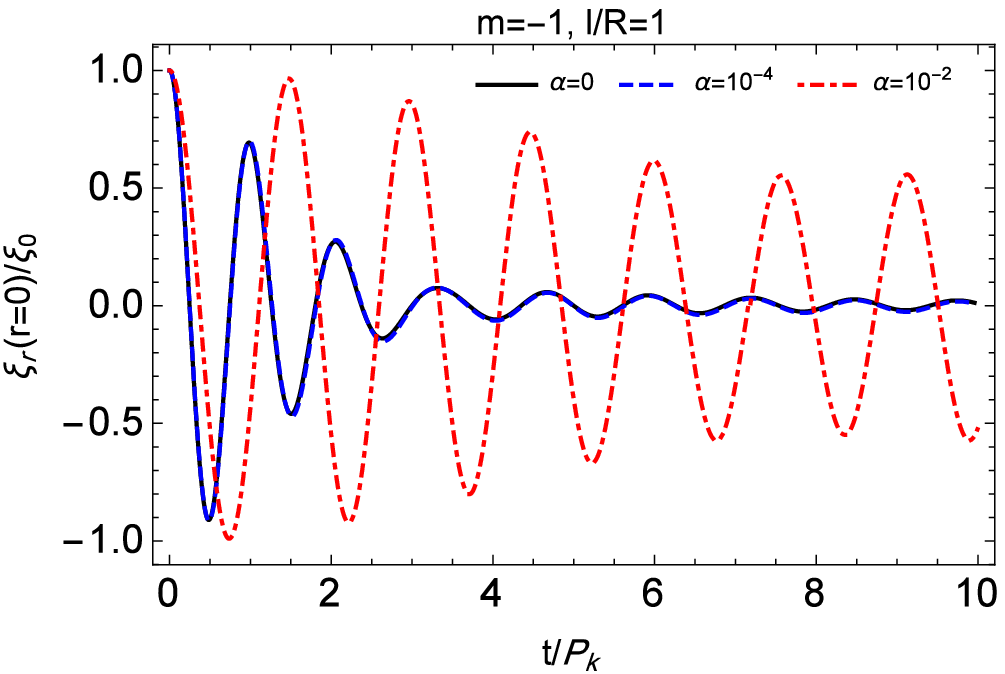}\\
    \end{tabular}
    \caption {Temporal evolution of $\xi_r(r=0)/\xi_0$ for $\alpha=0$ (solid line), $\alpha=10^{-4}$ (dashed line) and $\alpha=10^{-2}$ (dot-dashed line) for the model I. Here, $l/R=0.2$ (left panels), $l/R=1$ (right panels), $m=+1$ (top panels), $m=-1$ (bottom panels). Other auxiliary parameters are as in Fig. \ref{omega_MN}.}
    \label{xir_MN}
 \end{figure}

\begin{figure}
\centering
    \begin{tabular}{cc}
        \includegraphics[width=75mm]{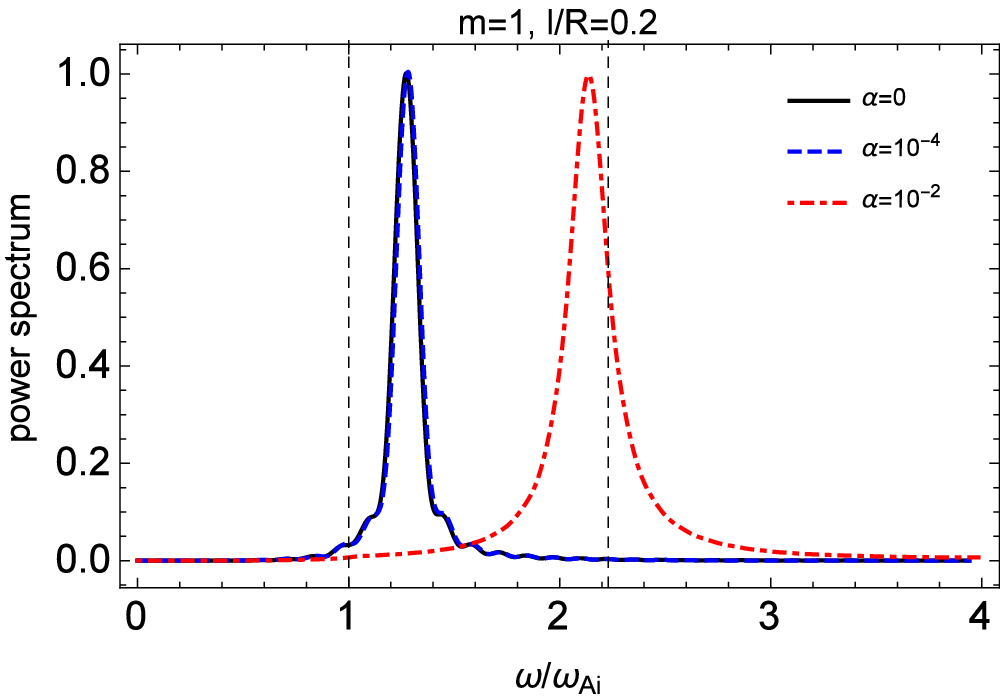}& \includegraphics[width=75mm]{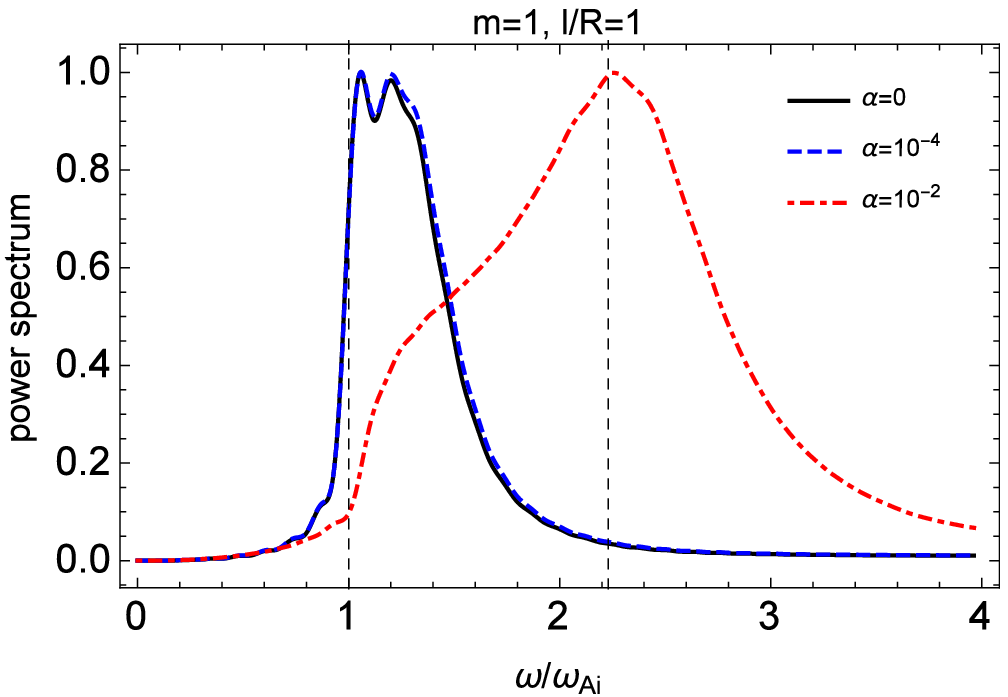}\\
        \includegraphics[width=75mm]{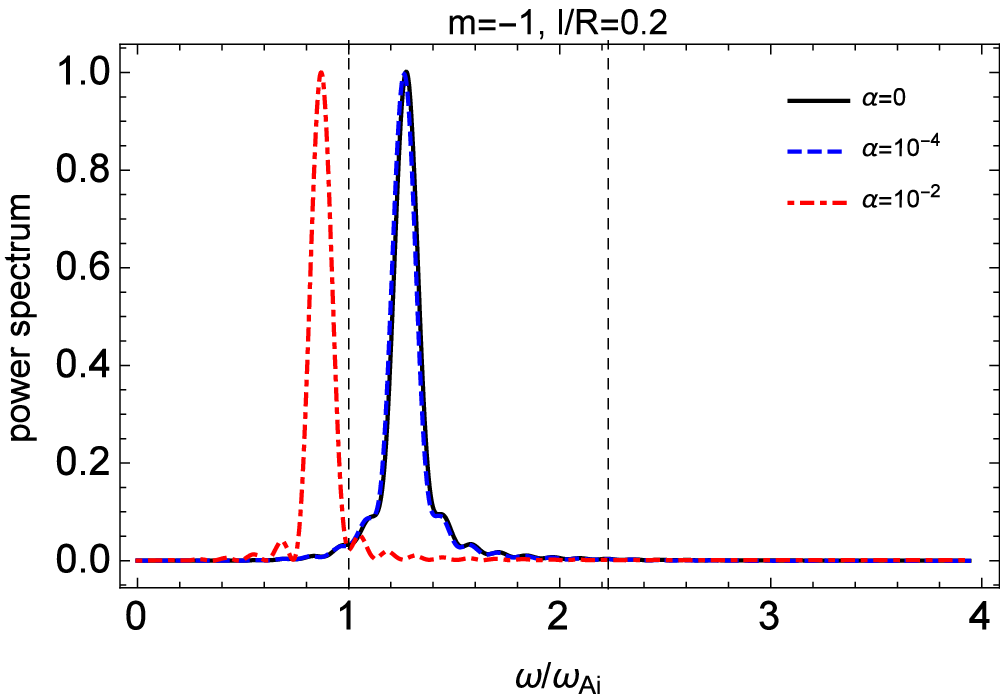}& \includegraphics[width=75mm]{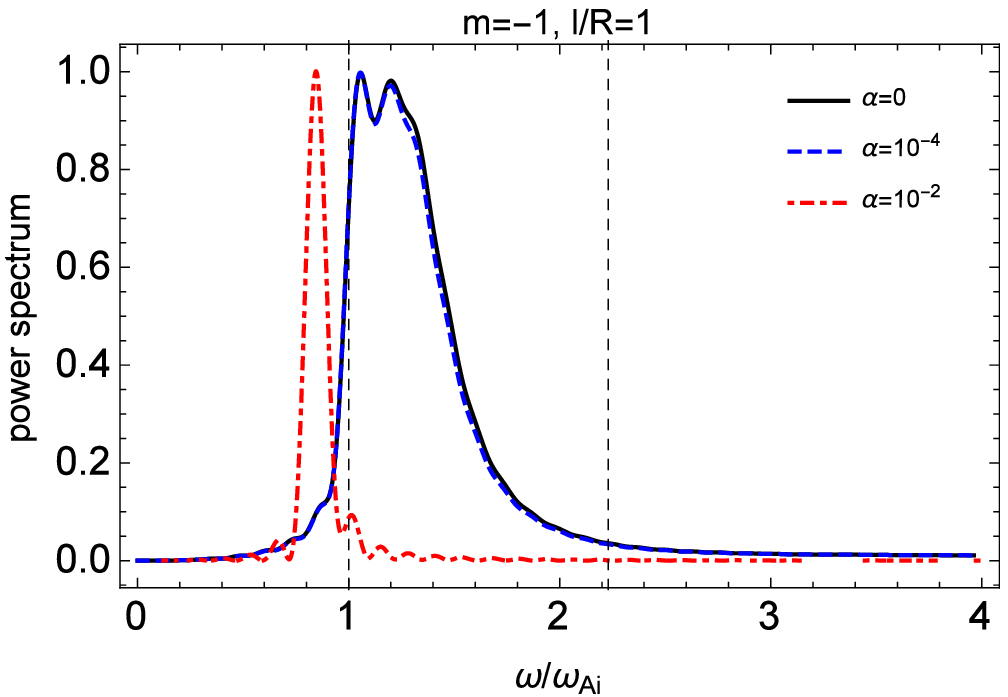}\\
    \end{tabular}
    \caption {Power spectrum of $\xi_r(r=0)$ for the model I for $l/R=0.2$ (left panels), $l/R=1$ (right panels), $m=+1$ (top panels), $m=-1$ (bottom panels). The left and right vertical dashed lines represents the interior and exterior Alfv\'{e}n frequencies, respectively. Other auxiliary parameters are as in Fig. \ref{omega_MN}.}
    \label{power_MN}
 \end{figure}

\begin{figure}
\centering
    \begin{tabular}{ll}
        \includegraphics[width=75mm]{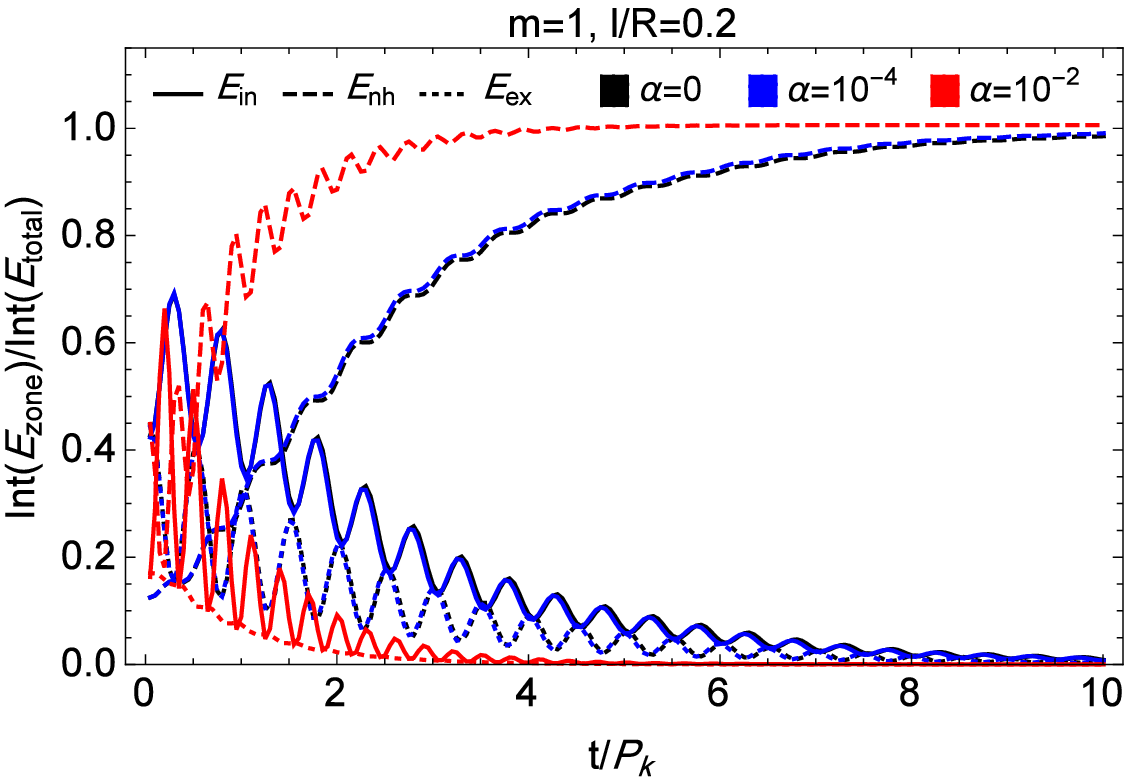}& \includegraphics[width=75mm]{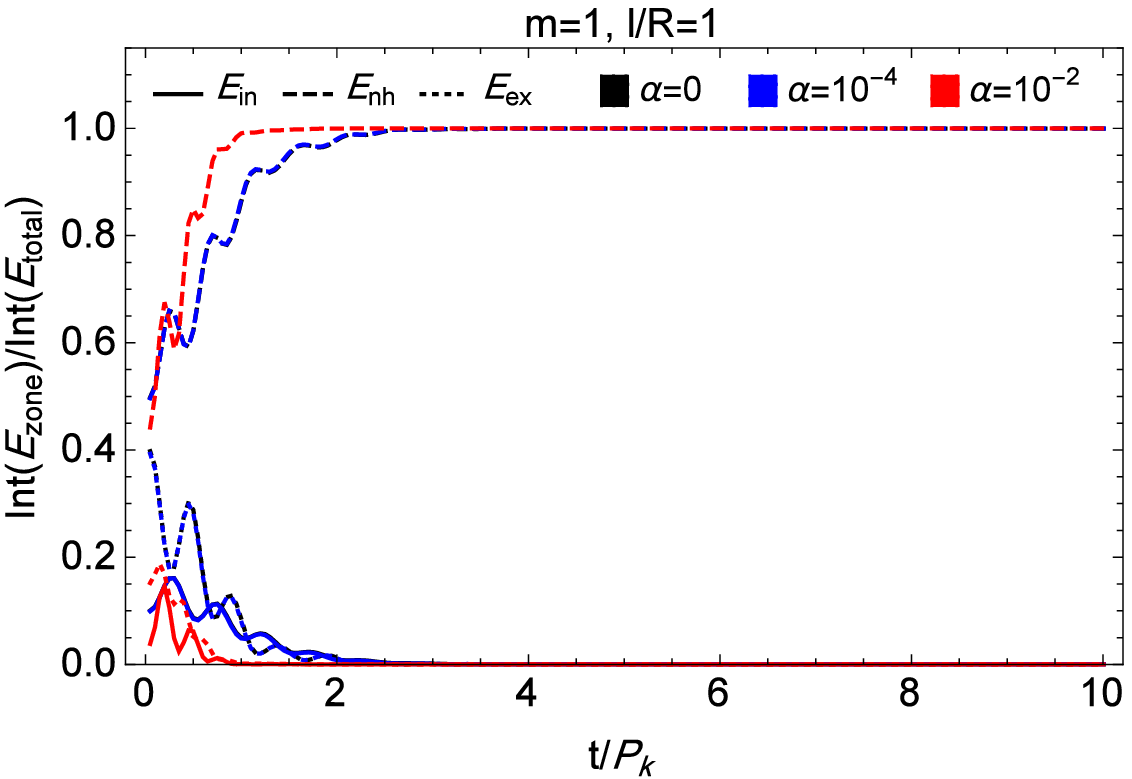}\\
        \includegraphics[width=75mm]{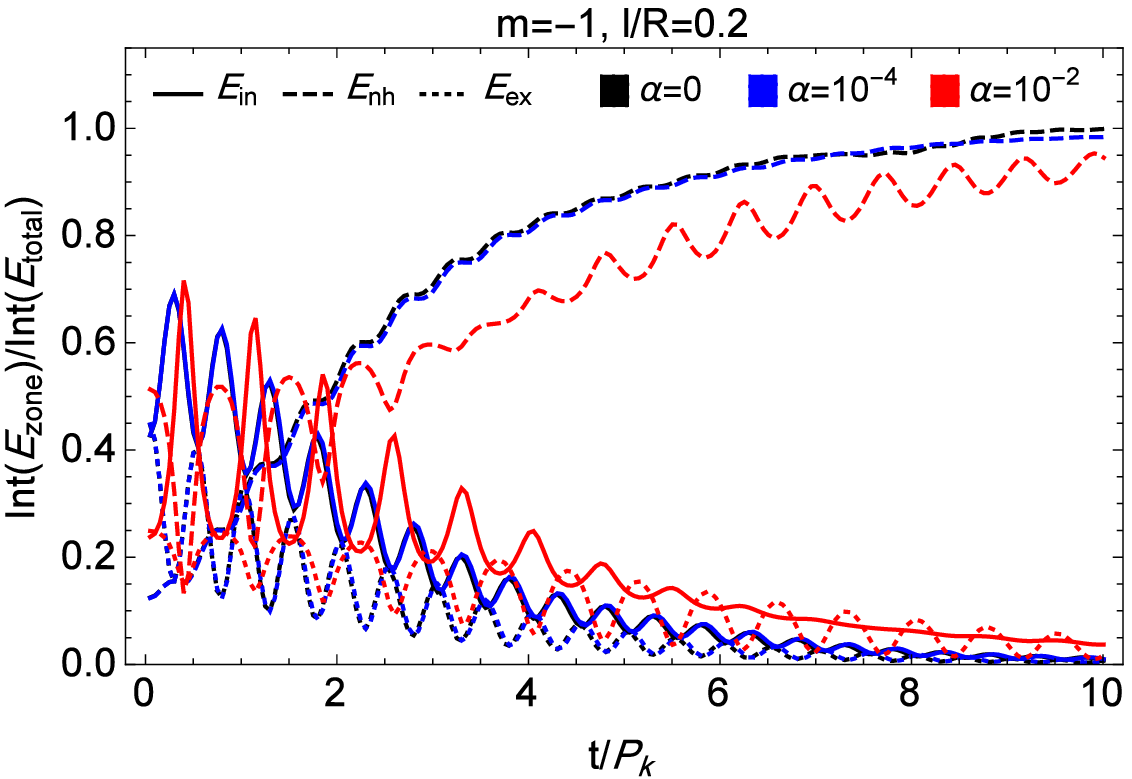}& \includegraphics[width=75mm]{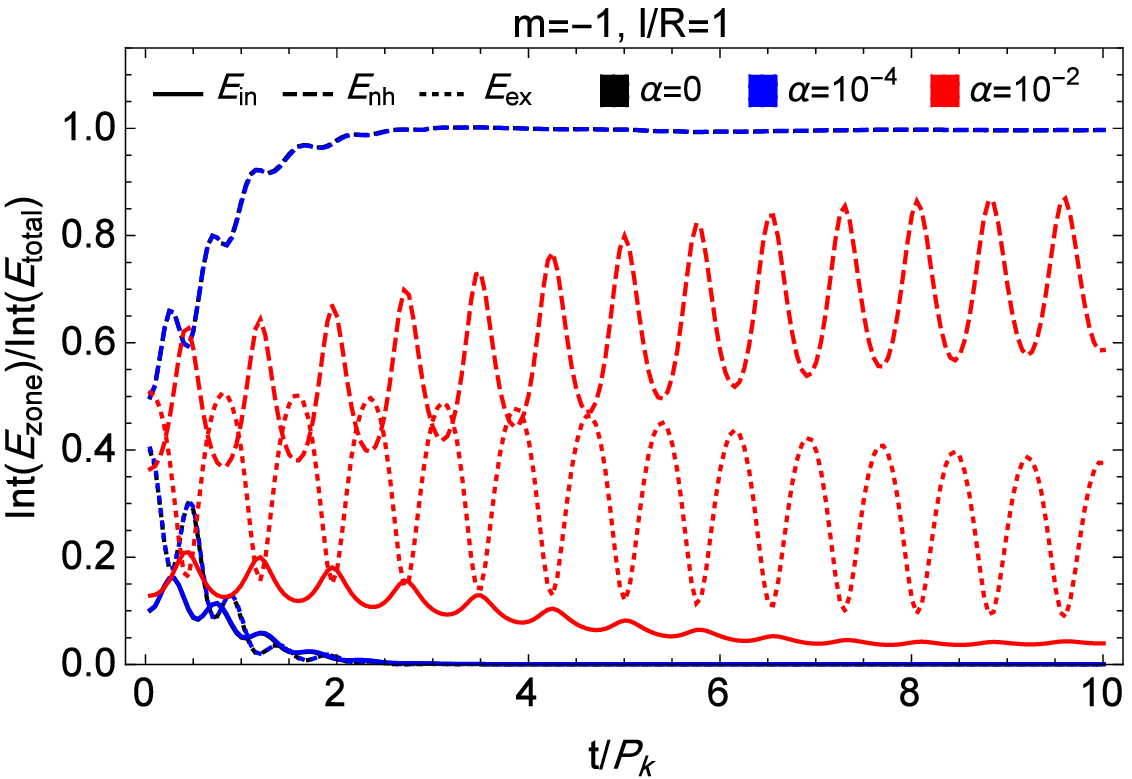}\\
    \end{tabular}
    \caption {Integrated energy of the interior ($E_{\rm {in}}$), nonhomogeneous ($E_{\rm {nh}}$) and exterior ($E_{\rm {ex}}$) regions of the loop as a function of time for the model I with $\alpha=0,~10^{-4}$ and $10^{-2}$ and $l/R=0.2$ (left panels), $l/R=1$ (right panels), $m=+1$ (top panels), $m=-1$ (bottom panels). Note that the results for $\alpha=0$ and $10^{-4}$ are very close together. Other auxiliary parameters are as in Fig. \ref{omega_MN}.}
    \label{intE_MN}
 \end{figure}

Here, we solve Eq. (\ref{eig}) for $\alpha=0,~10^{-4},~10^{-2}$. The twist parameters considered here are not large enough to allow the flux tube to be kink unstable (see Eq. \ref{alphamax}). The results for $k_z R=\pi/100$, $m=\pm1$, $l/R=0.2$ (thin layer) and $l/R=1$ (thick layer) are plotted in Figs. \ref{omega_MN}-\ref{intE_MN}. In the case of $m=-1$ and $\alpha=10^{-2}$ we set $N=300$, but in other cases we set $N=101$. Figure \ref{omega_MN} shows the background Alfv\'{e}n frequency, $\omega_{A}(r)$, and the corresponding discrete eigenfrequencies, $\omega_n$, in the region $r_1<r<r_2$ for three values of $\alpha=0,~10^{-4}$ and $10^{-2}$. Note that the results for $\alpha=0$ and $10^{-4}$ overlap with each other. Here, we consider the small value of the twist parameter, $\alpha=10^{-4}$, to show that how the results for the twisted magnetic field converge to the results of the untwisted magnetic field. The case with $\alpha = 0$ corresponds to the model used by ST2015.


Figure \ref{cn_MN} illustrates the normalized values of $|c_n|$ versus their corresponding eigenfrequencies, $\omega_n$. The figure reveals that for $m=+1$ and $l/R=0.2,~1$, by increasing the twist parameter, $\alpha$, the peaks of the diagrams become wider and shift toward the larger frequencies. The situation for $m=-1$ is different. As illustrated in the bottom panels of Fig. \ref{cn_MN}, for $m=-1$, the peaks become narrower and shift to the smaller frequencies. Since $|c_n|$ is the amplitude of the $n$th Alfv\'{e}n discrete mode, the increase/decrease in the width of the frequency distribution implies that with increasing the twist parameter, the number of Alfv\'{e}n discrete modes that have main contribution to the total displacements increases/decreases. As shown later, the wider frequency distribution results in an enhanced efficiency and a narrower frequency distribution results in a reduced efficiency of the phase-mixing process compared to the case with no twist.

Figures \ref{xi_MN_m(1)_lbyR(0.2)} to \ref{xi_MN_m(-1)_lbyR(1)} display the temporal evolution of different components of the Lagrangian displacement for (i) $m=+1$ \& $l/R=0.2$, (ii) $m=+1$ \& $l/R=1$, (iii) $m=-1$ \& $l/R=0.2$ and (iv) $m=-1$ \& $l/R=1$, respectively. As illustrated in Figs. \ref{xi_MN_m(1)_lbyR(0.2)} and \ref{xi_MN_m(1)_lbyR(1)}, corresponding to positive $m$, in the presence of magnetic twist, at a given time, the perturbations are more phase-mixed than those in the case of untwisted magnetic field ($\alpha=0$). Figures \ref{xi_MN_m(-1)_lbyR(0.2)} and \ref{xi_MN_m(-1)_lbyR(1)}, corresponding to negative $m$, show that at a given time, the perturbations are less phase-mixed than in the case of untwisted magnetic field. Thus, in model I for $k_zR=\pi/100$ and $m=+1/-1$, the small spatial scales due to phase-mixing in the nonuniform layer develop faster/slower than in the case of a straight field.

Figure \ref{xir_MN} shows the temporal evolution of $\xi_r/\xi_0$ at $r=0$ for $m=\pm1$ and $l/R=0.2,~1$. As shown in this figure, for $m=+1/-1$, in the presence of twisted magnetic field, $\xi_r$ decays faster/slower than that of in the case of no twist for both $l/R=0.2$ and 1. Note that in Fig. \ref{cn_MN} for a given $l/R$, the frequency distribution of the Alfv\'{e}n continuum modes for $m=+1/-1$ is wider/narrower for larger values of the twist parameter. This is consistent with the results of ST2015 who showed that when the distribution of the frequencies of the Alfv\'{e}n continuum modes has a wider peak, there should be a larger damping rate.

Figure \ref{power_MN} shows the power spectrum of $\xi_r(r=0)$ in the time interval $t\in[0,10 P_k]$ for $\alpha=0,~10^{-4}$ and $10^{-2}$. As illustrated in the figure, for $l/R=0.2$ and 1, by increasing the twist parameter, the peak frequency of the power spectrum increases for $m=+1$ and decreases for $m=-1$. This behavior is consistent with the results of Terradas \& Goossens (2012) and Ruderman (2015) who showed that for $k_z>0$ and $m=+1/-1$, by increasing the magnetic twist in a coronal flux tube, the MHD kink frequency increases/decreases.

To illustrate the flux of the total (kinetic plus magnetic) energy from the internal and external regions to the inhomogeneous region, we calculate the integrated energy in each region as
 \begin{eqnarray}\label{Eint}
    \nonumber E_{\rm {in}}=\int_0^{r_1}\frac{1}{2}\left(\rho\left|\frac{\partial \boldsymbol{\xi}}{\partial t}\right|^2+\frac{1}{\mu}\left|\delta \mathbf{B}\right|^2\right)r~{\rm d}r,\\
    E_{\rm {nh}}=\int_{r_1}^{r_2}\frac{1}{2}\left(\rho\left|\frac{\partial \boldsymbol{\xi}}{\partial t}\right|^2+\frac{1}{\mu}\left|\delta \mathbf{B}\right|^2\right)r~{\rm d}r,\\
    \nonumber E_{\rm {ex}}=\int_{r_2}^{\infty}\frac{1}{2}\left(\rho\left|\frac{\partial \boldsymbol{\xi}}{\partial t}\right|^2+\frac{1}{\mu}\left|\delta \mathbf{B}\right|^2\right)r~{\rm d}r,
\end{eqnarray}
where $\delta \mathbf{B}$ is calculated from Eq. (\ref{deltaB}). Here, $E_{\rm {in}}$, $E_{\rm {nh}}$ and $E_{\rm {ex}}$ are the integrated energies in the internal, nonhomogeneous and external regions, respectively. Note that in order to compute the third integral of Eq. (\ref{Eint}) we must replace the upper limit of the integral, $\infty$, with a sufficiently large radius (here $r/R=20$) where the amplitudes of the perturbations are approximately zero. Figure \ref{intE_MN} shows the integrated energy in these three regions as a function of time for $\alpha=0,~10^{-4},~10^{-2}$. It is clear from this figure that for both $l/R=0.2,~1$, when the twist parameter increases, for $m=+1/-1$ the rate of energy transfer from the internal and external regions to the inhomogeneous region increases/decreases. Hence, for $m=+1/-1$ the efficiency of the resonant absorption increases/decreases as the twist parameter becomes larger.\\\\

\subsection{Model II: continuous magnetic field}\label{con}

Following Terradas \& Goossens (2012), in order to have a continuous magnetic field we consider a parabolic profile for the azimuthal component of the magnetic field (in the region $r_1<r<r_2$) as
\begin{equation}\label{Bphi2}
    B_{0\varphi}(r)=A(r-r_1)(r_2-r).
\end{equation}
The corresponding gas pressure can be obtained in a similar way to the one followed for the model (\ref{Bdis}) as
\begin{equation}\label{gasp2}
    p(r)=\left\{\begin{array}{lll}
    p_0,&r\leqslant r_1,\\
    p_0+\frac{A^2}{2\mu_0}\Big(\frac{3}{2}r^4-\frac{10}{3}(r_1+r_2)r^3\\
        \left.+2(r_1^2+r_2^2+4r_1r_2)r^2\right.\\
        \left.-6(r_1^2r_2+r_1r_2^2)r-\frac{1}{6}r_1^4+\frac{4}{3}r_2r_1^3+4r_1^2r_2^2+2r_1^2r_2^2\ln(\frac{r}{r_1})\right),&r_1<r<r_2,\\\\
    p_0-\frac{A^2}{2\mu_0}\left(\frac{1}{6}(r_2^4-r_1^4)+\frac{4}{3}(r_2r_1^3-r_1r_2^3)\right),&r\geqslant r_2.
      \end{array}\right.
\end{equation}
\begin{figure}
\centering
    \begin{tabular}{cc}
        \includegraphics[width=70mm]{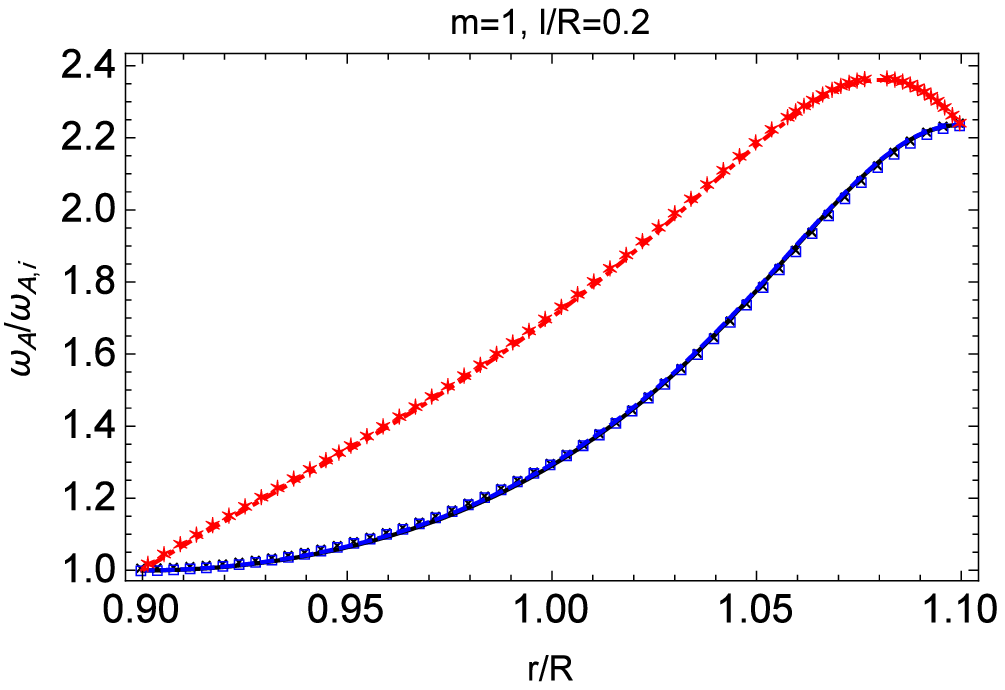}& \includegraphics[width=70mm]{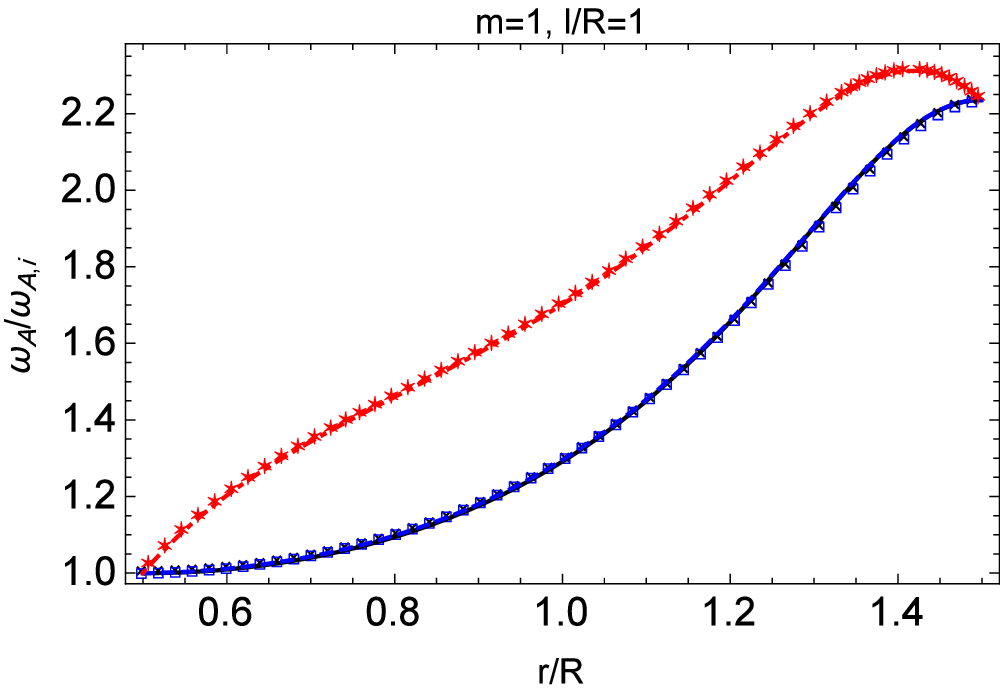}\\
        \includegraphics[width=70mm]{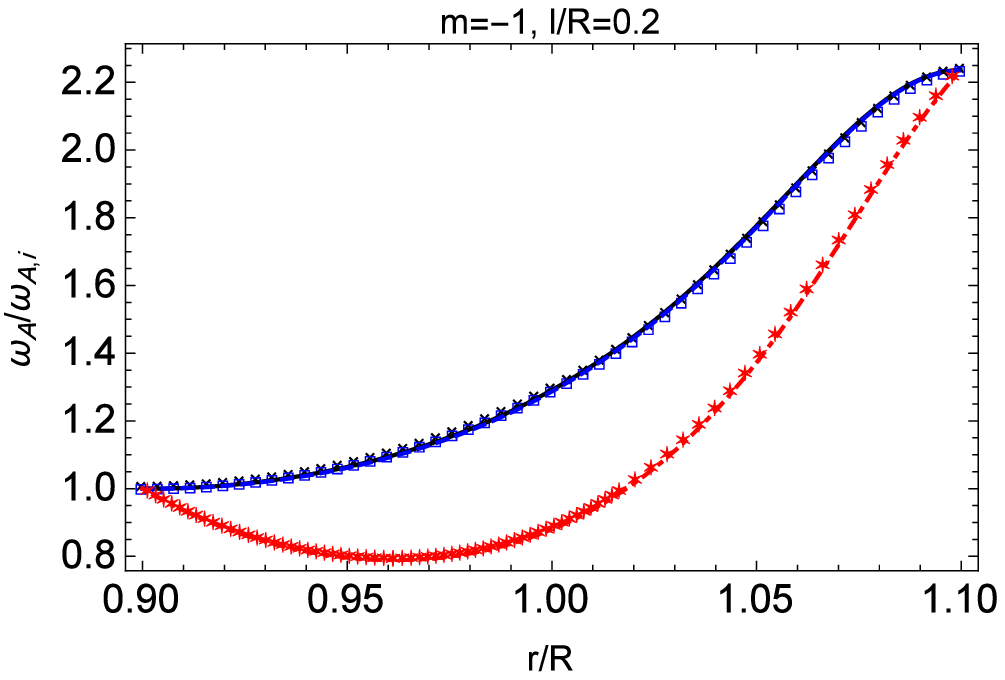}& \includegraphics[width=70mm]{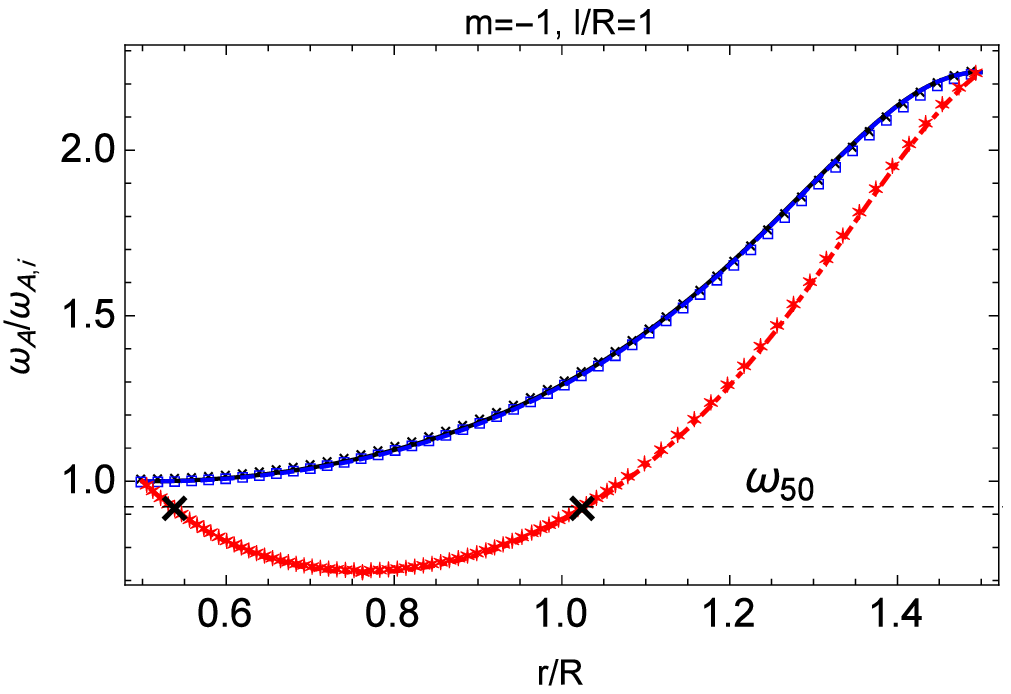}\\
    \end{tabular}
    \caption {Background Alfv\'{e}n frequency $\omega_A(r)$ in the annulus region ($r_1<r<r_2$) for model II. Here, $k_z R=\pi/100$ and $\alpha=0$ (solid line), $\alpha=10^{-4}$ (blue dashed line), $\alpha=10^{-2}$ (red dot-dashed line). The crosses, squares and asterisks, correspond to the discrete eigenfrequencies for $\alpha=0$, $10^{-4}$ and $\alpha=10^{-2}$, respectively. Here, $N=101$, but for convenience we show only multiples of 2. Note that the results for $\alpha=0$ and $10^{-4}$ overlap with each other. Here, $l/R=0.2$ (left panels), $l/R=1$ (right panels), $m=+1$ (top panels), $m=-1$ (bottom panels). The horizontal dashed line in the right panel denotes $\omega_{50}$ for $\alpha=10^{-2}$ and the big crosses are the locations where $\omega_A(r)=\omega_{50}$.}
    \label{omega_T2}
 \end{figure}
\begin{figure}
    \centering
    \includegraphics[width=70mm]{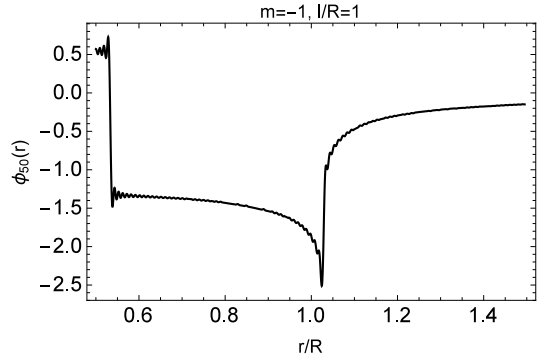}
    \caption {Eigenfunction of Alfv\'{e}n discrete mode, $\phi_{50}(r)$, for $m=-1$, $l/R=1$ and $\alpha=10^{-2}$ in the model II. Other Auxiliary parameters are as in Fig. \ref{omega_T2}.}
    \label{phi50_T2_m(-1)}
\end{figure}
\begin{figure}
\centering
    \begin{tabular}{cc}
       \includegraphics[width=70mm]{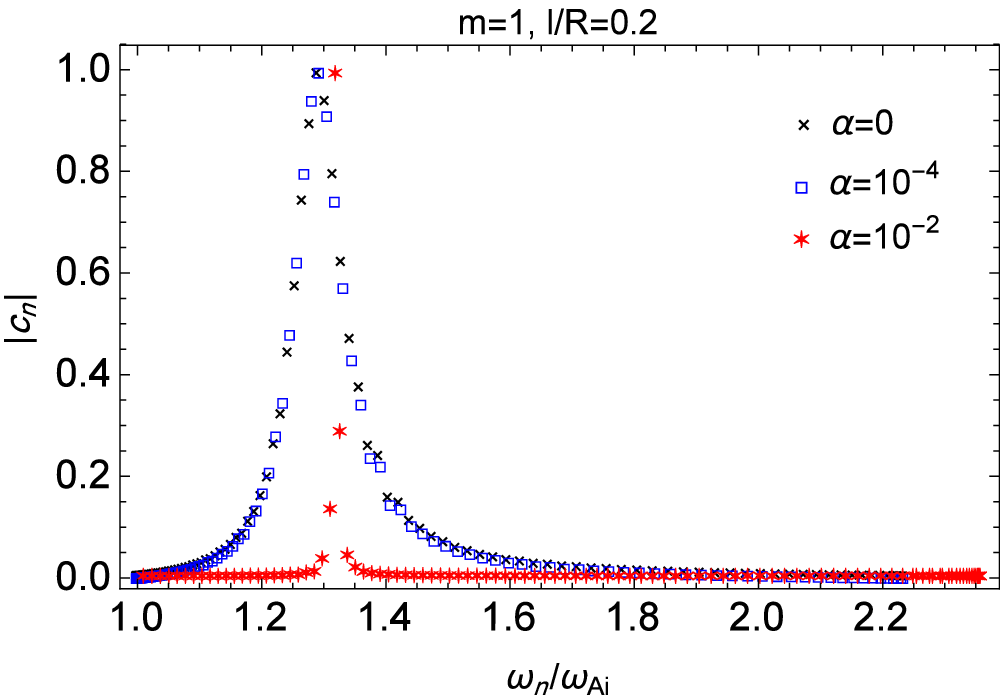}& \includegraphics[width=70mm]{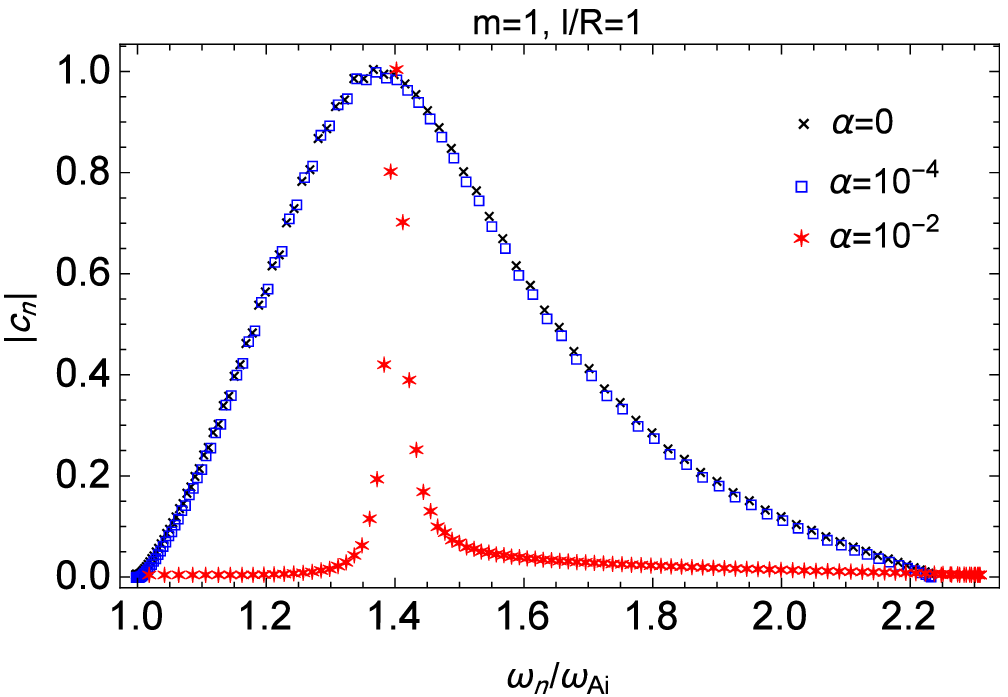}\\
       \includegraphics[width=70mm]{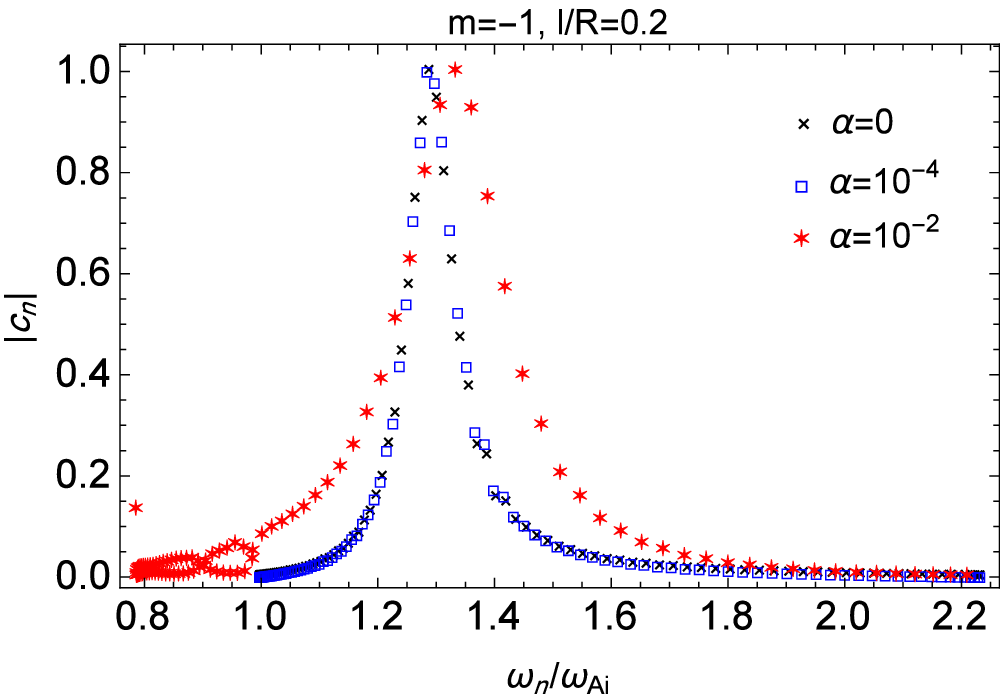}& \includegraphics[width=70mm]{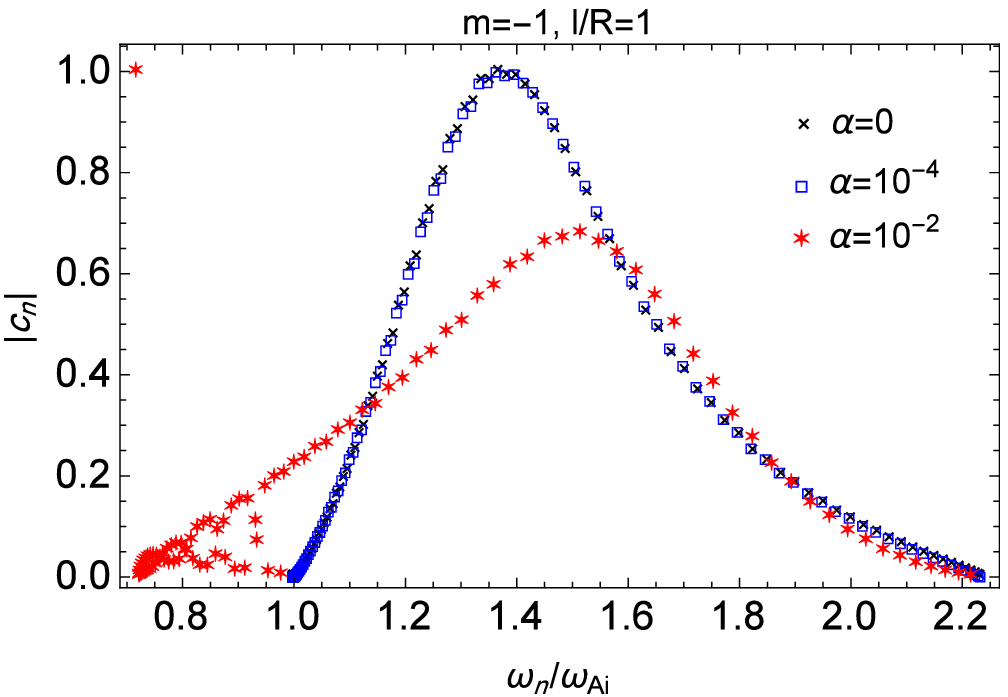}\\
    \end{tabular}
    \caption {Same as Fig. (\ref{cn_MN}) but for the model II.}
    \label{cn_T2}
\end{figure}
\begin{figure}
  \centering
  \begin{tabular}{ccc}
    \includegraphics[width=50mm]{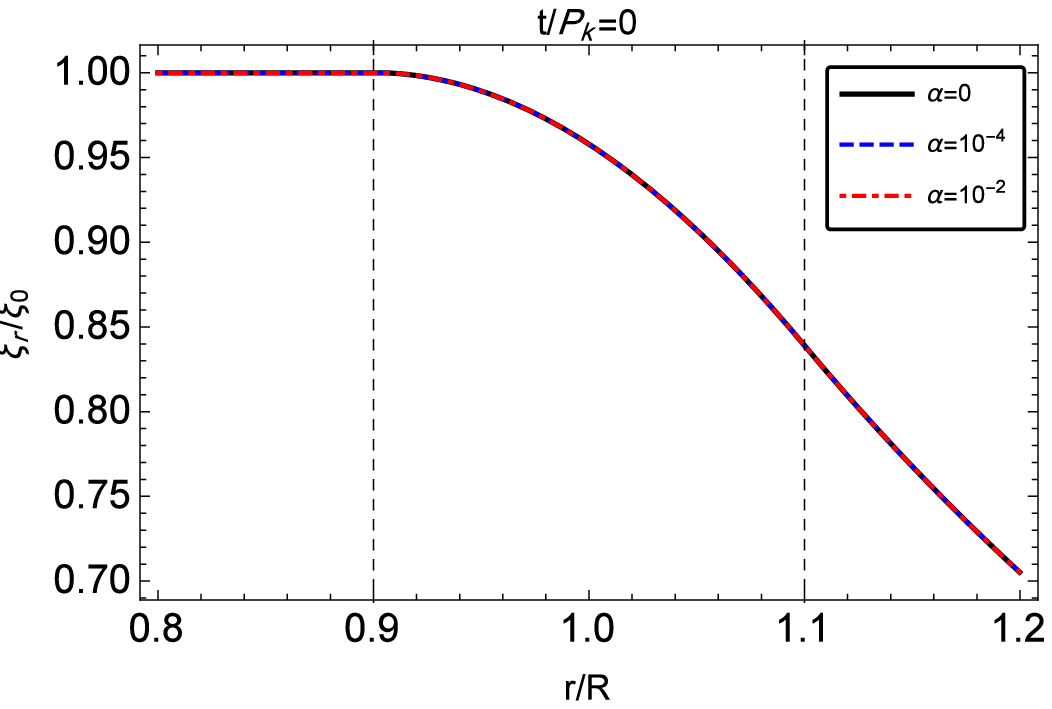}& \includegraphics[width=50mm]{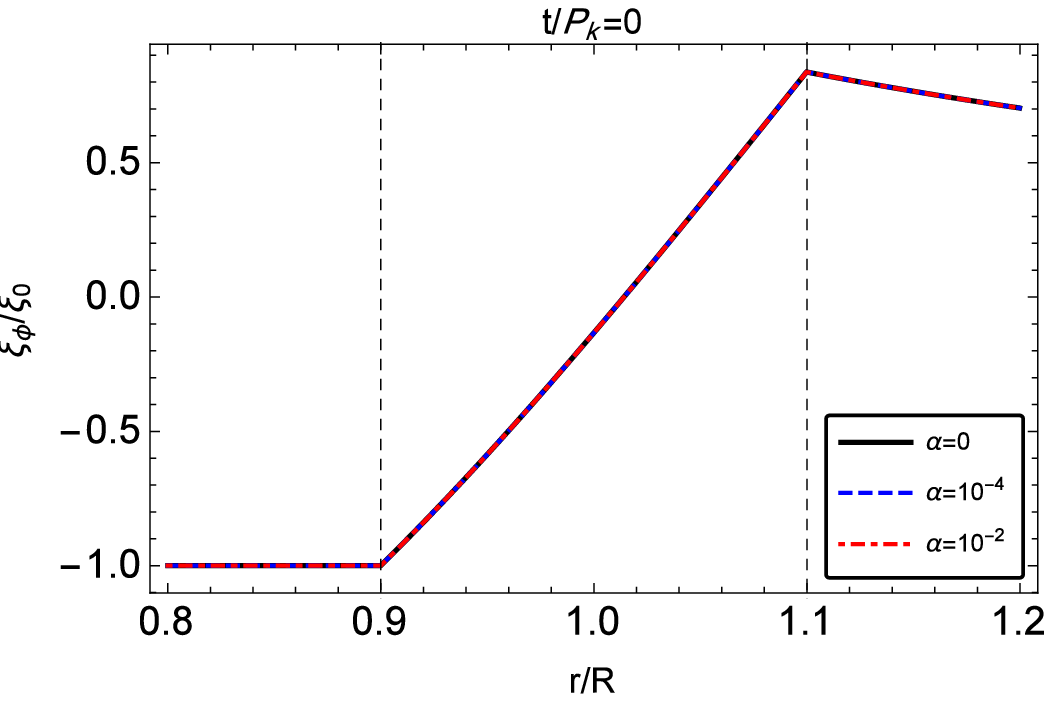}& \includegraphics[width=50mm]{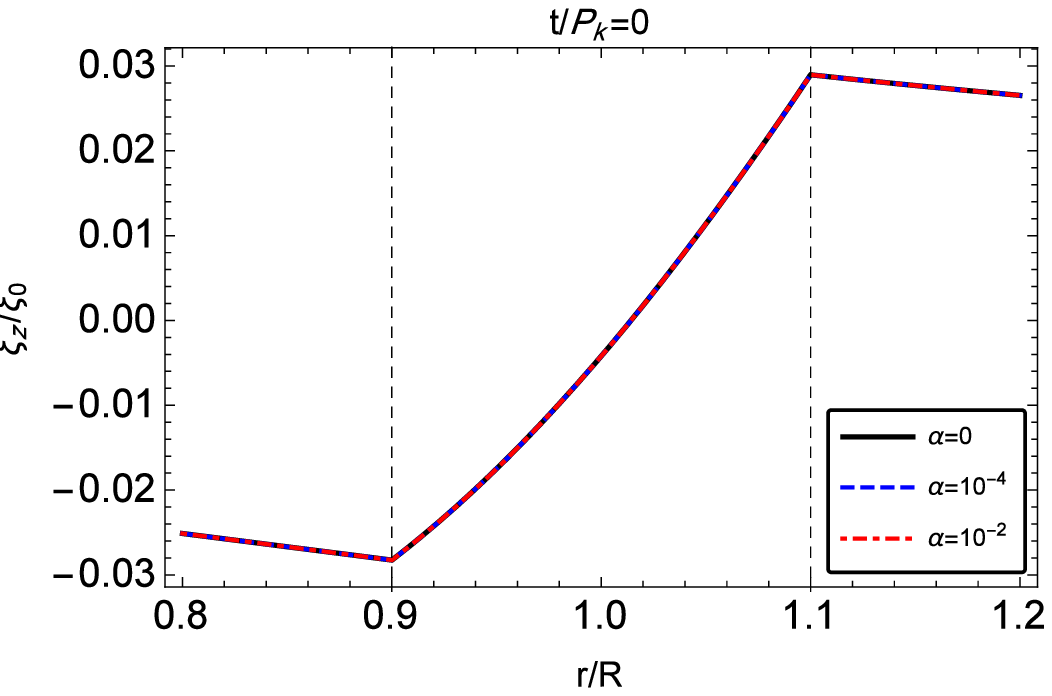}\\
    \includegraphics[width=50mm]{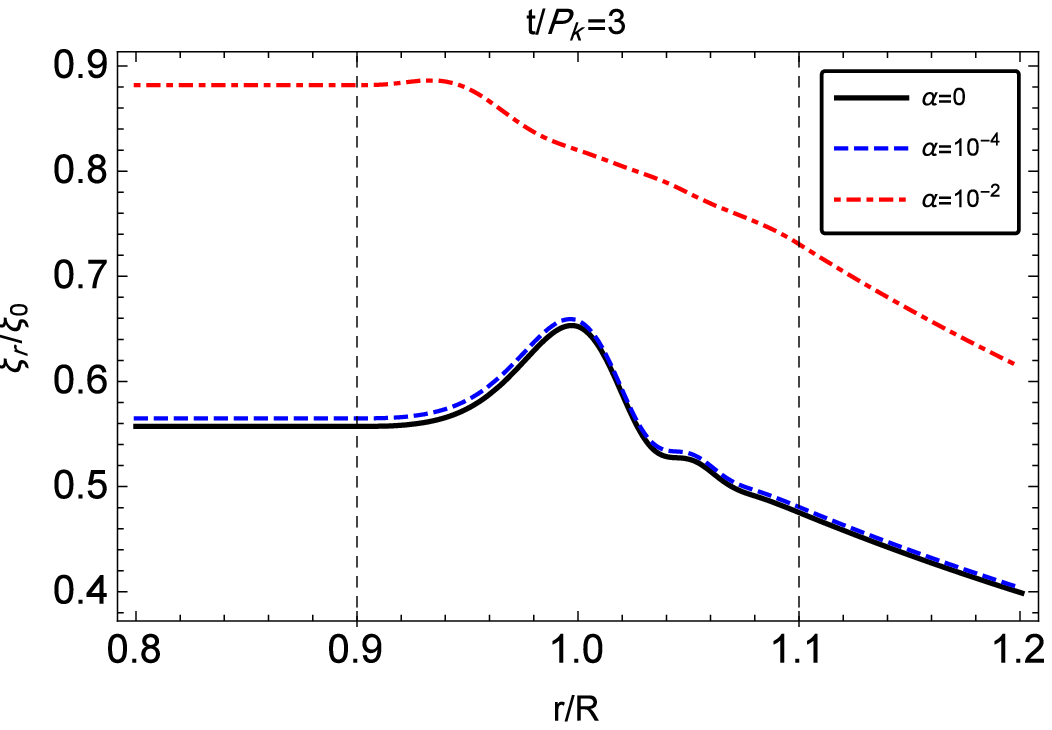}& \includegraphics[width=50mm]{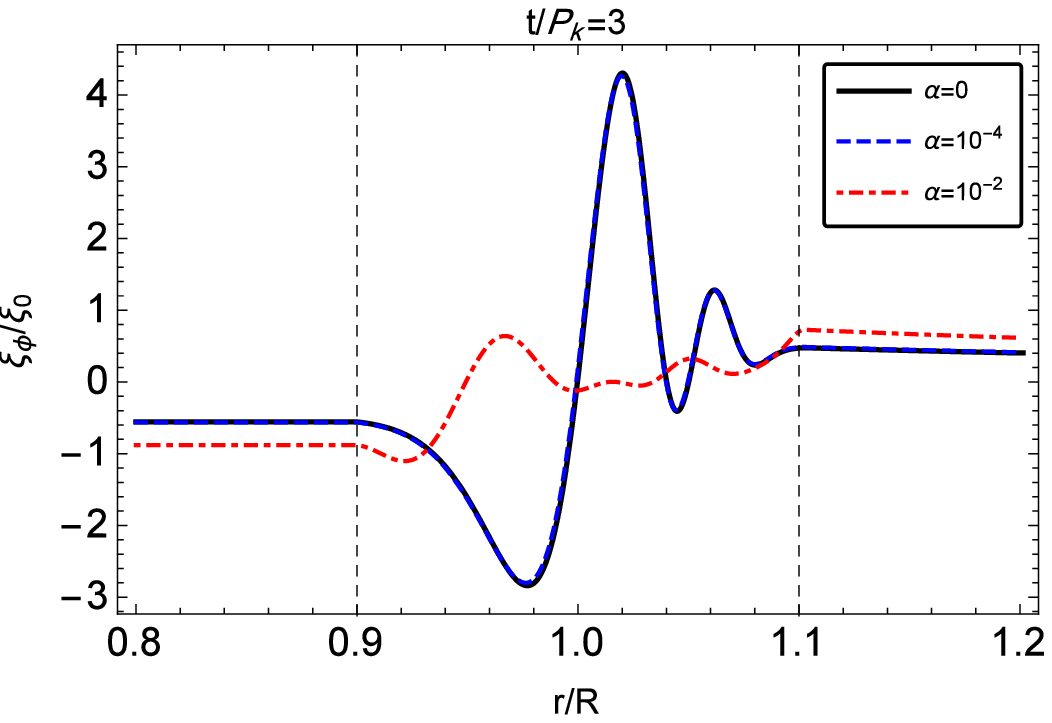}& \includegraphics[width=50mm]{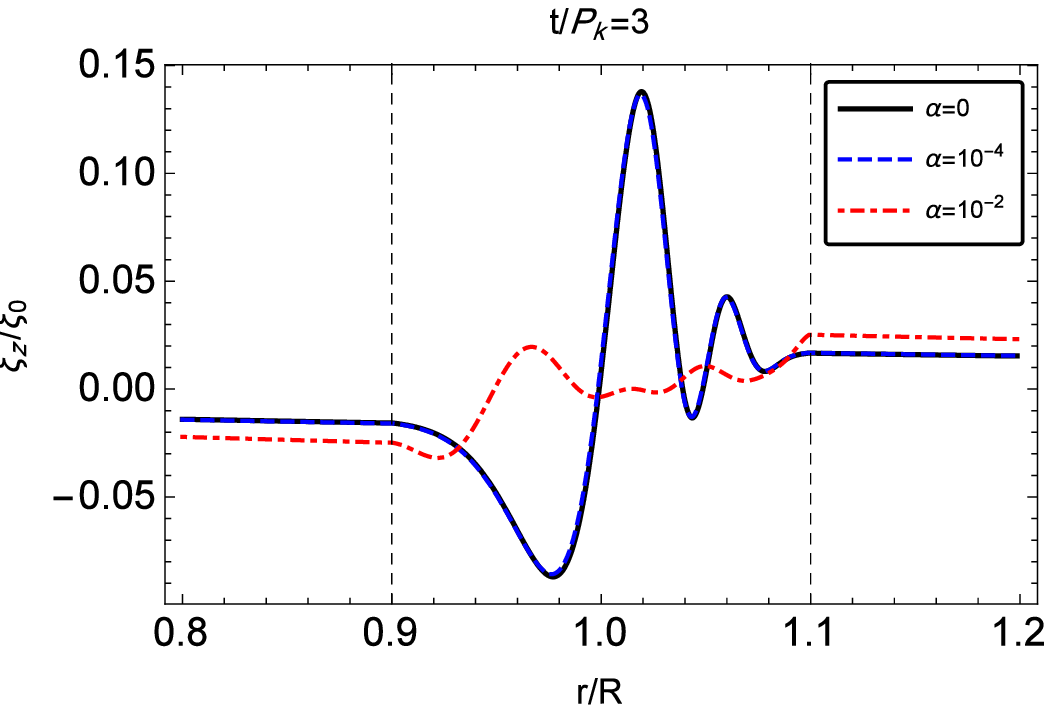}\\
    \includegraphics[width=50mm]{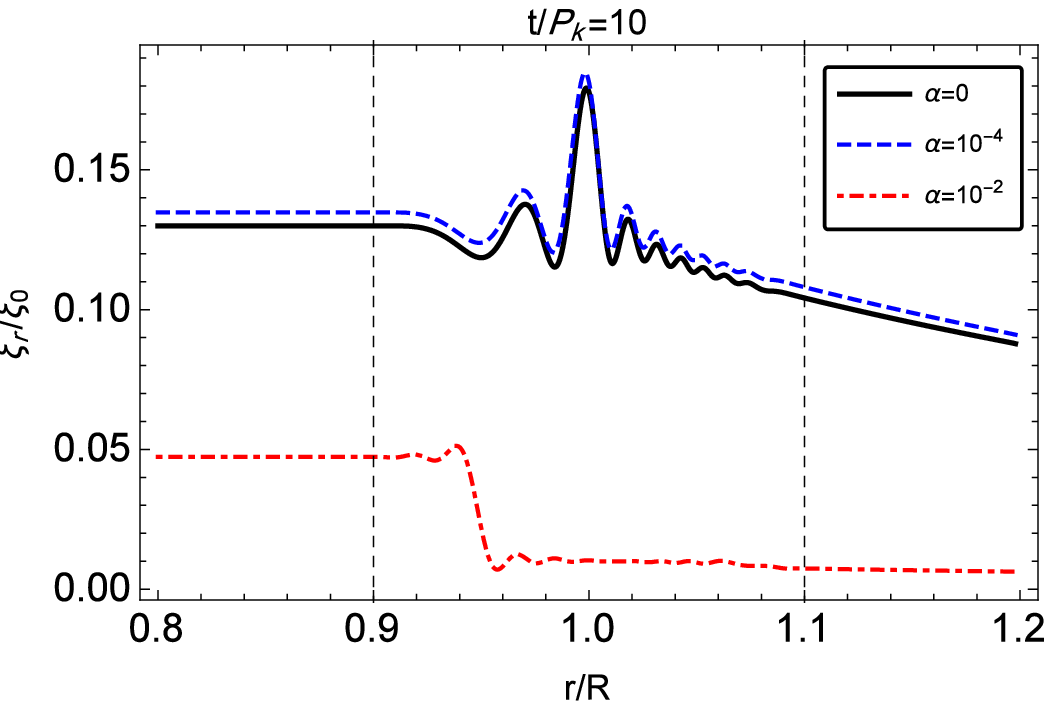}& \includegraphics[width=50mm]{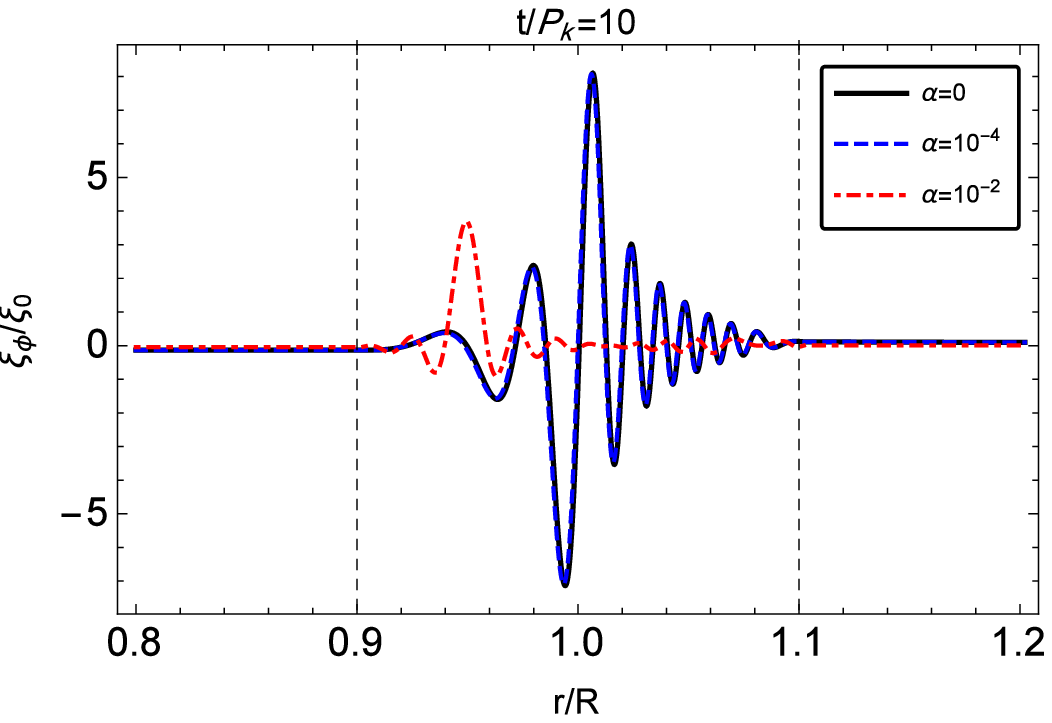}& \includegraphics[width=50mm]{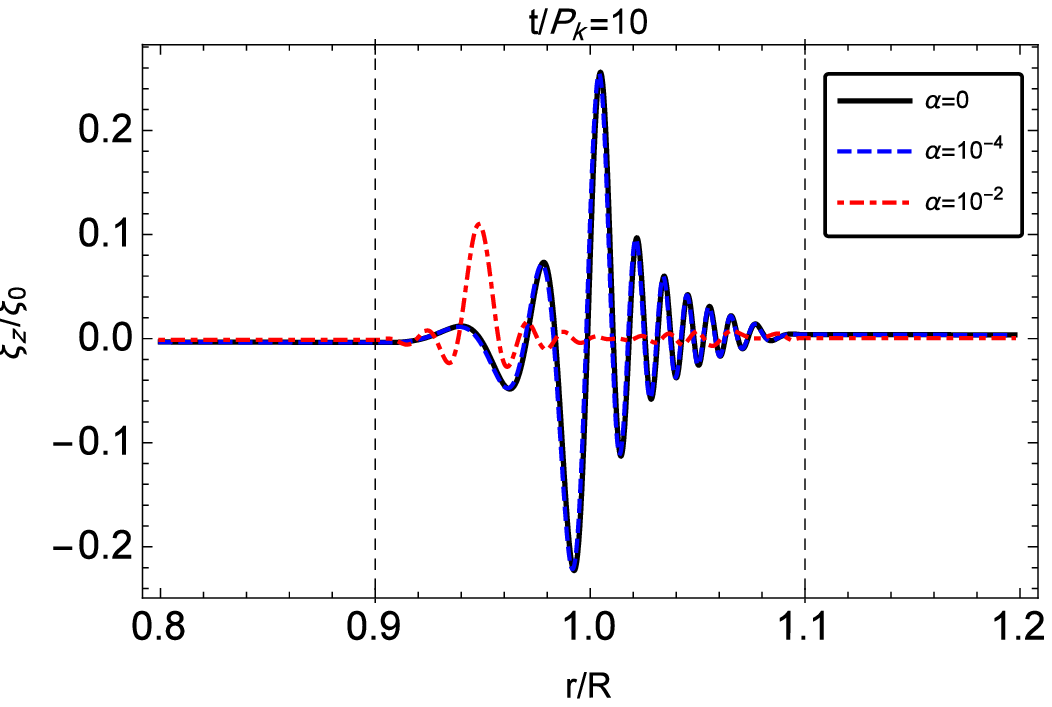}\\
  \end{tabular}
\caption {Temporal evolution of different components of the Lagrangian displacement, $\xi_r$ (left), $\xi_\varphi$ (middle)
and $\xi_z$ (right) for $\alpha=0$ (solid line), $\alpha=10^{-4}$ (blue dashed line) and $\alpha=10^{-2}$ (red dot-dashed line) for model II with $l/R=0.2$ and $m=+1$. Here $t/P_k=0$ (top), 3 (middle) and 10 (bottom). The left and right vertical dashed lines denote $r_1$ and $r_2$, respectively.}
    \label{xi_T2_m(1)_lbyR(0.2)}
\end{figure}
\begin{figure}
  \centering
  \begin{tabular}{ccc}
    \includegraphics[width=50mm]{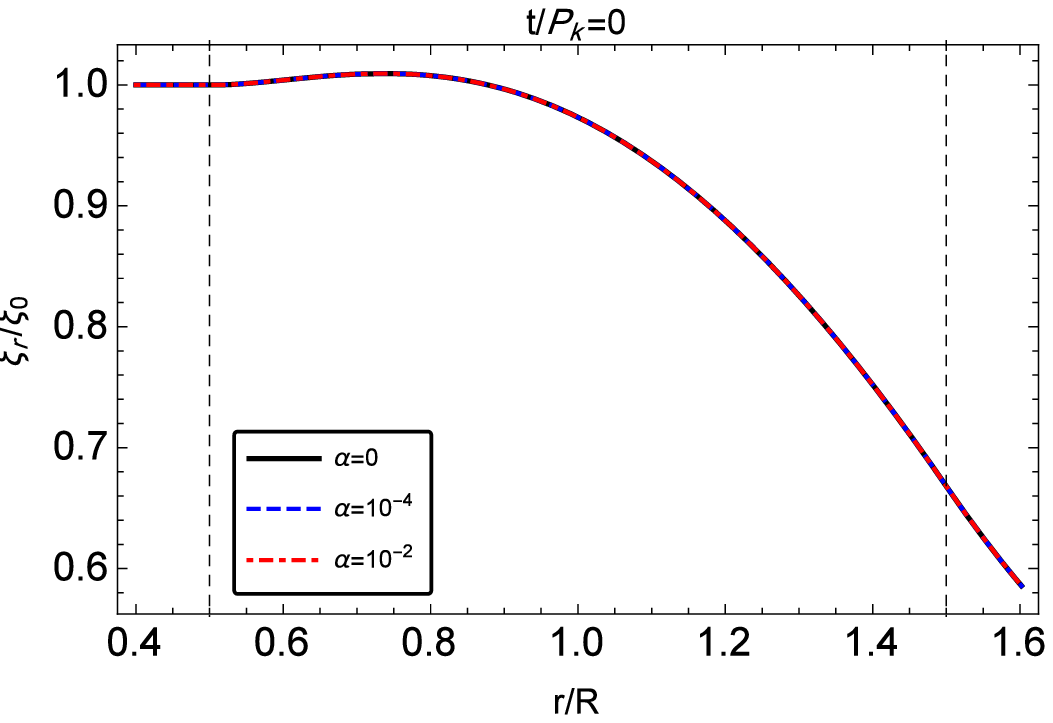}& \includegraphics[width=50mm]{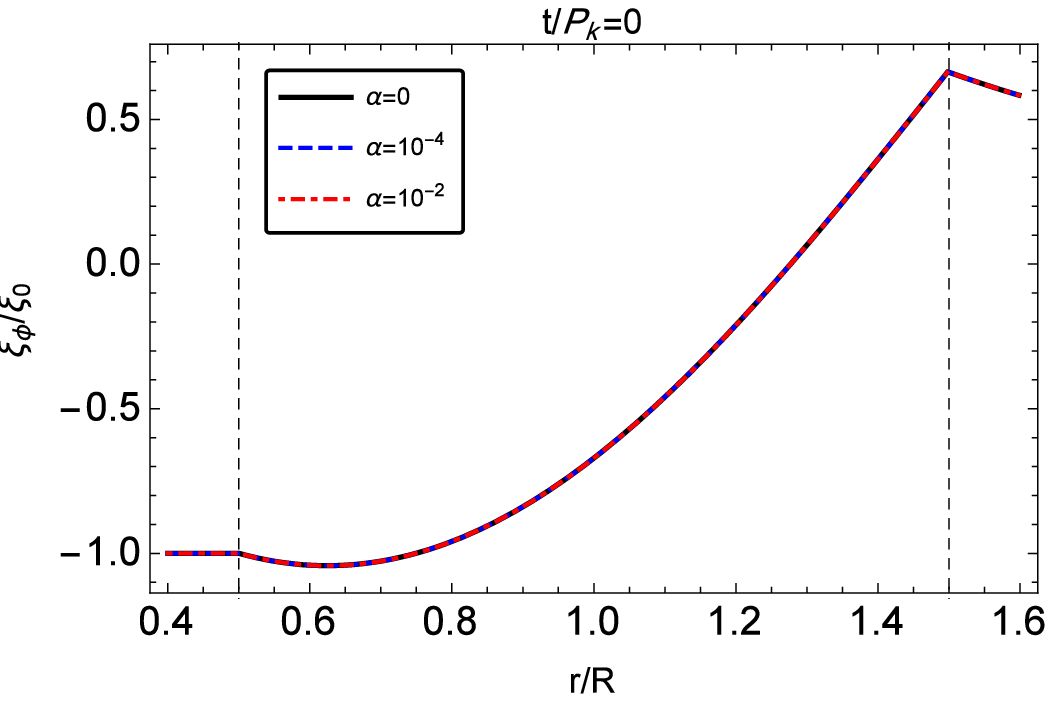}& \includegraphics[width=50mm]{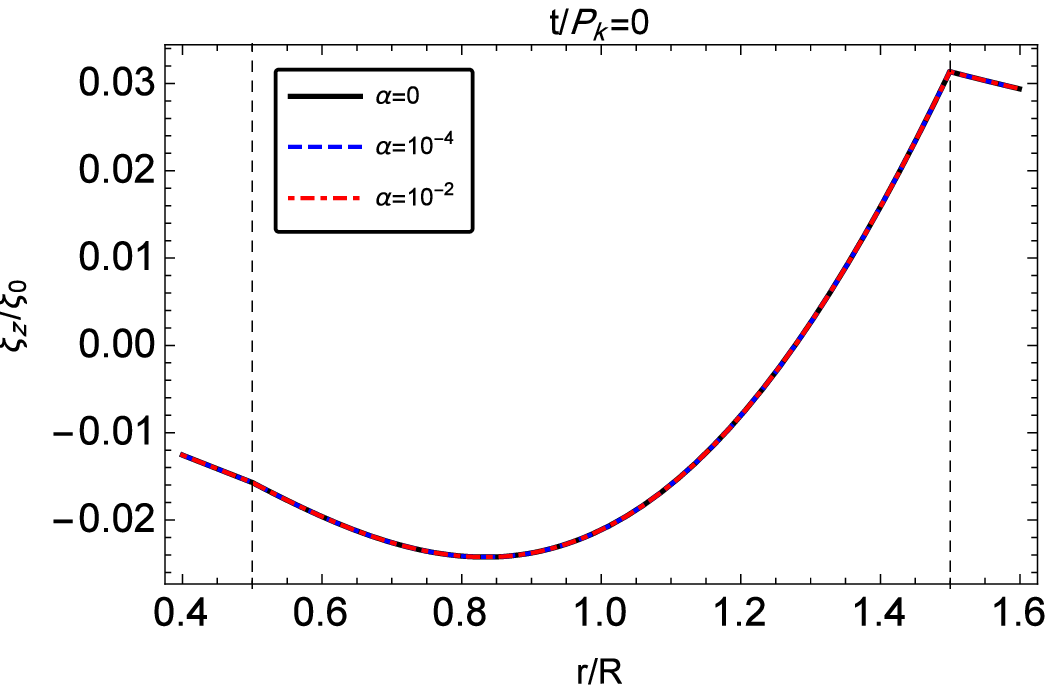}\\
    \includegraphics[width=50mm]{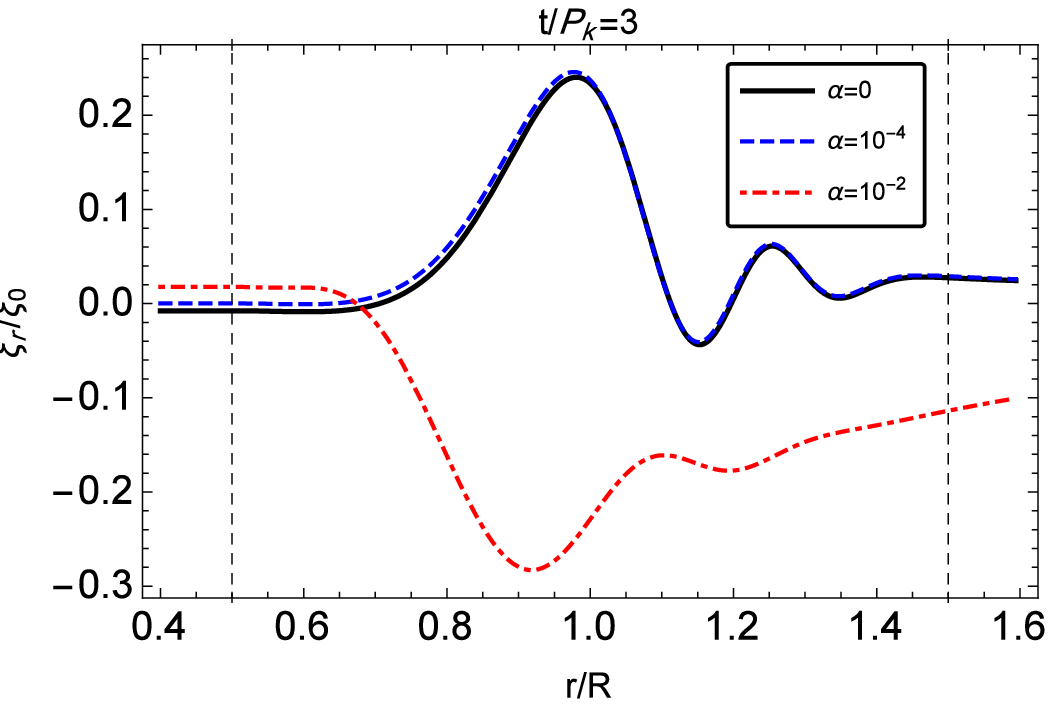}& \includegraphics[width=50mm]{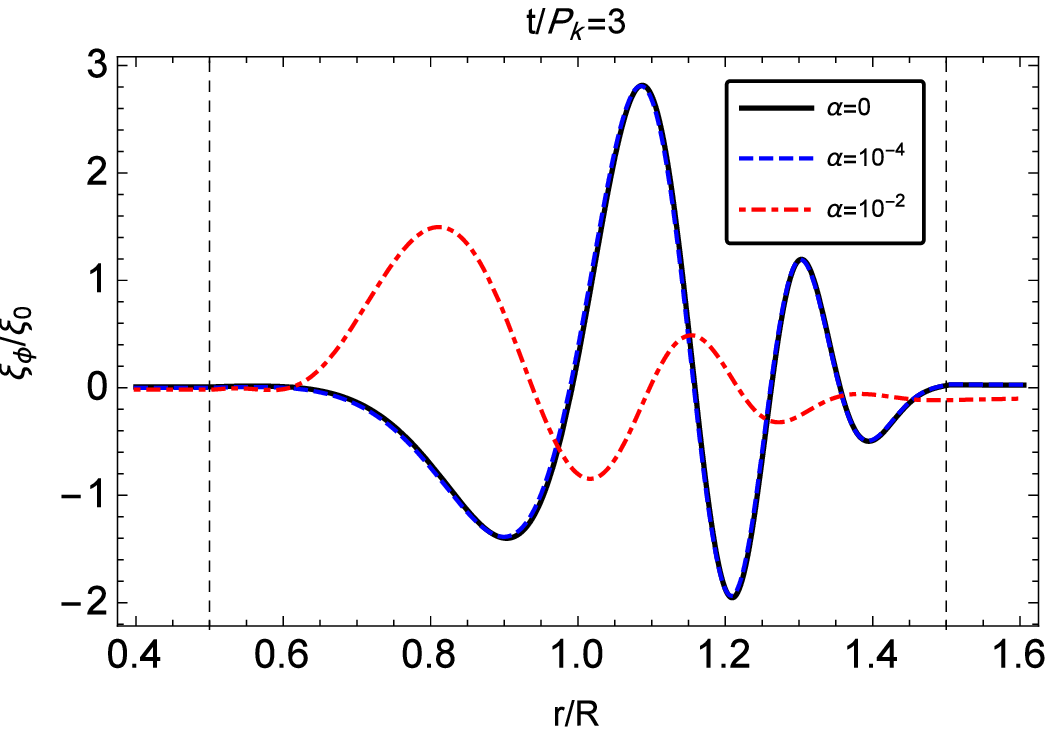}& \includegraphics[width=50mm]{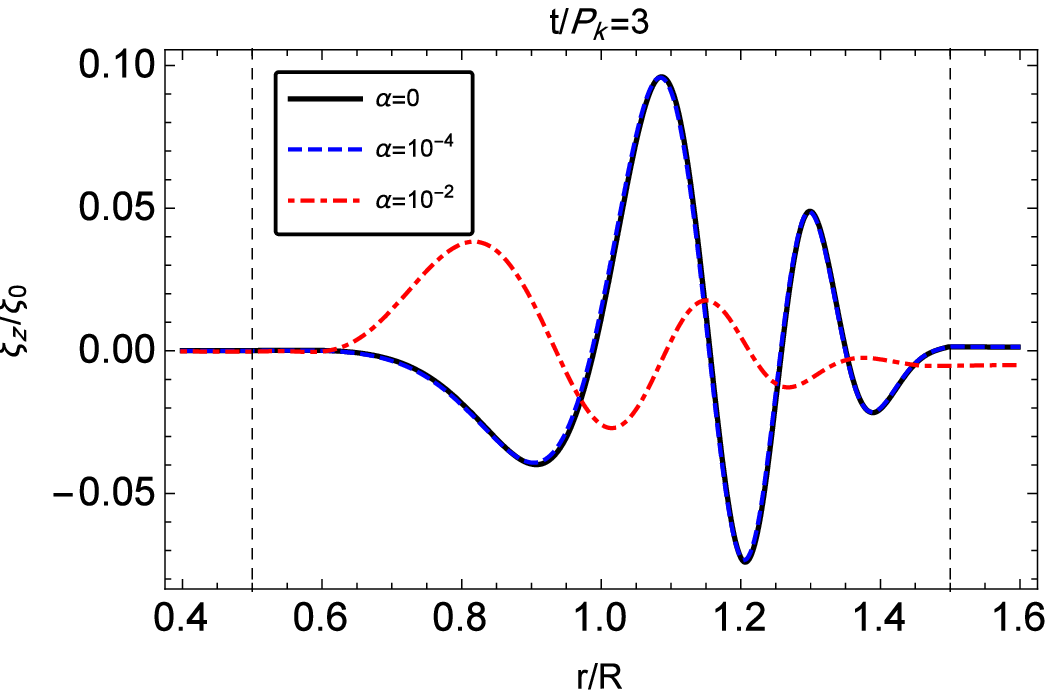}\\
    \includegraphics[width=50mm]{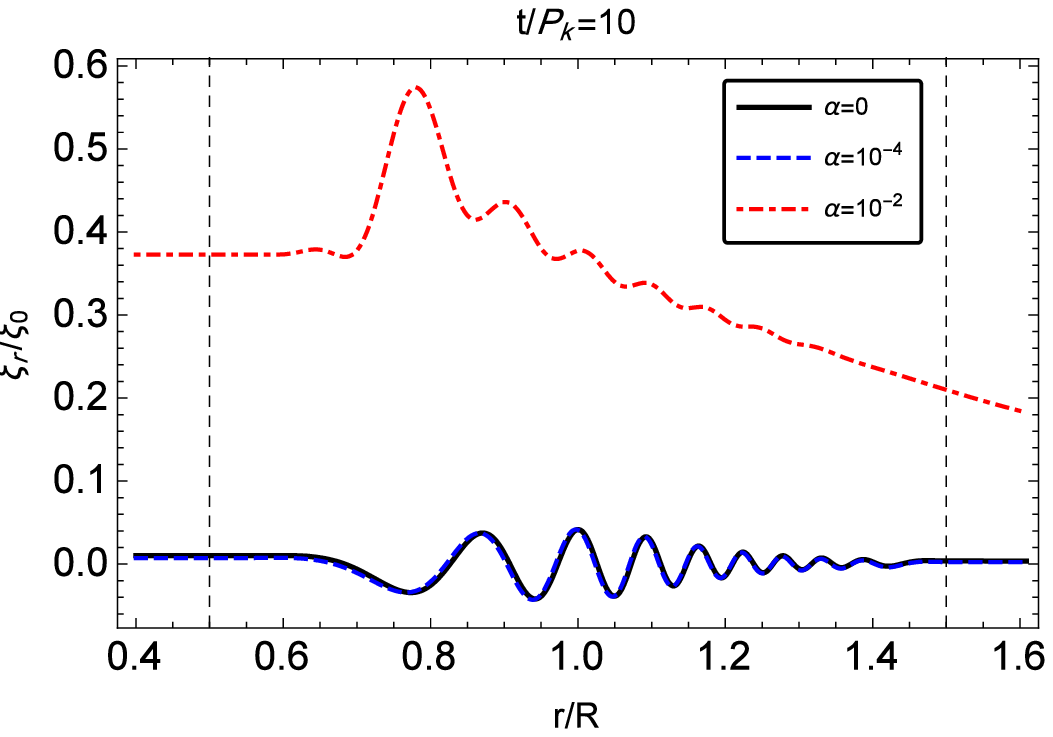}& \includegraphics[width=50mm]{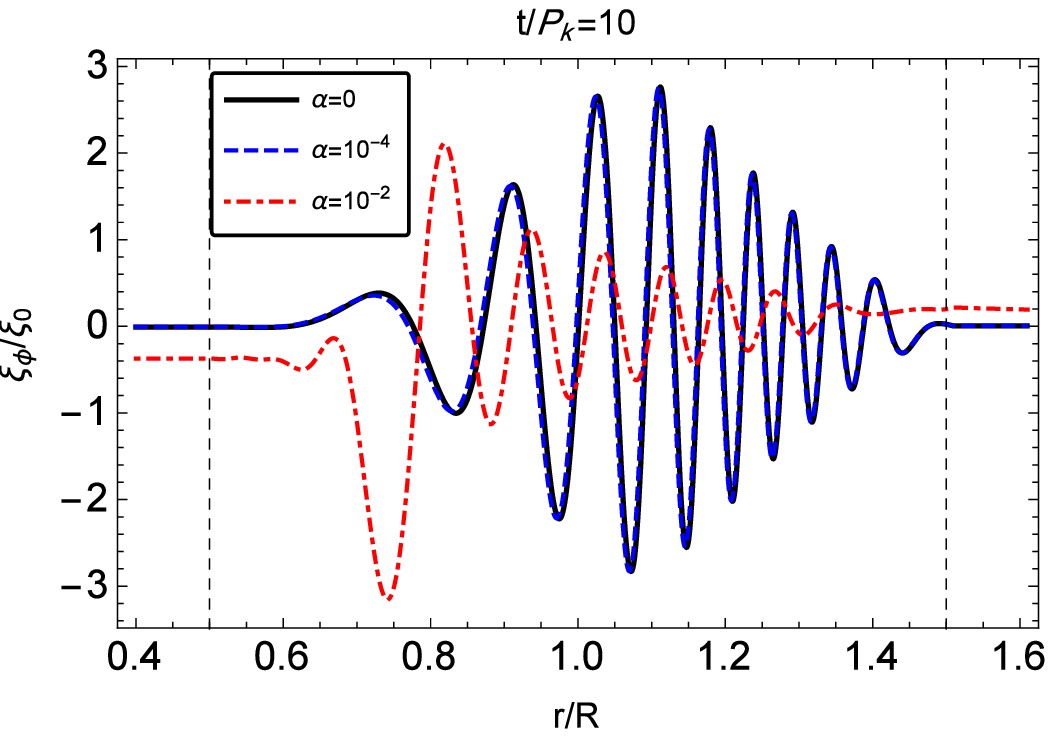}& \includegraphics[width=50mm]{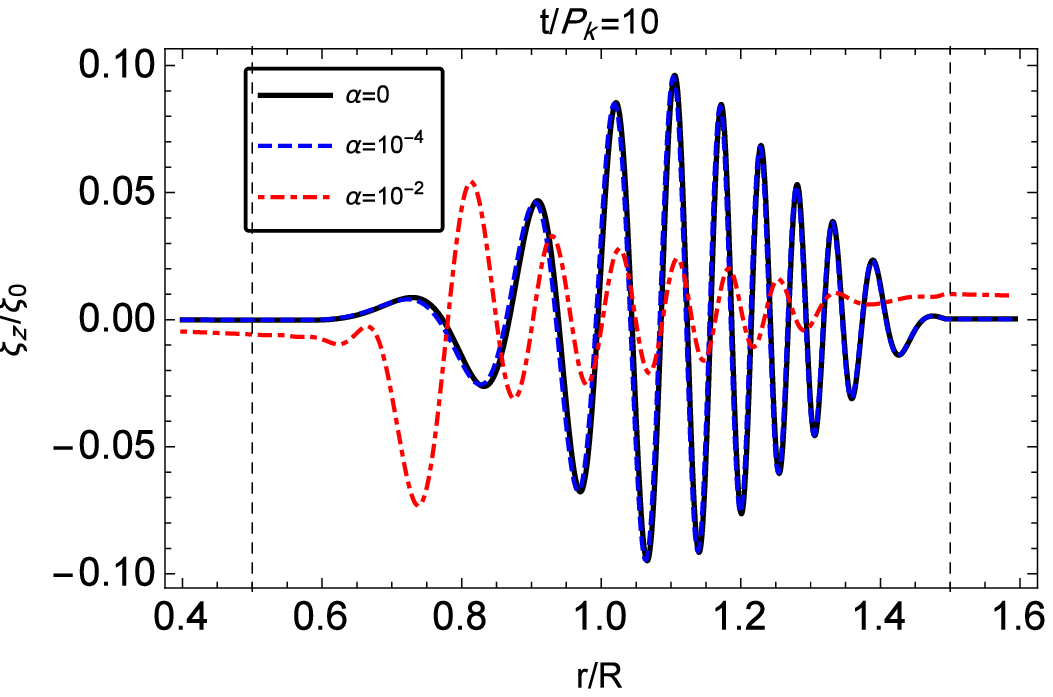}\\
  \end{tabular}
\caption {Same as Fig. \ref{xi_T2_m(1)_lbyR(0.2)}, but for $l/R=1$.}
    \label{xi_T2_m(1)_lbyR(1)}
\end{figure}
\begin{figure}
  \centering
  \begin{tabular}{ccc}
    \includegraphics[width=50mm]{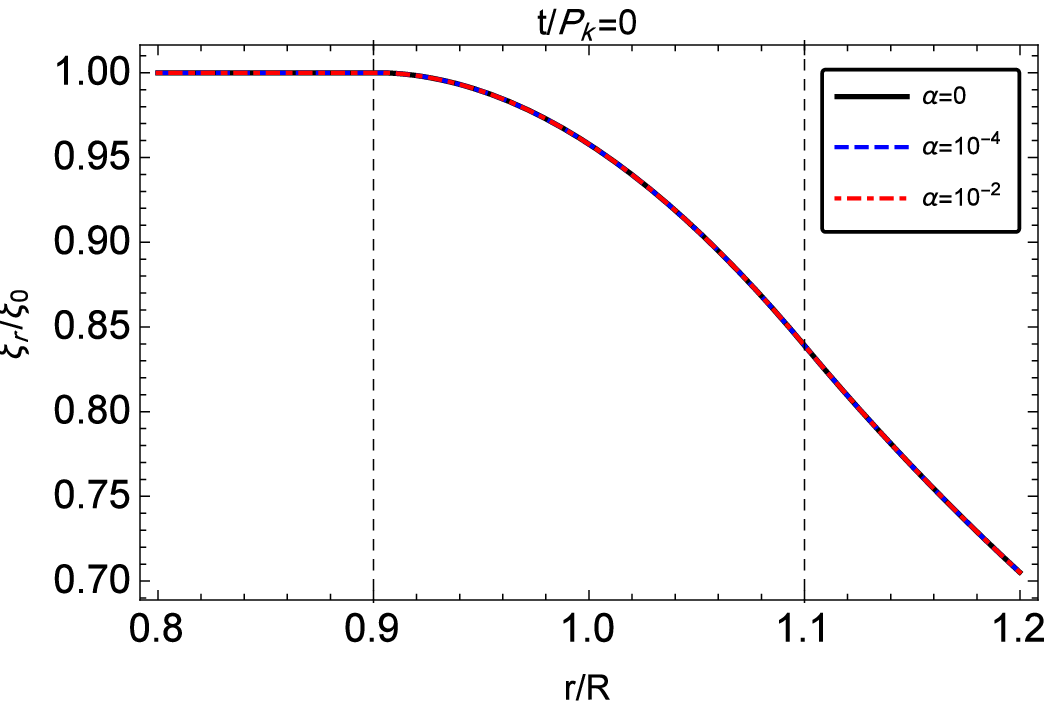}& \includegraphics[width=50mm]{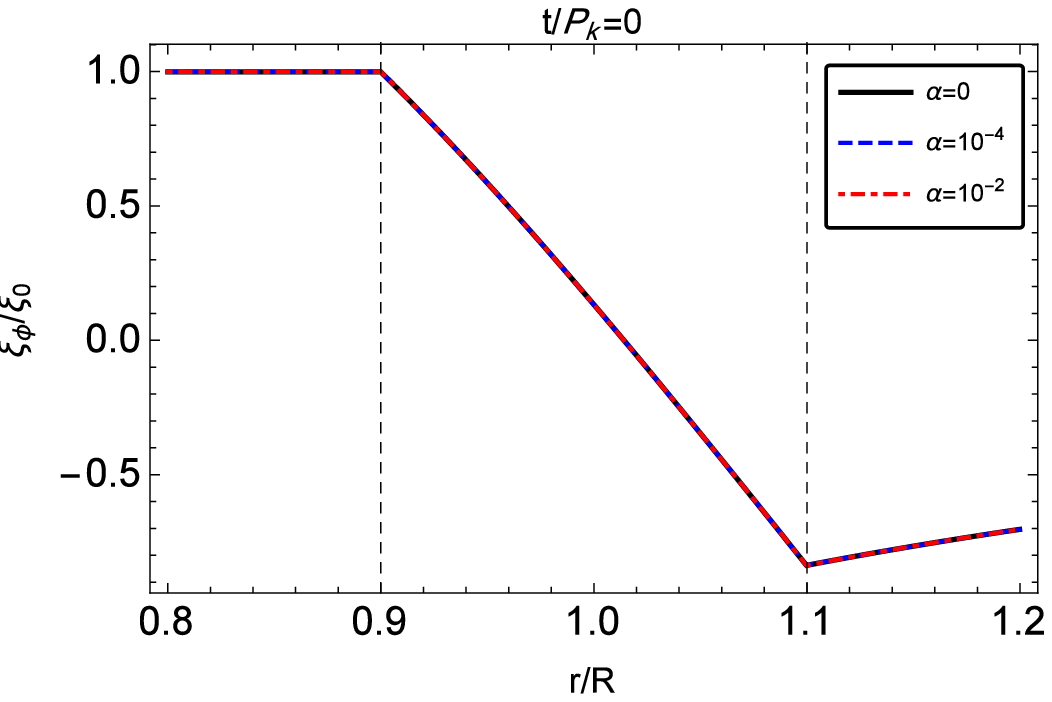}& \includegraphics[width=50mm]{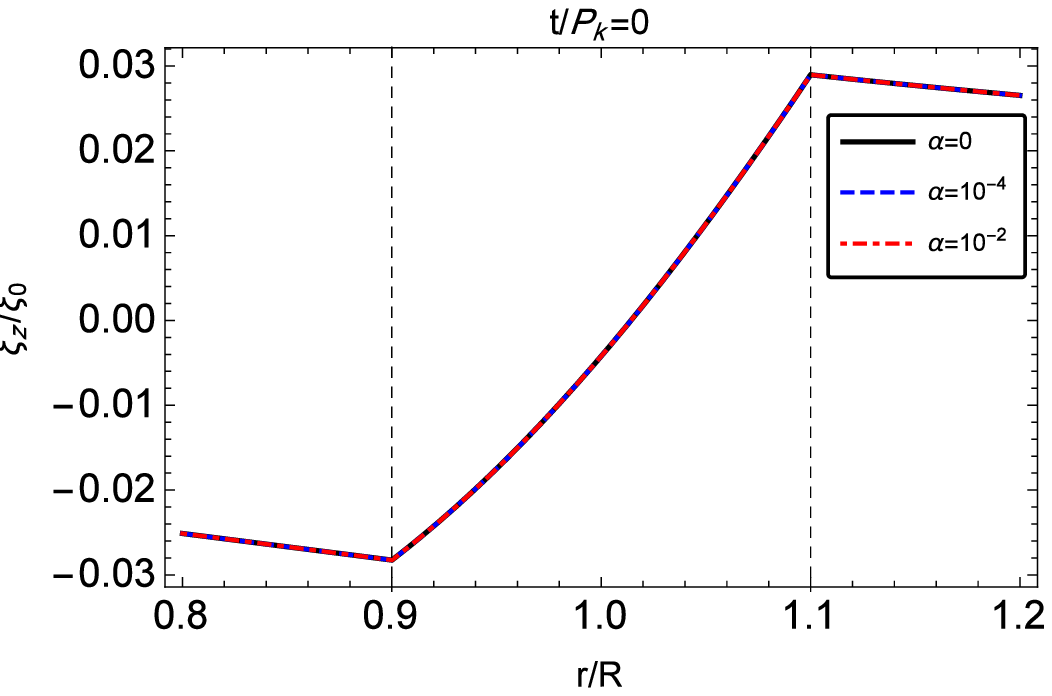}\\
    \includegraphics[width=50mm]{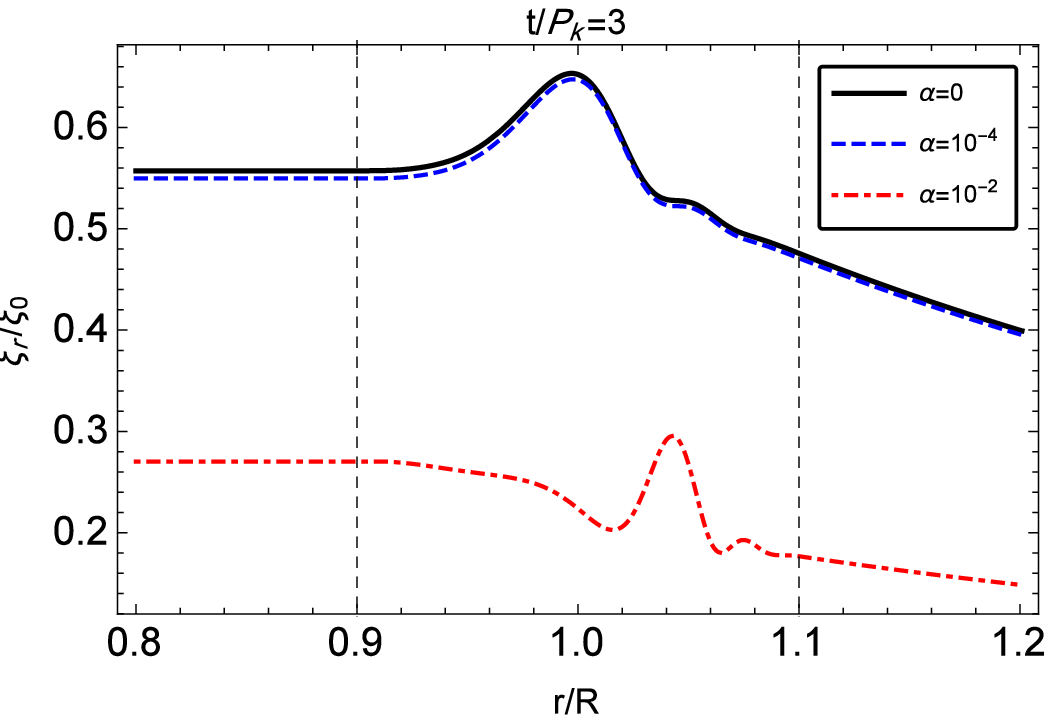}& \includegraphics[width=50mm]{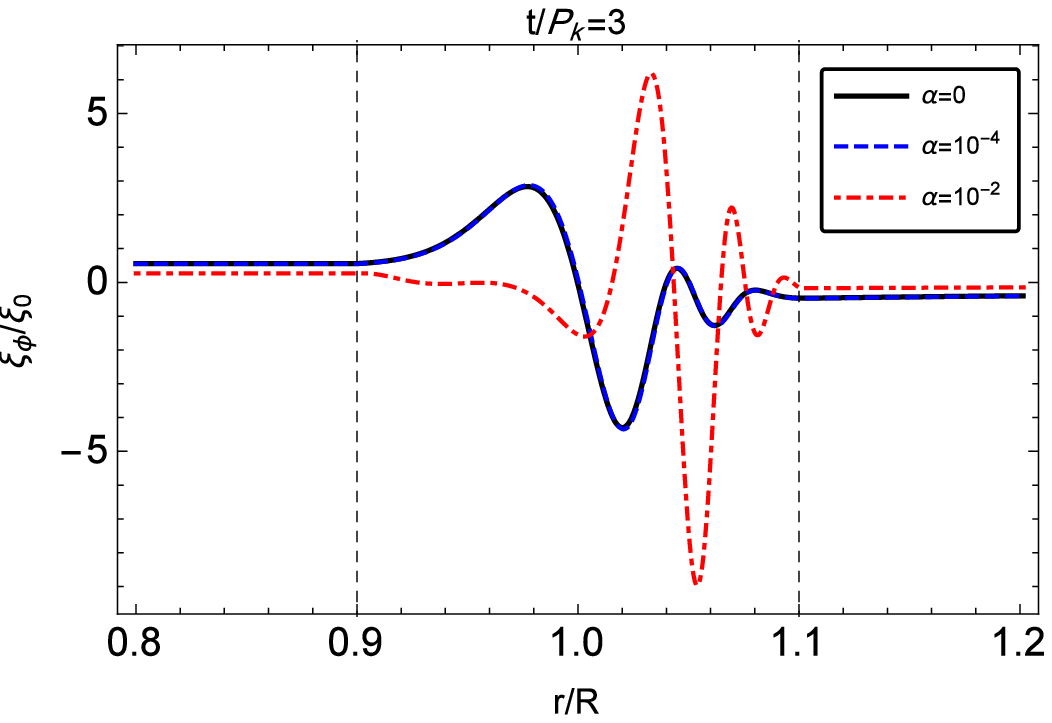}& \includegraphics[width=50mm]{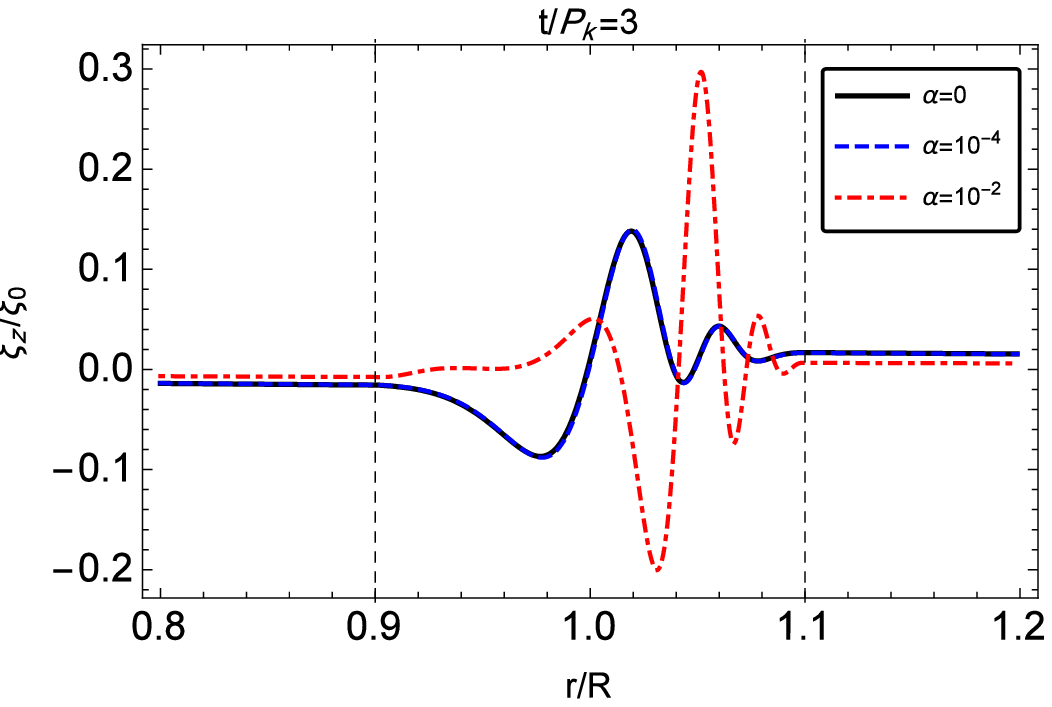}\\
    \includegraphics[width=50mm]{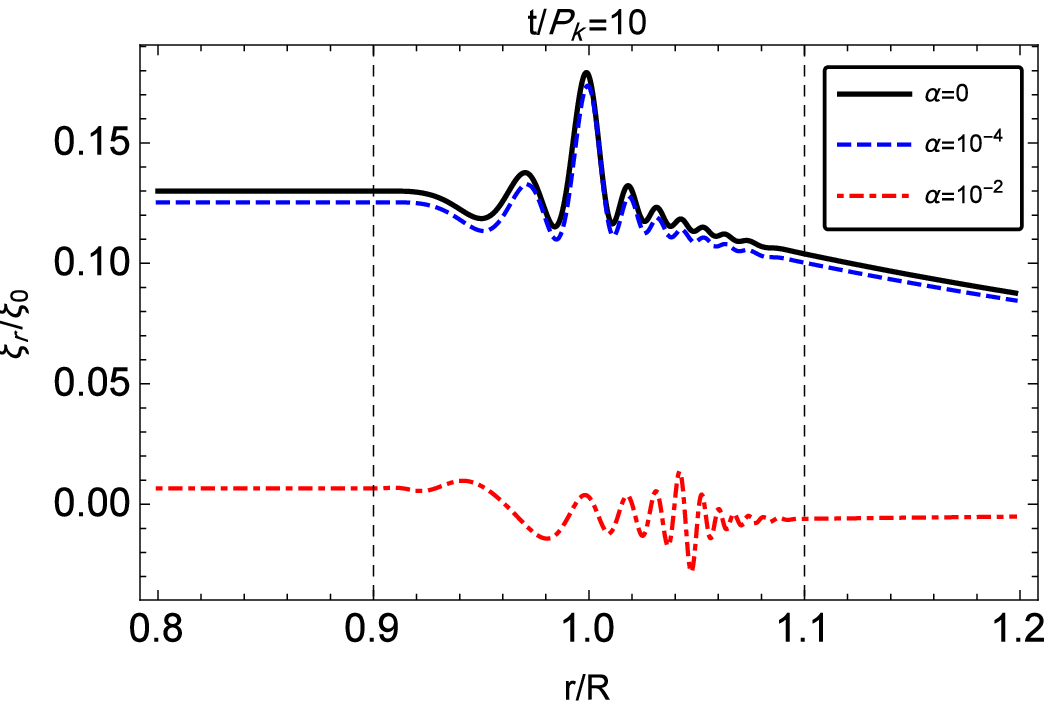}& \includegraphics[width=50mm]{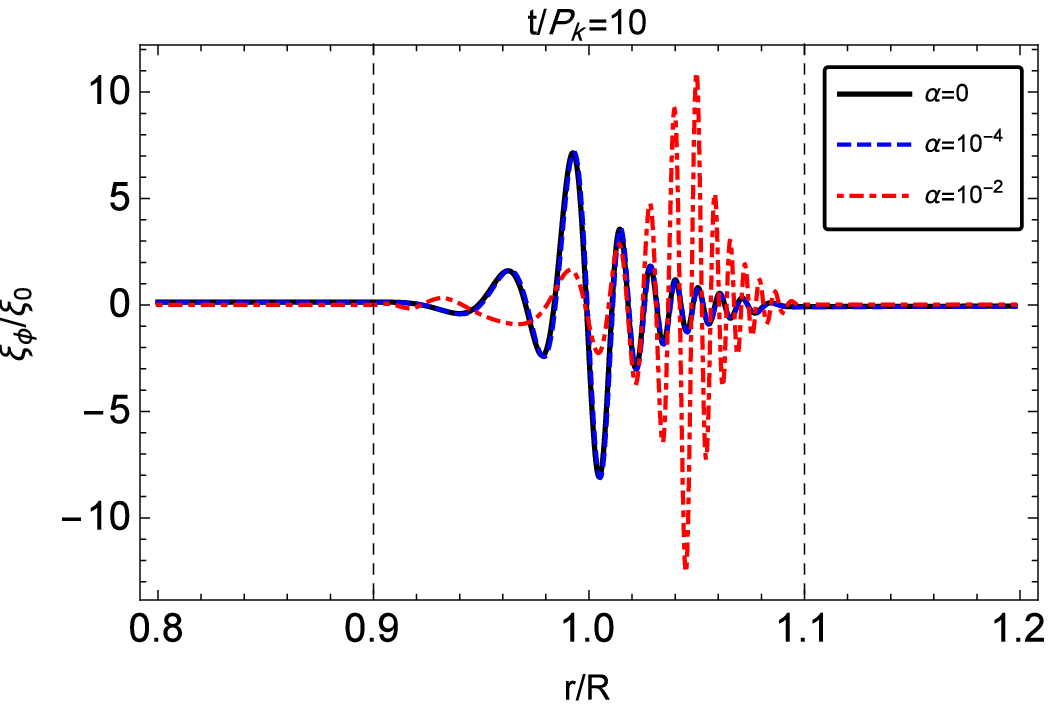}& \includegraphics[width=50mm]{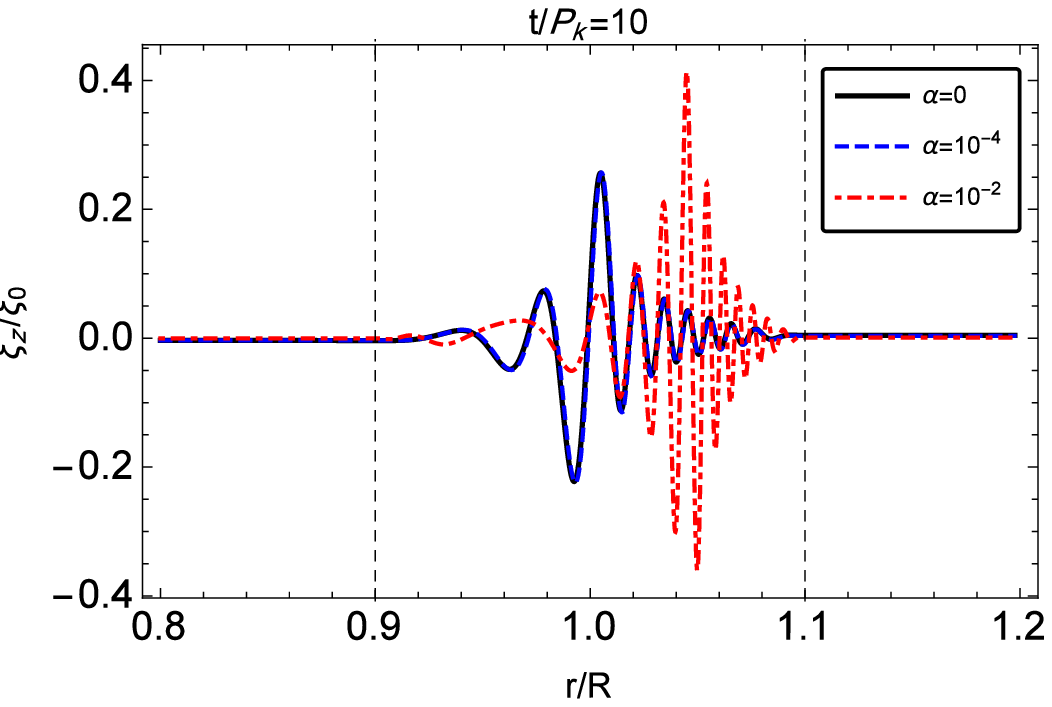}\\
  \end{tabular}
    \caption {Same as Fig. \ref{xi_T2_m(1)_lbyR(0.2)}, but for $m=-1$.}
    \label{xi_T2_m(-1)_lbyR(0.2)}
\end{figure}
\begin{figure}
  \centering
  \begin{tabular}{ccc}
    \includegraphics[width=50mm]{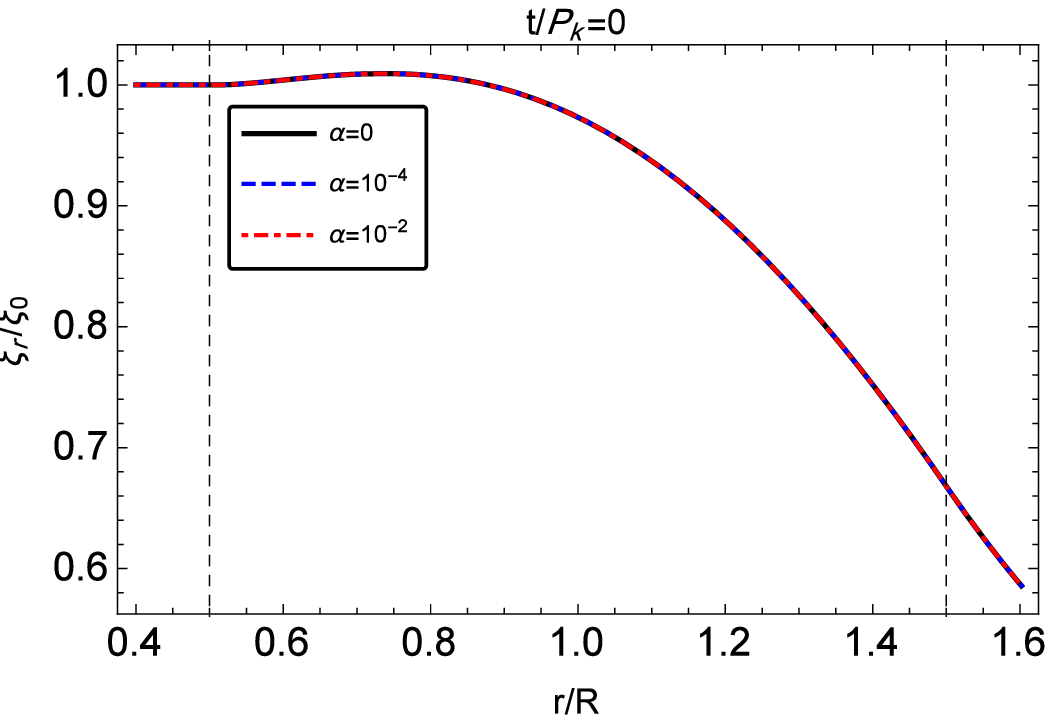}& \includegraphics[width=50mm]{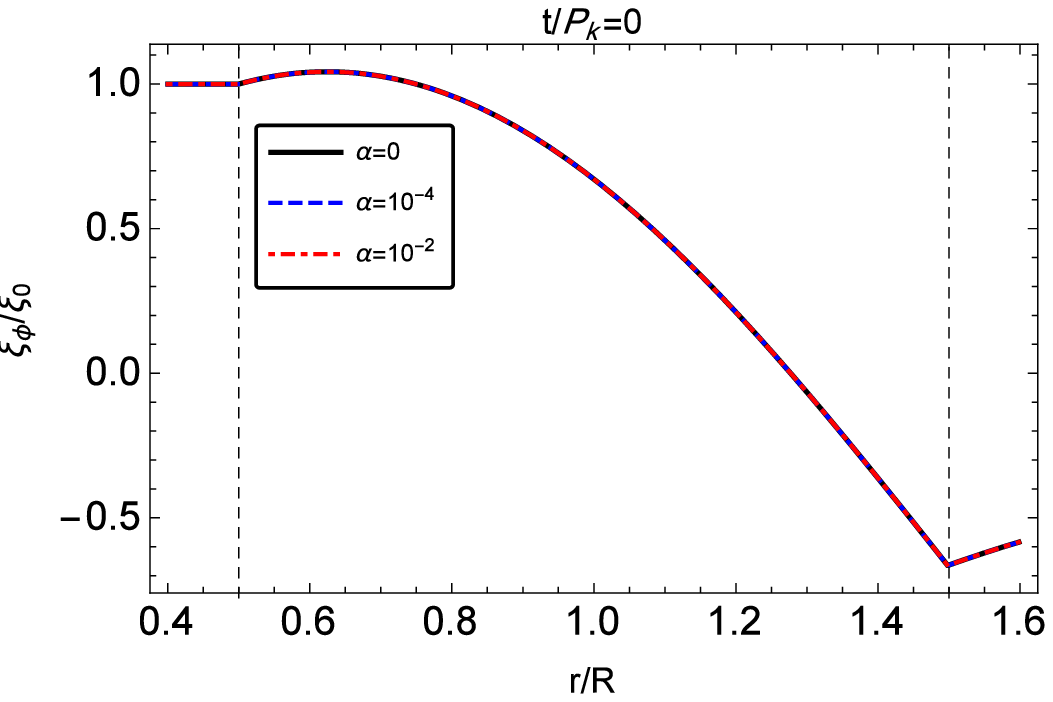}& \includegraphics[width=50mm]{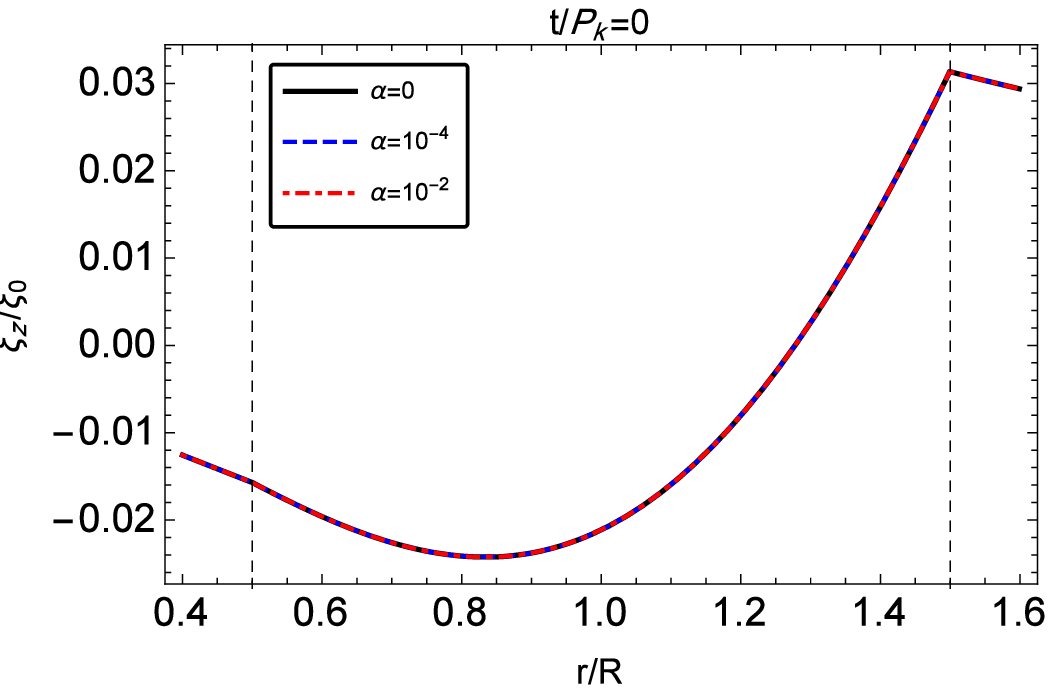}\\
    \includegraphics[width=50mm]{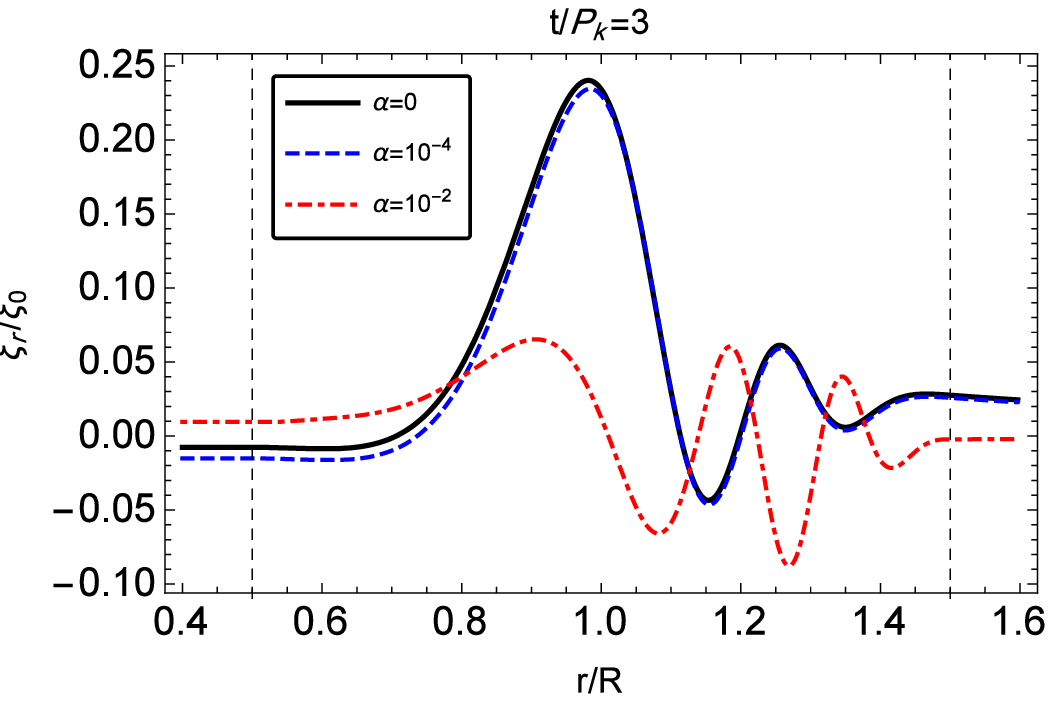}& \includegraphics[width=50mm]{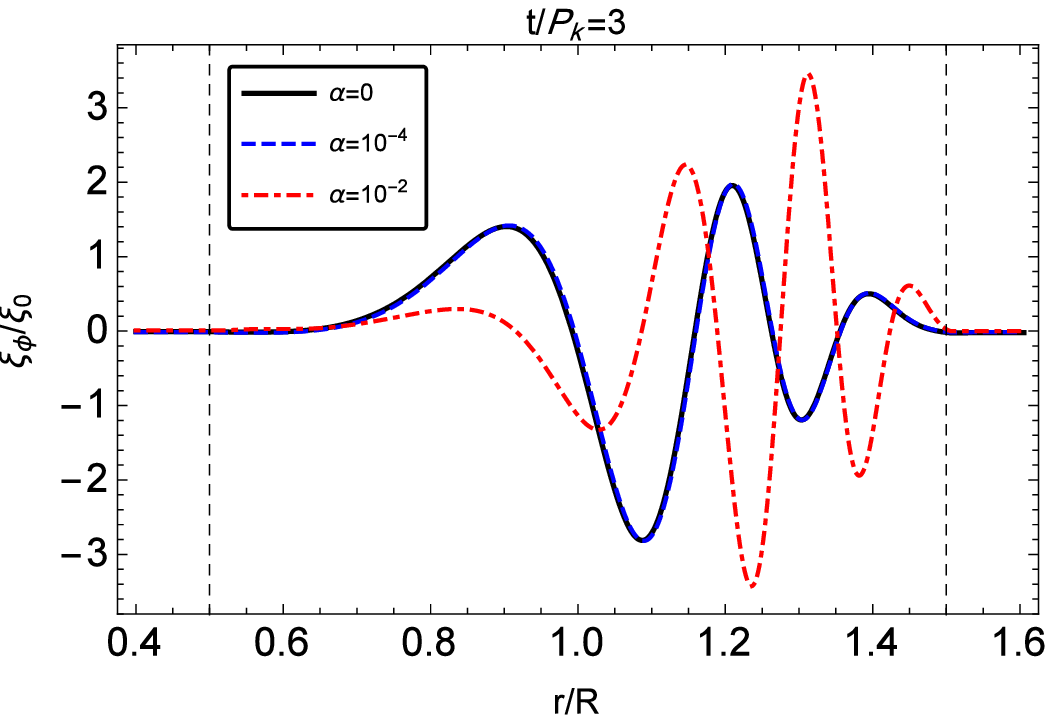}& \includegraphics[width=50mm]{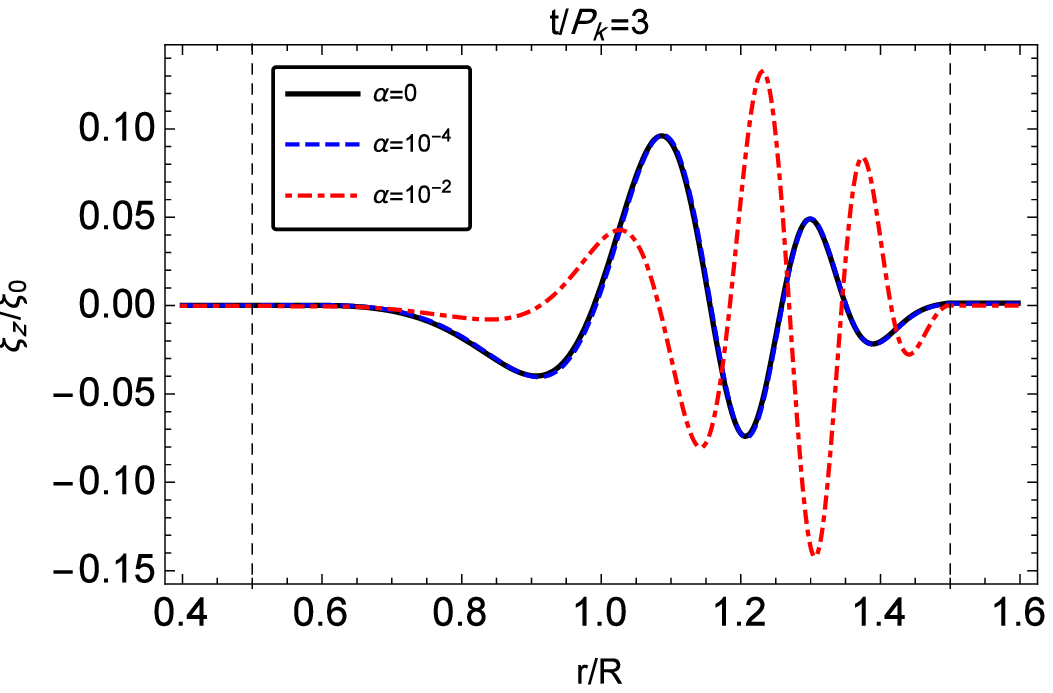}\\
    \includegraphics[width=50mm]{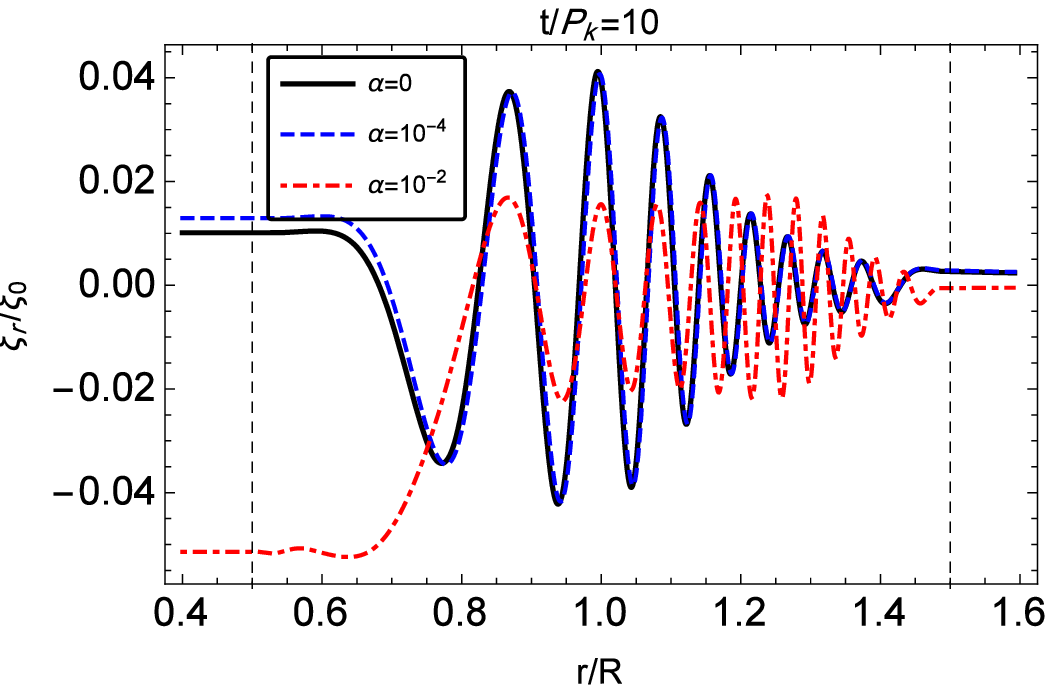}& \includegraphics[width=50mm]{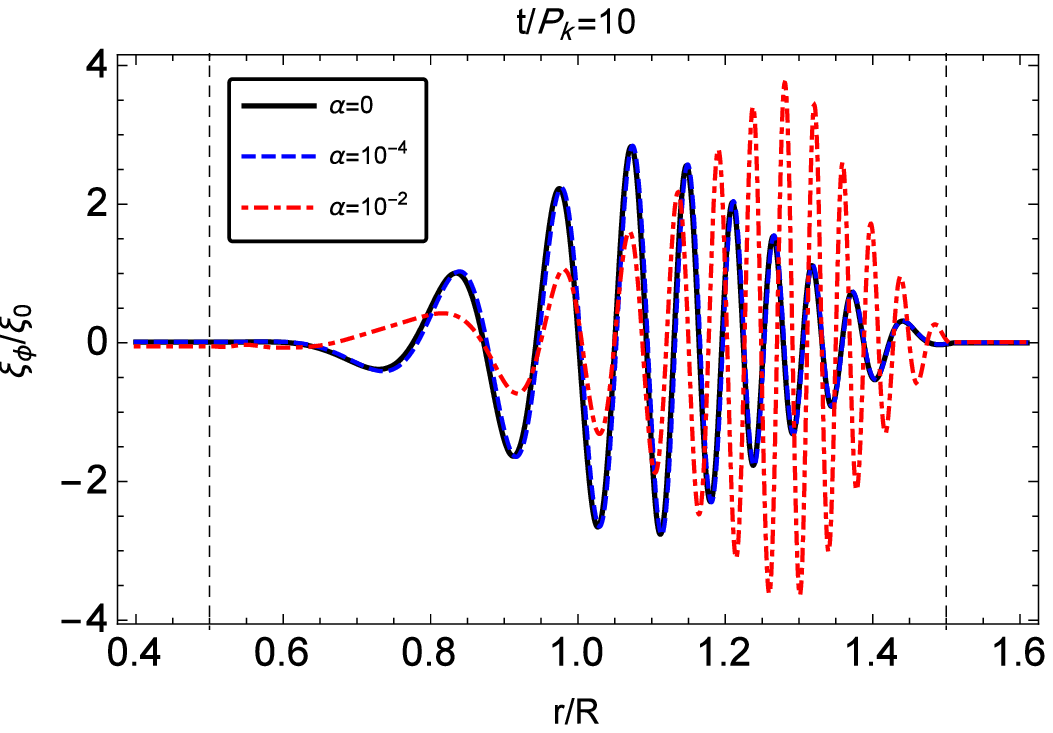}& \includegraphics[width=50mm]{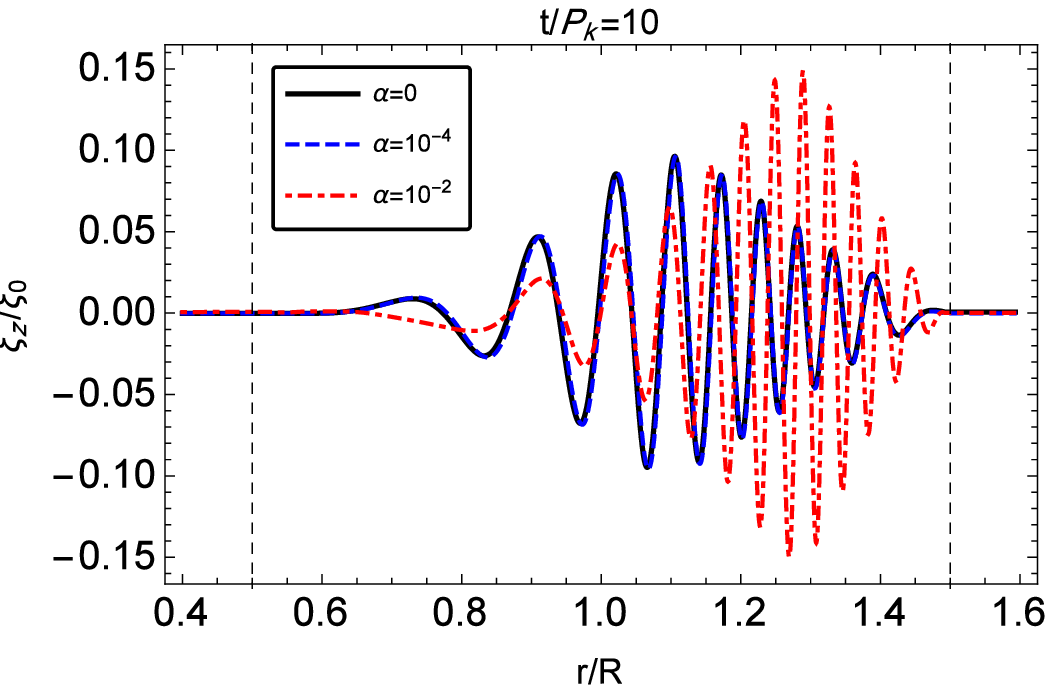}\\
  \end{tabular}
\caption {Same as Fig. \ref{xi_T2_m(1)_lbyR(0.2)}, but for $m=-1$ and $l/R=1$.}
    \label{xi_T2_m(-1)_lbyR(1)}
\end{figure}
\begin{figure}
\centering
    \begin{tabular}{cc}
        \includegraphics[width=75mm]{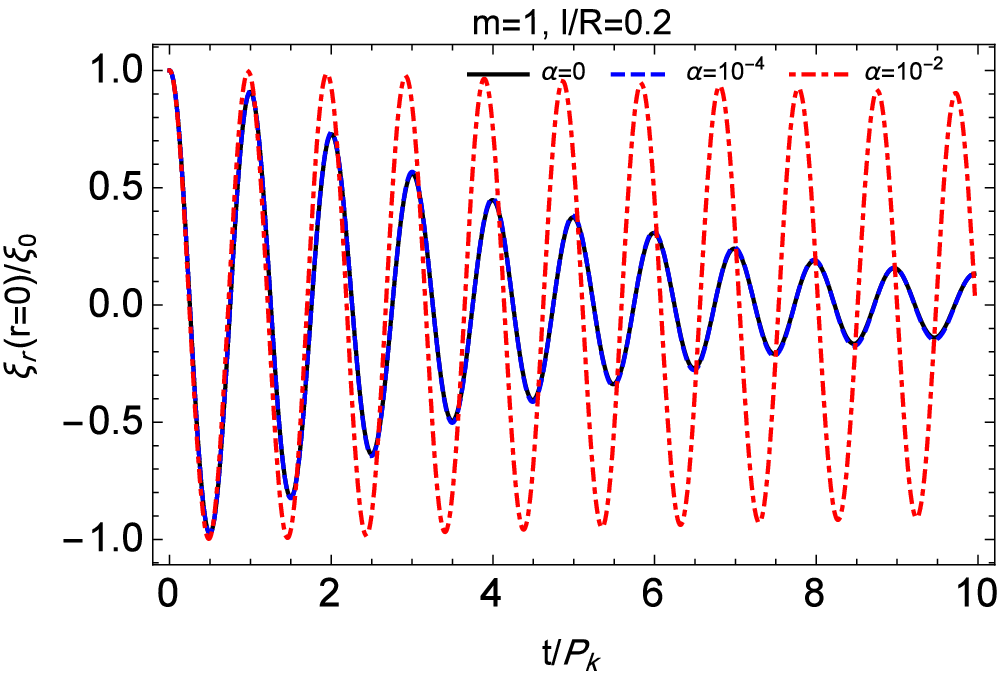}& \includegraphics[width=75mm]{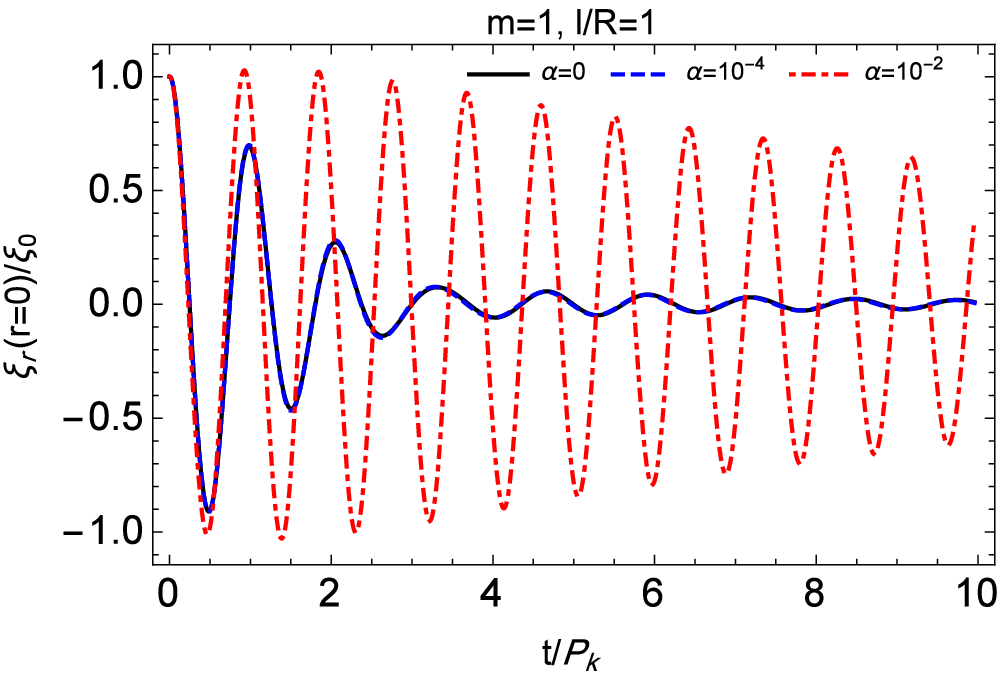}\\
        \includegraphics[width=75mm]{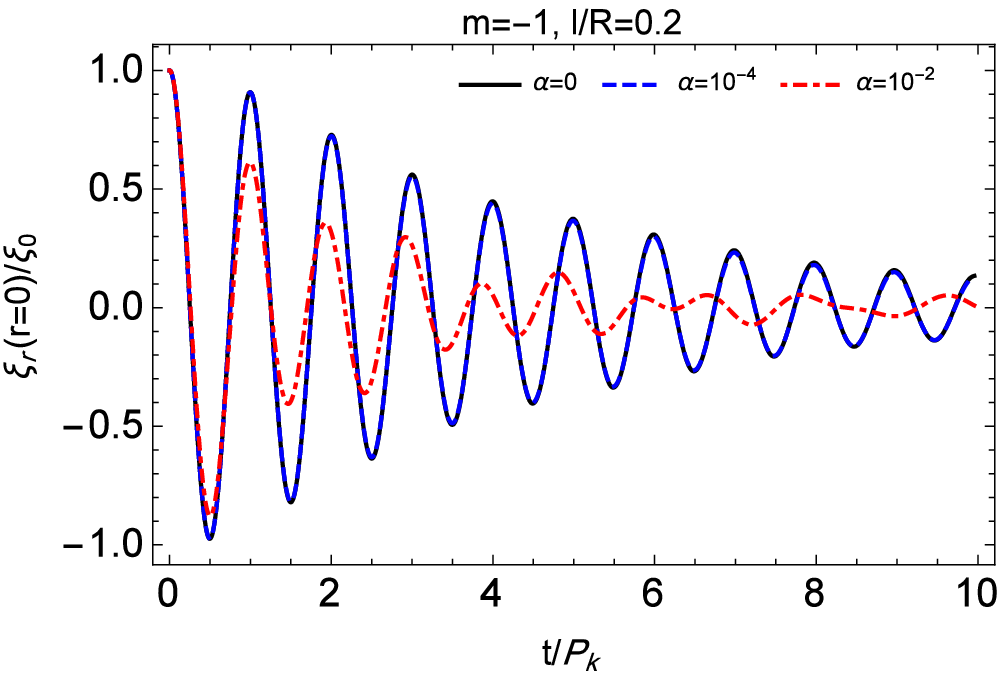}& \includegraphics[width=75mm]{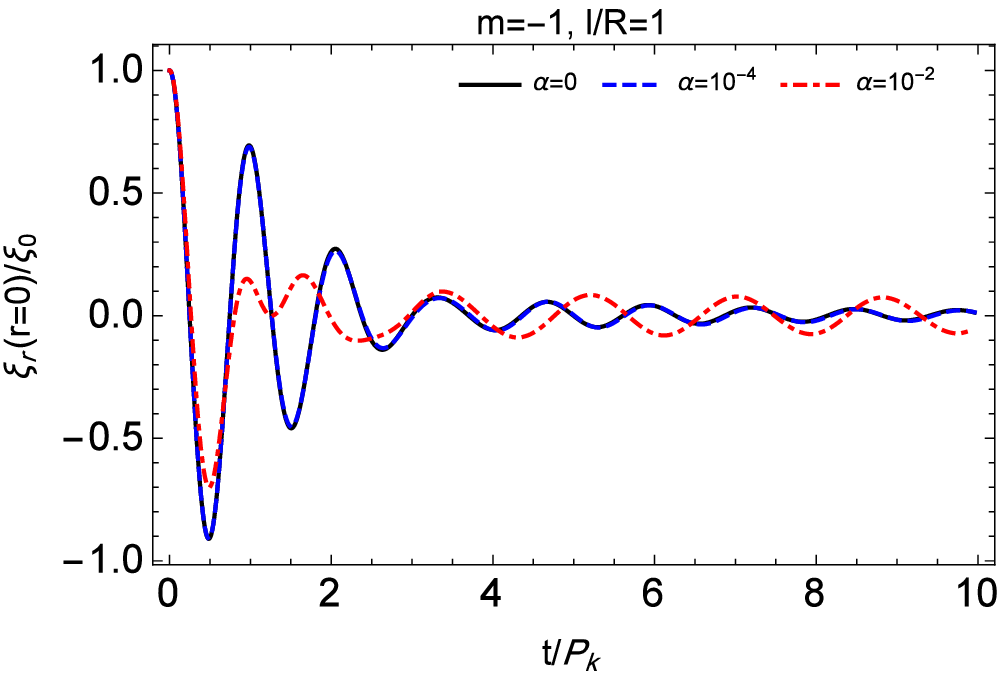}\\
    \end{tabular}
    \caption{Same as Fig. \ref{xir_MN}, but for the model II.}
    \label{xir_T2}
 \end{figure}
\begin{figure}
\centering
    \begin{tabular}{cc}
        \includegraphics[width=75mm]{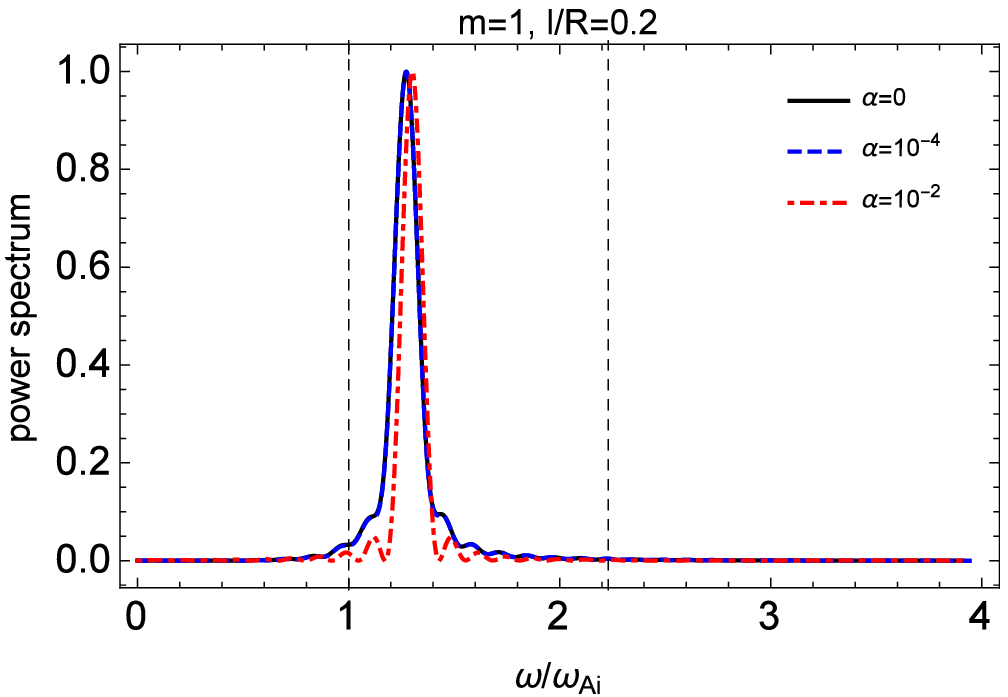}& \includegraphics[width=75mm]{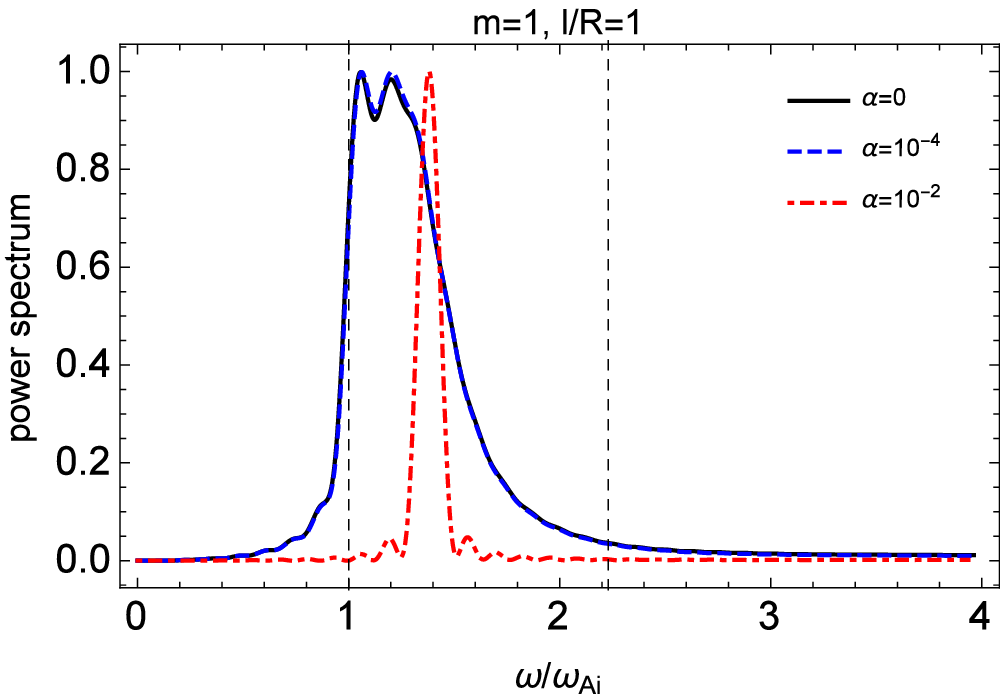}\\
        \includegraphics[width=75mm]{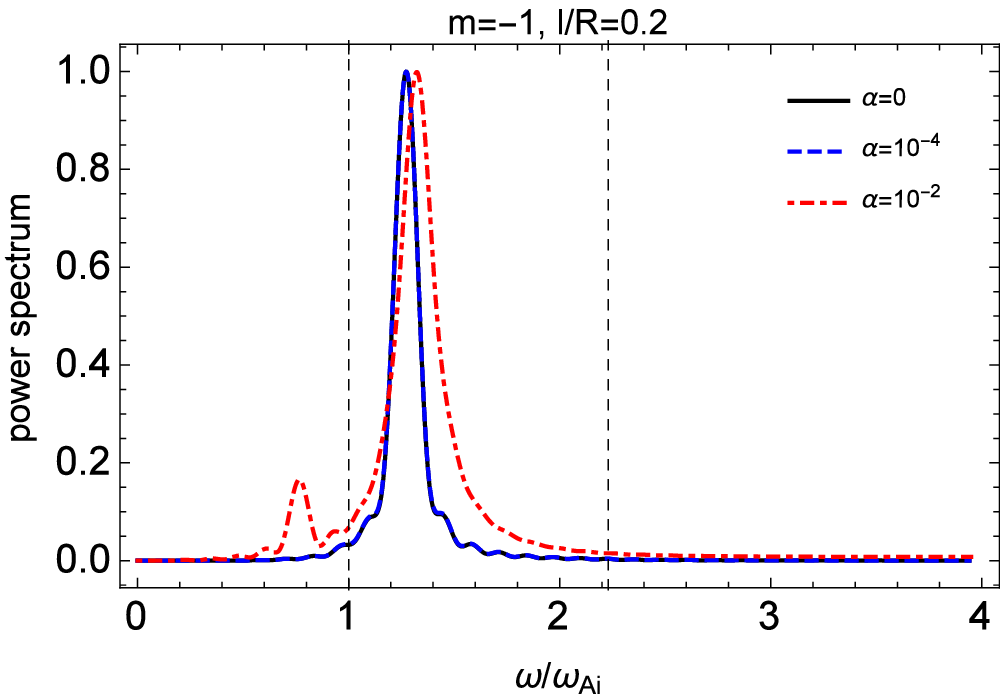}& \includegraphics[width=75mm]{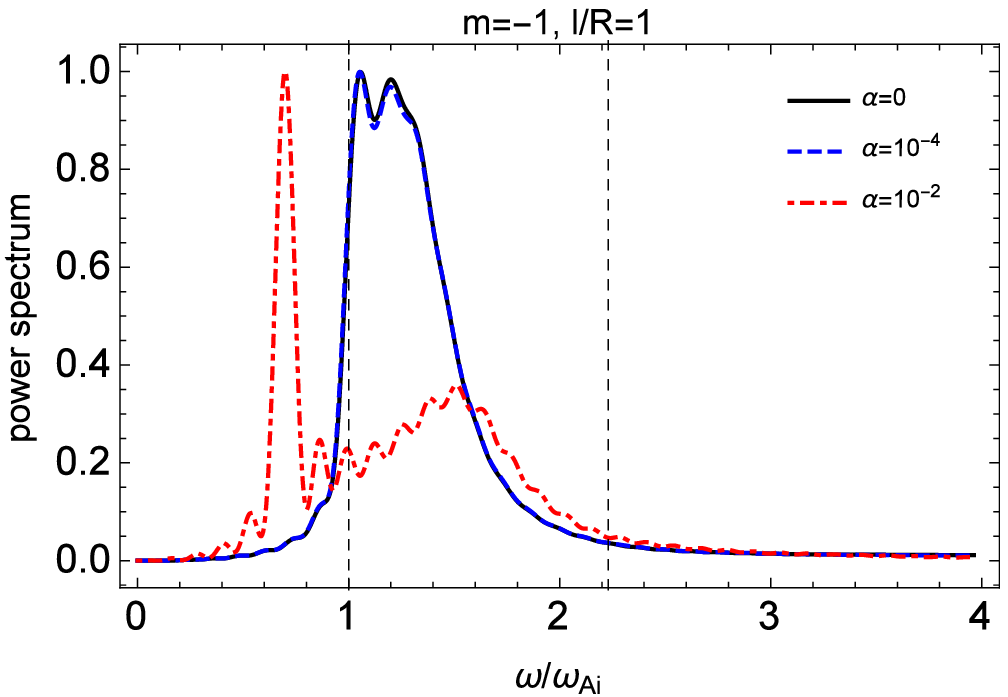}\\
    \end{tabular}
    \caption{Same as Fig. \ref{power_MN}, but for the model II.}
    \label{power_T2}
 \end{figure}
\begin{figure}
\centering
    \begin{tabular}{cc}
        \includegraphics[width=75mm]{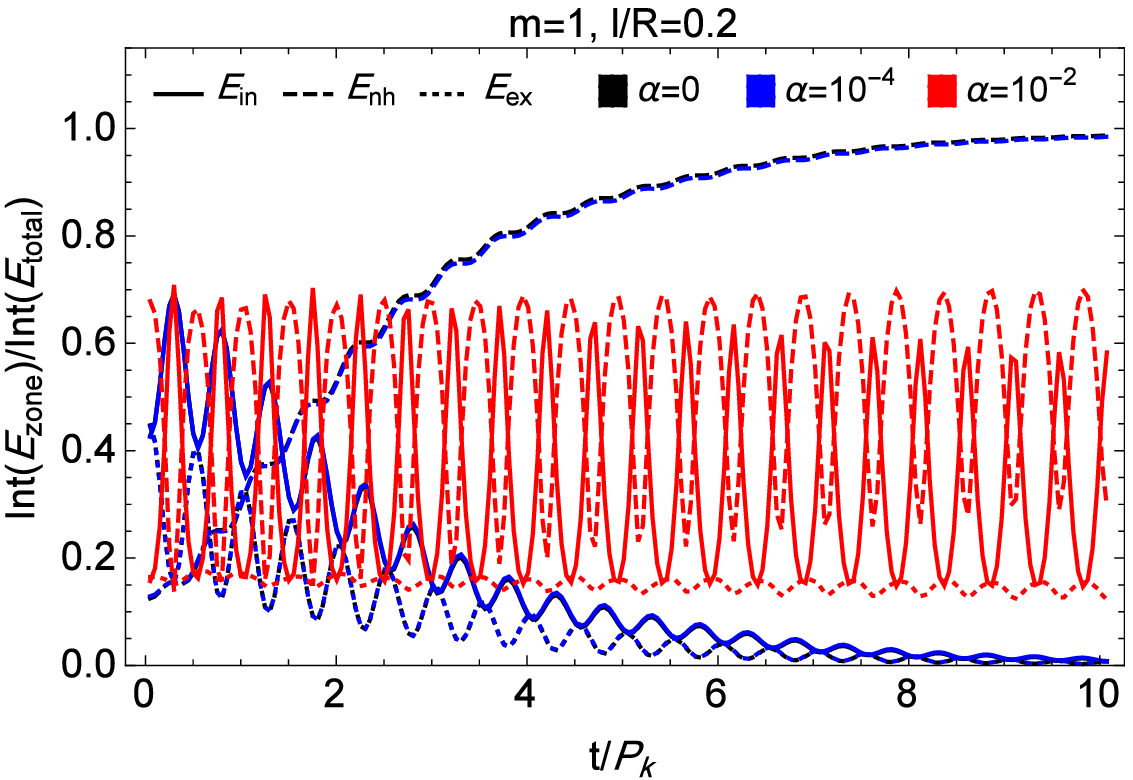}& \includegraphics[width=75mm]{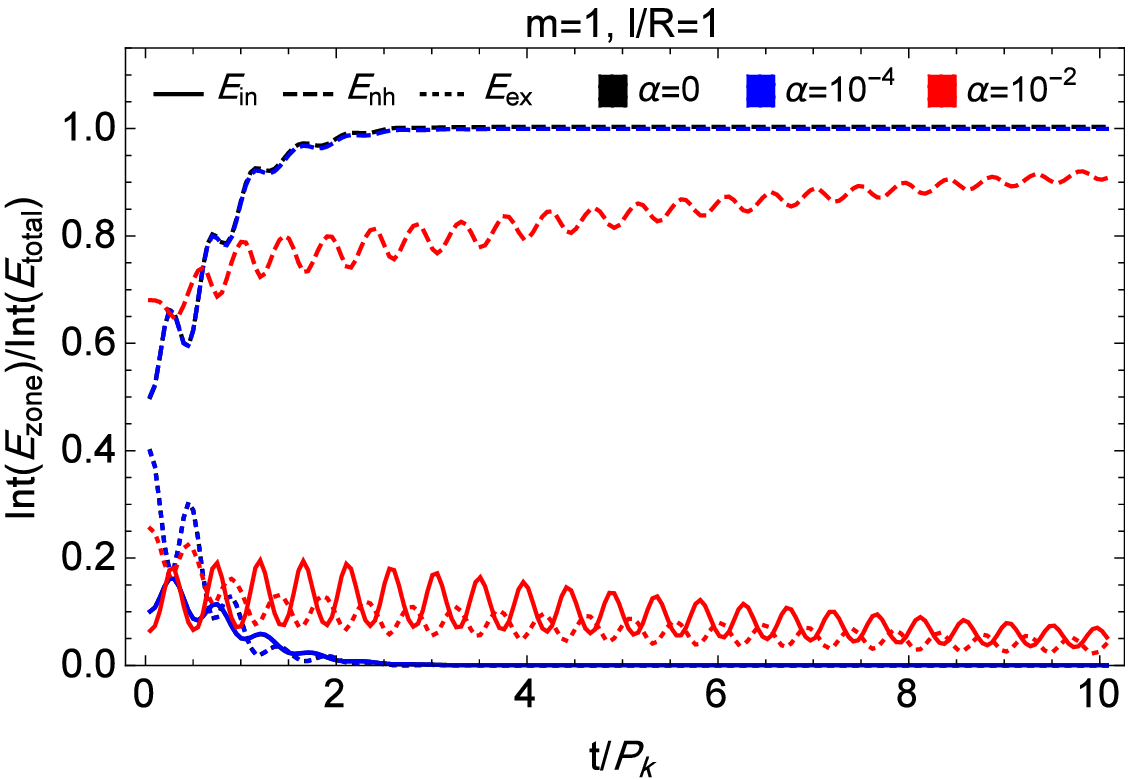}\\
        \includegraphics[width=75mm]{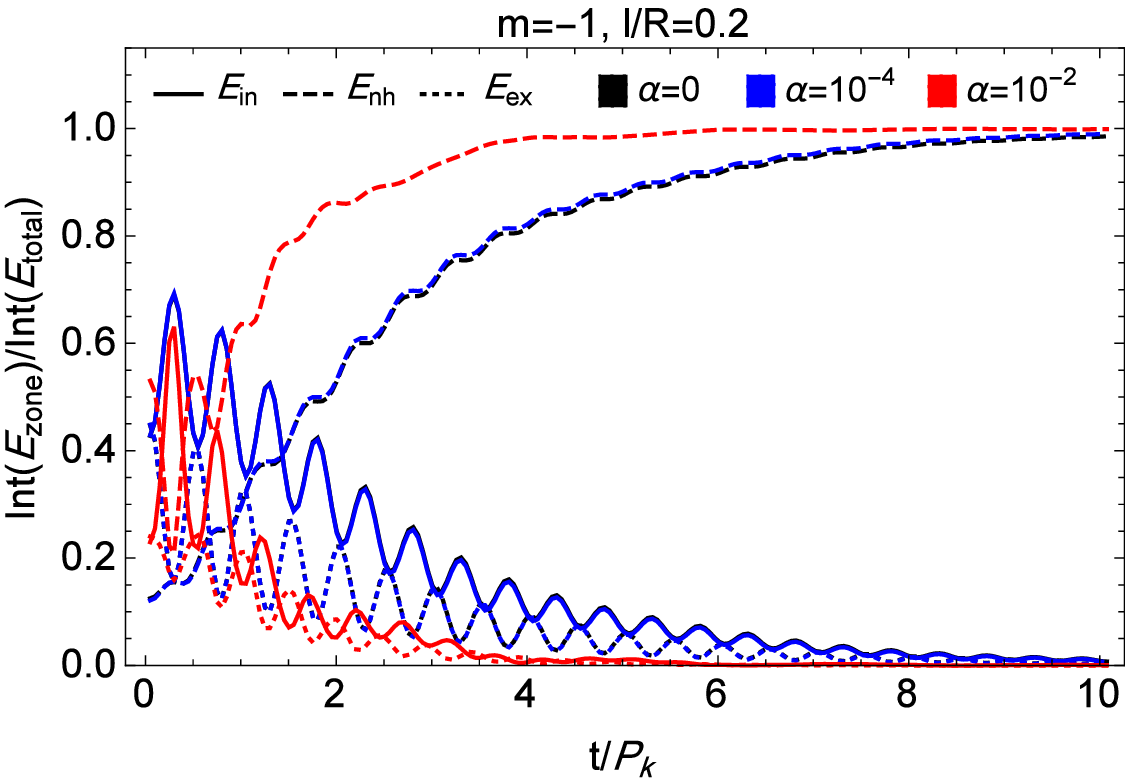}& \includegraphics[width=75mm]{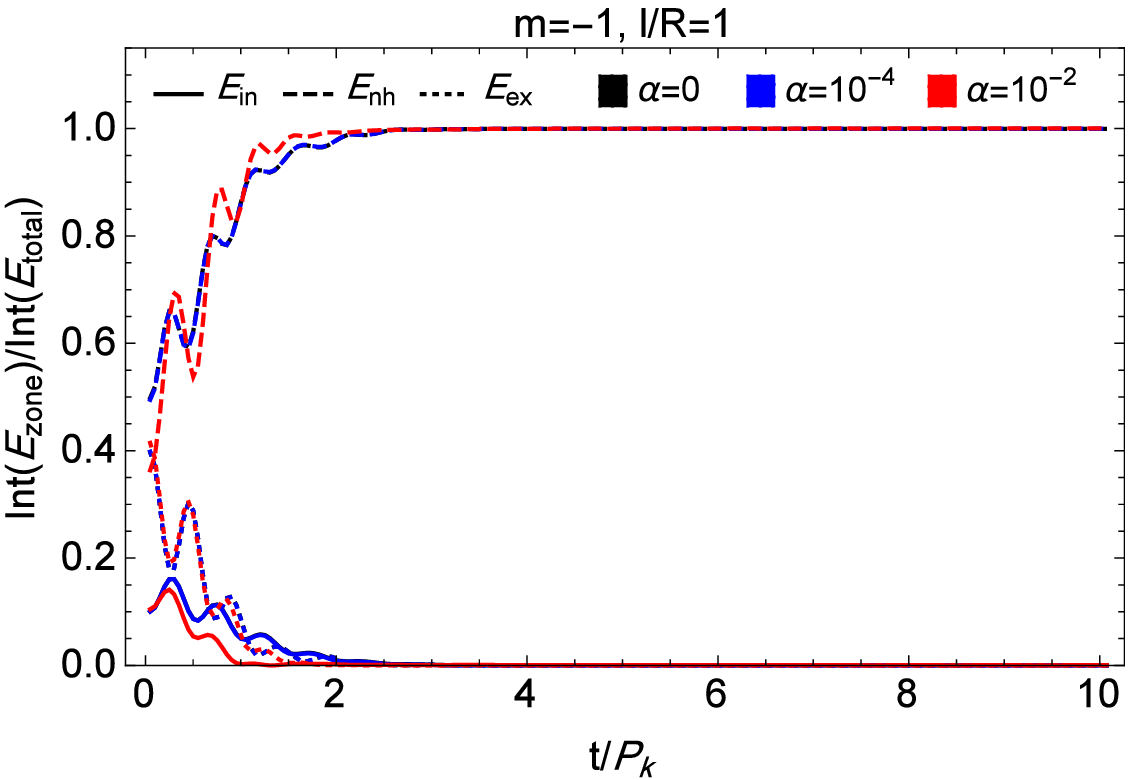}
    \end{tabular}
    \caption{Same as Fig. \ref{intE_MN}, but for the model II.}
    \label{intE_T2}
 \end{figure}
Here, we solve Eq. (\ref{eig}) for $\alpha=0, 10^{-4}, 10^{-2}$, and $N=101$. Figures \ref{omega_T2}-\ref{intE_T2} show the results for $m=\pm1$ and $l/R=0.2,~1$. As illustrated in Fig. \ref{omega_T2}, the background Alfv\'{e}n frequency for model II is not a monotonic function of $r/R$ for both $m=\pm1$. As a result, for the $\omega_n$s values that match the background Alfv\'{e}n frequency at two positions, there should be two singularities in the corresponding $\phi_n(r)$ function. For instance, the dotted line in the bottom right panel of Fig. \ref{omega_T2} displays the value of $\omega_{50}$ for $m=-1$, $l/R=1$ and $\alpha=10^{-2}$ that matches the corresponding background Alfv\'{e}n frequency at two points (big crosses). Figure \ref{phi50_T2_m(-1)} shows that the $\phi_{50}(r)$ eigenfunction of the Alfv\'{e}n discrete  mode becomes singular at two locations where the $\omega_{50}$ matches the background Alfv\'{e}n frequency.

Figure \ref{cn_T2} shows the normalized values of $|c_n|$ versus their corresponding eigenfrequencies, $\omega_n$. The figure presents that for the both $l/R=0.2,~1$, when $m=+1$, with increasing the twist parameter, the peaks of diagrams shift slightly toward larger frequencies and the frequency distribution of the Alfv\'{e}n continuum modes becomes narrower. For $m=-1$ and $\alpha=10^{-2}$ there are two peaks in the diagrams: a wide peak that is shifted to the larger frequencies and a narrow peak located at the beginning of the plots that is shifted to the smaller frequencies. Note that the frequency of the narrow peak corresponds to the minimum value of the Alfv\'{e}n frequency in the nonuniform region (see the bottom panel of Fig. \ref{omega_T2}). Although the narrow peaks are not well resolved in the plots and are illustrated by only one value of $c_n$'s (look at the single discontinuous red asterisk on the left side of the plots), the effect of this peak can be seen in the power spectrum of $\xi_r(r=0)$ (see the bottom panel of Fig. \ref{power_T2}). Note that the height of the narrow peak is smaller and larger than that of the wide peak for $l/R=0.2,~1$, respectively.

Figures \ref{xi_T2_m(1)_lbyR(0.2)}-\ref{xi_T2_m(-1)_lbyR(1)} show the temporal evolution of different components of the Lagrangian displacement for (i) $m=+1$ \& $l/R=0.2$, (ii) $m=+1$ \& $l/R=1$, (iii) $m=-1$ \& $l/R=0.2$ and (iv) $m=-1$ \& $l/R=1$, respectively. As illustrated in these figures, in the presence of magnetic twist, at a given $t/P_k$, for $m=+1/-1$, the perturbations are less/more phase-mixed than those in the case of untwisted magnetic field ($\alpha=0$).

Figure \ref{xir_T2} exhibits the temporal evolution of the radial component of the displacement on the loop axis for $m=\pm1$ and $l/R=0.2,~1$. The figure shows that for $m=+1/-1$, when the twist parameter increases, the decay rate of the perturbations decreases/increases. As illustrated in the bottom panel of Fig. \ref{xir_T2}, for $m=-1$, when the twist parameter increases there are two phases of oscillations for both $l/R=0.2,~1$. In the first phase the perturbations have slightly larger frequencies and decay faster than those in the case of untwisted magnetic field. In the second phase, the perturbations are almost decayless and have larger frequencies than those of in the case of untwisted magnetic field. These phases correspond to the wide and narrow peaks in the $c_n$'s distribution (see the bottom panel of Fig. \ref{cn_T2}). Figure \ref{power_T2} shows the power spectrum of $\xi_r(r=0)$ for $m=\pm1$ and $l/R=0.2,~1$. It is clear in the figure that for $m=-1$ and $l/R=0.2,~1$, when the twist parameter increases, there are two frequencies corresponding to two oscillating phases in Fig. \ref{xir_T2} and two peaks of the $c_n$'s distribution in Fig. \ref{cn_T2}. Figures \ref{xir_T2} and \ref{power_T2} show that for $m=+1$ and $l/R=0.2,~1$, when the twist parameter increases, the loop axis oscillates with a frequency higher than the case with no twist.

Figure \ref{intE_T2} illustrates the integrated total (kinetic plus magnetic) energy calculated by Eq. (\ref{Eint}) in the internal, nonhomogeneous and external regions as a function of time for $m=\pm1$ and $l/R=0.2,~1$. It is clear from the figure that, for $m=-1$ and $l/R=0.2$ (bottom left panel), the magnetic twist enhances the rate of the energy flux toward the nonhomogeneous region but for $m=-1$ and $l/R=1$ (bottom right panel), there is not a big difference between the twisted and untwisted case except for the internal energy. As illustrated in the figure, for $m=+1$, the rate of energy transfer to the nonhomogeneous region decreases for both $l/R=0.2$ and 1.

\subsection{Comparison with the results of Terradas \& Goossens (2012)}\label{TG}

Terradas \& Goossens (2012) investigated the effect of twisted magnetic field, Eq. (\ref{Bphi2}), on the MHD kink waves in a coronal flux tube. They solved the linearized MHD eigenvalue problem numerically and showed that in a magnetically twisted flux tube, the frequency of kink waves depends on the propagation direction. For instance, for a given value of the longitudinal wavenumber, $k_z$, the quasi-mode frequency of the kink waves has different values depending on the sign of the azimuthal mode number $m=\pm1$. Terradas \& Goossens (2012) found that for $m=1$ and $m=-1$, the corresponding frequencies are larger and smaller than that in the case without twist, respectively. Now, we are interested in recovering this result in our work. To this aim, following Terradas \& Goossens (2012), we take $k_zR=\pi/50$ and $\rho_{\rm i}/\rho_{\rm e}=3$. We also set $r_1=0.5$ and $r_2=1.5$ in Eq. (\ref{Bphi2}). In the model of Terradas \& Goossens (2012), because of the existence of the discontinuous piecewise plasma density, the background Alfv\'{e}n frequency is discontinuous at $r=R$. Hence, to avoid of this discontinuity, we replace and approximate the piecewise density profile, Eq. (1) in Terradas \& Goossens (2012), with a sinusoidal density profile that has a thin transitional layer $l/R=0.1$.

\begin{figure}
\centering
    \includegraphics[width=100mm]{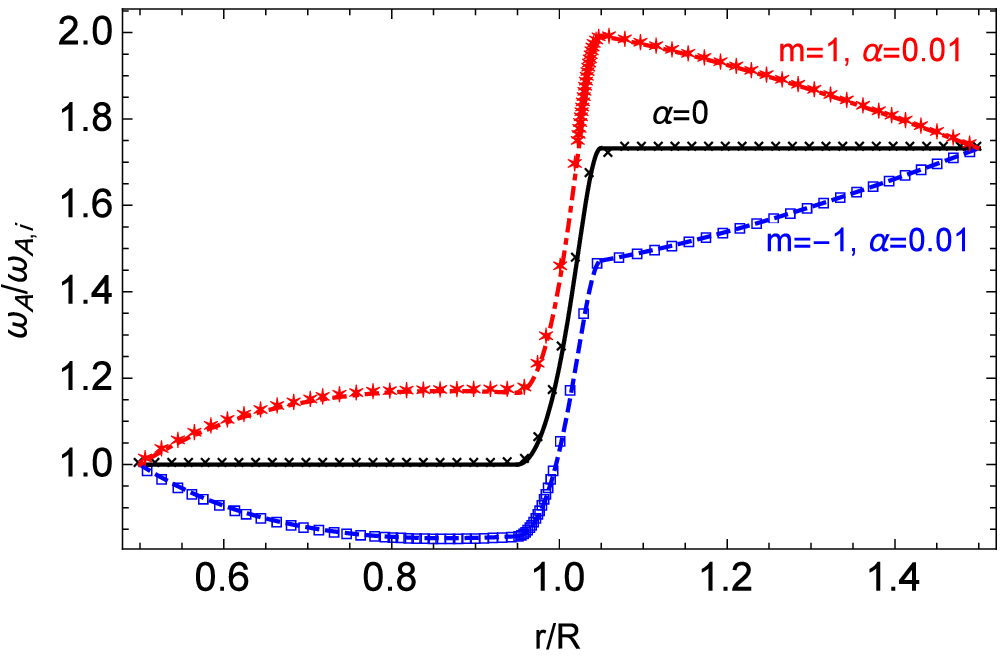}
    \caption {Background Alfv\'{e}n frequency and the corresponding eigenfrequencies of the model of Terradas \& Goossens (2012). Note that here we have approximated the piecewise step function density profile of Terradas \& Goossens (2012) with a continuous density profile that has a sharp variation in a thin layer of thickness $0.1R$.}
    \label{omega_TG}
 \end{figure}
\begin{figure}
\centering
    \includegraphics[width=100mm]{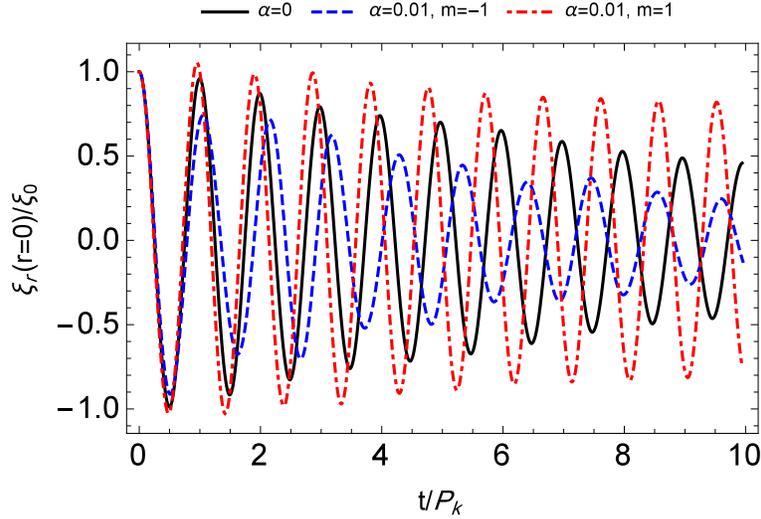}
    \caption{Temporal evolution of $\xi_r(r=0)/\xi_0$ for $\alpha=0$ (solid line), $\alpha=0.01$ \& $m=-1$ (blue dashed line) and $\alpha=0.01$ \& $m=1$ (red dot-dashed line).}
    \label{xir_TG}
 \end{figure}
\begin{figure}
\centering
    \begin{tabular}{cc}
        \includegraphics[width=100mm]{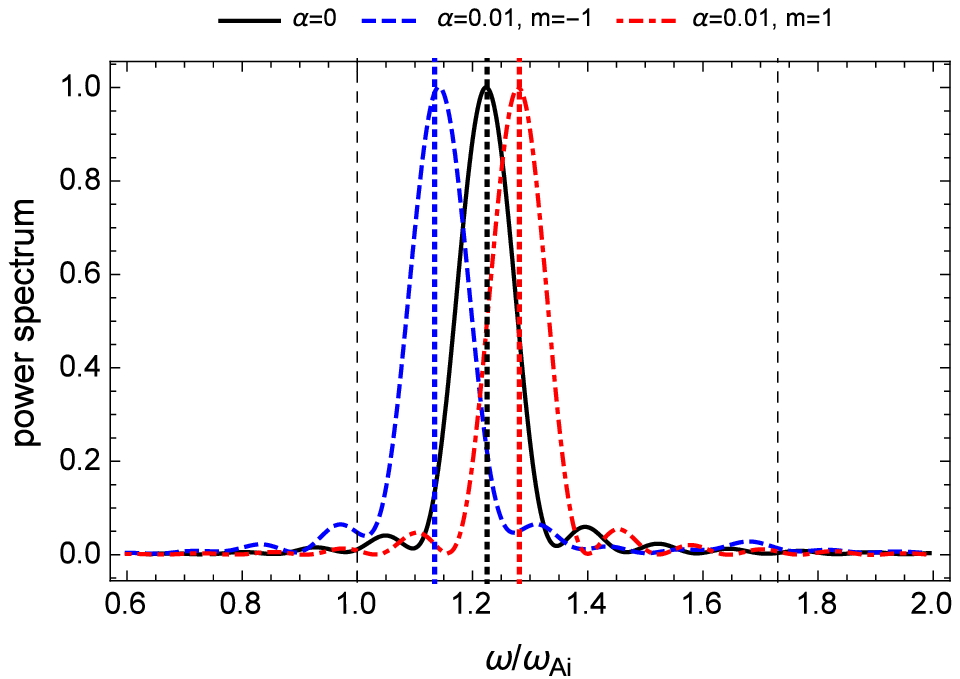}
    \end{tabular}
    \caption{Power spectrum of $\xi_r(r=0)$ for $\alpha=0$ (solid line), $\alpha=0.01$ \& $m=-1$ (blue dashed line) and $\alpha=0.01$ \& $m=1$ (red dot-dashed line). The vertical black, blue and red dotted lines are the frequencies obtained by Terradas \& Goossens (2012) for $\alpha=0$, $\alpha=0.01$ \& $m=-1$ and $\alpha=0.01$ \& $m=1$, respectively. The left and right vertical dashed lines are the interior and exterior Alfv\'{e}n frequencies, respectively.}
    \label{power_TG}
 \end{figure}

In the following, we solve Eq. (\ref{eig}) for $m=\pm1$, $\alpha=0,~0.01$, and $N=101$ and compare our results with those obtained by Terradas \& Goossens (2012). Figure \ref{omega_TG} shows the background Alfv\'{e}n frequency and the corresponding eigenfrequencies of the Alfv\'{e}n modes. Figure \ref{xir_TG} exhibits the temporal evolution of the radial component of the displacement on the loop axis. As illustrated in Fig. \ref{xir_TG}, the decay rate of the perturbations in the presence of magnetic twist for $m=-1$ and $m=+1$ is higher and lower than that of in the case without twist, respectively. Figure \ref{power_TG} represents the power spectrum of $\xi_r(r=0)$ in the time interval $t\in[0,10P_k]$ for $\alpha=0,~0.01$ and $m=\pm1$. As shown in this figure, in the presence of magnetic twist, the peak frequency of the power spectrum for $m=1$ and $m=-1$ is larger and smaller than that in the case of untwisted magnetic field, respectively. Note that the middle vertical dotted line represents the so-called kink frequency, Eq. (\ref{omegakink}), obtained for an untwisted thin magnetic flux tube with a piecewise plasma density profile. The left and right vertical dotted lines represent the frequencies of the MHD kink waves obtained by Terradas \& Goossens (2012) for $m=-1$ and $m=+1$, respectively, with the twist parameter $\alpha=0.01$. Therefore, Fig. \ref{power_TG} clearly shows that the result of Terradas \& Goossens (2012) is completely recovered in our work.

\section{Conclusions}\label{Conclusions}

Here, we investigated the effect of twisted magnetic field on the phase-mixing and resonant absorption of the propagating MHD kink waves in coronal flux tubes. The mathematical approach used in this paper is based on the work of Cally (1991) (in Cartesian coordinates) and ST2015 (in cylindrical coordinates). We solved an initial-value problem using the linear ideal MHD equations. Hence, our results cannot be extended to the large times when the strong phase-mixing develops in the system. The reason is that in this limit, due to the strong phase-mixing of the Alfv\'{e}n waves, the viscous and resistive dissipation mechanisms become significant even though the dissipation coefficients are small in the corona (see, e.g. Heyvaerts \& Priest 1983; Karami \& Ebrahimi 2009).

Following ST2015, in order to find the temporal and spatial behaviour of the kink perturbations, using a modal expansion technique, we solved the linear incompressible MHD equations in the nonuniform region of a coronal flux tube that has both radial density variation and magnetic field twist. In order to simplify the MHD equations, we used the thin tube approximation (i.e. $k_z R\ll 1$) in our analysis. Thus, our results are only applicable to the propagating kink MHD waves in the limit of long wavelengths, i.e., when the wavelengths of the waves are much larger than the thickness of the loop.

We considered two types of twisted magnetic fields containing the discontinuous and continuous ones to investigate how different magnetic field profiles affect the resonant absorption and phase-mixing of the MHD kink waves in coronal loops. Also, we examined the effect of magnetic twist on the kink waves in the cases of both thin and thick nonuniform layers. In order to prevent the kink instability, the amount of the magnetic twist in the flux tube must be restricted with a maximum value that is obtained from a stability analysis. Investigating a stability analysis for the models considered here is beyond the scope of the present work. Instead, we considered the twist parameters small enough in order to be in the range of stability obtained in previous works.

One of the interesting effects of the twisted magnetic fields in coronal flux tubes is the asymmetry of the phase speed of the MHD kink waves with respect to the propagation direction (see, e.g., Terradas \& Goossens 2012; Ruderman 2015). Hence, in order to investigate this effect, we considered two propagation directions (i) $k_z R>0$ \& $m=+1$ and (ii) $k_z R>0$ \& $m=-1$ for the kink MHD wave in both continuous and discontinuous magnetic field models.\\

For the model I (discontinuous magnetic field), we found the following:

\begin{itemize}

\item  By increasing the magnetic twist parameter in the loop, the decay rate of the radial component of the Lagrangian displacement on the axis of the flux tube increases/decreases for $m=+1/-1$.

\item The power spectrum of $\xi_r(r=0)$ shows that by increasing the twist parameter, the effective frequency of the kink wave increases/decreases when $m=+1/-1$. Hence, in the presence of a twisted magnetic field, the frequency of the kink wave is asymmetric with respect to the propagation direction. This is in agreement with the result obtained by Terradas \& Goossens (2012) and Ruderman (2015).

\item When the twist parameter increases, for both $l/R=0.2$ and 1, a wider/narrower range of the Alfv\'{e}n continuum modes contributes to the total displacement of the kink waves for $m=+1/-1$.

\item The rate of phase-mixing of the perturbations increases/decreases for $m=+1/-1$ as the twist parameter increases in the loop. The reason is that when the twist parameter increases, for $m=+1/-1$ the slope of the profile of the Alfv\'{e}n frequency increases/decreases.

\item The rate of energy flux from the interior and exterior regions of the loop toward the nonuniform region increases/decreases for $m=+1/-1$ when the twist parameter increases. As the energy of the kink wave transfers to the nonuniform region, the amplitude of perturbations inside and outside the loop decreases. The energy mostly transfers to the azimuthal component of the perturbations in the nonuniform region which is subjected to phase-mixing owing to the existence of an inhomogeneous background Alfv\'{e}n frequency across the loop.
\end{itemize}

For the model II (continuous magnetic field), the results show the following:

\begin{itemize}

\item With increasing the twist parameter, for $m=+1$ and $l/R=0.2,1$, a narrower range of the Alfv\'{e}n continuum modes contributes to the total displacement of the kink waves. But for $m=-1$ and $l/R=0.2,1$ there are two peaks in the distribution profile of the Alfv\'{e}n continuum modes. In comparison with the case of untwisted magnetic field, one of these peaks is wider and shifts to higher frequencies and another one is narrower and shifts to lower frequencies.

\item  When the twist parameter increases, the decay rate of the radial component of the Lagrangian displacement on the axis of the flux tube decreases/increases for $m=+1/-1$.

\item For $m=+1$ and $l/R=0.2,1$, when the magnetic twist increases, the power spectrum of $\xi_r(r=0)$ becomes narrower and shifts toward higher frequencies. When $m=-1$, for both $l/R=0.2,~1$, the power spectrum splits into two peaks: a wider peak in higher frequencies and a narrower peak in lower frequencies with respect to the case of untwisted magnetic field.

\item The rate of phase-mixing of the perturbations decreases/increases for $m=+1/-1$ as the twist parameter increases in the loop.

\item The rate of energy flux from the interior and exterior regions of the loop toward the nonuniform region decreases/increases for $m=+1/-1$ when the twist parameter increases.

\end{itemize}
In the case of the discontinuous magnetic field model, by increasing the twist parameter for the both cases of thin and thick nonuniform layers, the peak frequency of the power spectrum increases/decreases for $m=+1/-1$. This is in agreement with that obtained using the quasi-mode approach (see, e.g., Karami \& Bahari 2010; Terradas \& Goossens 2012; Ebrahimi \& Karami 2016). However, for the continuous magnetic field model, the situation is different. In this case, when the twist parameter increases, for $m=-1$ the single peak of the power spectrum for $l/R=0.2$ splits into two peaks located at higher and lower frequencies than that of in the case of no twist. The reason is that the oscillation of the radial component of the Lagrangian displacement has two phases for $m=-1$. The first and second phases, respectively, have smaller and larger frequencies than those in the case of untwisted magnetic field. This result also holds in the case of thick nonuniform layer $l/R=1$. In this case, when the twist is absent, the power spectrum has two peaks. When the twist parameter increases, these peaks get away from each other and move toward higher and lower frequencies. These results point out the important effect that the particular twist model has on the behavior of kink waves. The ignorance of the actual twist profile in coronal loops turns out to be very important in this regard.

Applying the modal expansion approach to the model of Terradas \& Goossens (2012), we found that when the twist parameter increases, the peak frequency of the power spectrum for $m=1$ and $m=-1$ shifts toward the higher and lower frequencies, respectively. This is in well agreement with the result obtained by Terradas \& Goossens (2012).

As illustrated by Terradas \& Goossens (2012), in the presence of magnetic field twist, quasi-mode frequencies of MHD kink waves obtained for $k_z>0$ \& $m=\pm1$ are the same as for $k_z<0$ \& $m=\mp1$. To investigate this symmetry in our work, we obtained the results for $k_z<0$ \& $m=\pm1$ in both models. We have not included these results here for the sake of simplicity. We found that, besides the effective frequency of the kink waves, the whole properties of the evolution of the propagating kink waves are symmetric if we change the signs of $k_z$ and $m$, simultaneously. It is not straightforward to translate the present results for propagating waves to the case of standing oscillations line-tied at the ends of the tube. In the case of standing oscillations, the perturbations necessarily contain the two possible signs of $k_z$ and $m$. Hence, in the case of standing waves it is not simple to deduce the net effect that the effect of twist would have on the process of phase-mixing. However, the present results suggest that the effect of twist can be relevant for standing waves as well. Further investigation in this direction is needed.

It is worth to mentioning that during kink oscillations of coronal loops the Kelvin-Helmholtz instability (KHI) can occur around the boundary of the flux tube (Heyvaerts \& Priest 1983). The torsional motions, which are amplified in the inhomogeneous region of the flux tube, introduce velocity shears that are liable to be unstable to KHI. Since the observation of KHI has not been reported to date in coronal flux tubes, it is believed that some mechanism is able to suppress it. It is known that the existence of a component of a magnetic field aligned with the direction of the velocity shears has a stabilizing effect and can restrain the KHI (e.g. Chandrasekhar 1961). Soler et al. (2010) showed that a very small amount of magnetic twist, which is very likely and realistic in coronal flux tubes, can suppress the KHI in a cylindrical flux tube. Therefore, the lack of KHI can be one of the possible indirect confirmations of the existence of magnetic twist in coronal loops.

Twist of the magnetic field, even in a small amount, can have a significant impact on the generation of small scales and the energy cascade from the global kink motion to the small scales. This has implications concerning the efficiency of the phase-mixing process and the ability of the process to feed energy to the dissipative scales, where plasma heating takes place. In this paper we have considered a simple scenario to investigate the effect of twist, in order to pave the way for future works that should tackle the full nonlinear 3D problem.

\section*{Acknowledgements}

The authors thank the anonymous referee for very valuable comments. R.S. acknowledges the support from grant AYA2014-54485-P (AEI/FEDER, UE) and from the Ministerio de Econom\'{\i}a, Industria y Competitividad, and the Conselleria d'Innovaci\'{o} Recerca i Turisme del Govern Balear (Pla de ci\`{e}ncia, tecnologia, innovaci\'{o} i emprenedoria 2013-2017) for the Ram\'{o}n y Cajal grant RYC-2014-14970.

\end{document}